


\newif\iffigs\figstrue

\input harvmac

\iffigs
  \input epsf
\else
  \message{No figures will be included. See TeX file for more
information.}
\fi

\noblackbox
\def\CY{Calabi--Yau}
\def\zbar{{\bar z}}
\def\thetabar{{\bar \theta}}
\def\chibar{{\bar\chi}}
\def\taubar{{\bar\tau}}
\def\ibar{{\bar \imath}}

\def\phibar{{\bar\phi}}
\def\Phibar{{\overline\Phi}}
\def\sigmabar{{\bar\sigma}}
\def\alphad{{\dot\alpha}}
\def\CM{{\cal M}}
\def\CO{{\cal O}}
\def\CJ{{\cal J}}
\def\CK{{\cal K}}
\def\CT{{\cal T}}
\def\bN{{\bf N}}
\def\bM{{\bf M}}
\def\vn{{\vec n}}
\def\p{\partial}
\def\la {\langle}
\def\ra {\rangle}
\def\lal{\la \!\la }
\def\rar {\ra \!\ra }
\def\lall{\lal\!\la}
\def\rarr{\rar\!\ra}

\mathchardef\varLambda="0103
\mathchardef\varDelta="0101

\def\inbar{\,\vrule height1.5ex width.4pt depth0pt}
\def\IB{\relax{\rm I\kern-.18em B}}
\def\smallinbar{\,\vrule height1.1ex width.36pt depth0pt}
\def\IC{\relax\ifmmode\mathchoice
  {\hbox{$\inbar\kern-.3em{\rm C}$}}
  {\hbox{$\inbar\kern-.3em{\rm C}$}}
  {\,\hbox{$\scriptstyle\smallinbar\kern-.27em{\rm C}$}}
  {\,\hbox{$\scriptstyle\smallinbar\kern-.27em{\rm C}$}}
  \else{\hbox{$\inbar\kern-.3em{\rm C}$}}\fi}
\def\ID{\relax{\rm I\kern-.18em D}}
\def\IE{\relax{\rm I\kern-.18em E}}
\def\IF{\relax{\rm I\kern-.18em F}}
\def\IG{\relax\hbox{$\inbar\kern-.3em{\rm G}$}}
\def\IH{\relax{\rm I\kern-.18em H}}
\def\II{\relax{\rm I\kern-.18em I}}
\def\IK{\relax{\rm I\kern-.18em K}}
\def\IL{\relax{\rm I\kern-.18em L}}
\def\IM{\relax{\rm I\kern-.18em M}}
\def\IN{\relax{\rm I\kern-.18em N}}
\def\IO{\relax\hbox{$\inbar\kern-.3em{\rm O}$}}
\def\IP{\relax{\rm I\kern-.18em P}}
\def\IQ{\relax\hbox{$\inbar\kern-.3em{\rm Q}$}}
\def\IR{\relax{\rm I\kern-.18em R}}
\font\cmss=cmss10 \font\cmsss=cmss10 at 7pt
\def\IZ{\relax\ifmmode\mathchoice
{\hbox{\cmss Z\kern-.4em Z}}{\hbox{\cmss Z\kern-.4em Z}}
{\lower.9pt\hbox{\cmsss Z\kern-.4em Z}} {\lower1.2pt\hbox{\cmsss
Z\kern-.4em Z}}\else{\cmss Z\kern-.4em Z}\fi}
\def\IGa{\relax\hbox{${\rm I}\kern-.18em\Gamma$}}
\def\IPi{\relax\hbox{${\rm I}\kern-.18em\Pi$}}
\def\ITh{\relax\hbox{$\inbar\kern-.3em\Theta$}}
\def\IOm{\relax\hbox{$\inbar\kern-3.00pt\Omega$}}
\def\CP#1{\relax\hbox{$\hbox{\IP}^{#1}$}}
\def\WCP#1#2{\relax\hbox{$\hbox{\IP}_{#1}^{#2}$}}

\def\Res{{\rm Res}}
\def\mult{{\rm mult}}
\def\ztim{\mathrel{\mathop \otimes_{\IZ}}}
\def\boldone{\relax{\rm 1\kern-.35em 1}}

\Title{\vbox{\baselineskip12pt\hbox{DUKE-TH-94-78}
\hbox{IASSNS-HEP-94/82}\hbox{hep-th/9412236}}}
{\vbox{\centerline{Summing the Instantons: Quantum Cohomology and}
\vskip2pt\centerline{Mirror Symmetry in Toric Varieties}
}}

\centerline{
David R. Morrison$^{a,b}$ and
M. Ronen Plesser$^c$
}

\vskip.3in {\ninepoint
\centerline{$^a$ School of Mathematics, Institute for Advanced Study,
Princeton, NJ 08540}
\centerline{$^b$ Department of Mathematics, Box 90320, Duke University, Durham,
NC 27708}
\centerline{$^c$ School of Natural Sciences, Institute for Advanced Study,
Princeton, NJ 08540}
}

\vskip.5in
\centerline{\bf Abstract}
\bigskip
We use the gauged linear sigma model introduced by Witten to calculate
instanton expansions for correlation functions in topological sigma
models with target space a toric variety $V$ or a Calabi--Yau
hypersurface $M\subset V$. In the linear model
the instanton moduli spaces are relatively simple objects and the
correlators are explicitly computable; moreover, the instantons can be
summed, leading to explicit solutions for both kinds of models.  In the case of
smooth $V$, our results reproduce and clarify an algebraic
solution of the $V$ model
due to Batyrev.
In addition,  we find an algebraic relation
determining the solution for $M$ in terms of that for $V$.
Finally, we propose a modification of the linear model
which computes instanton expansions about any limiting point in the
moduli space. In the smooth case this leads to a (second) algebraic
solution of the $M$ model. We use this description to prove some
conjectures about mirror symmetry, including the previously conjectured
``monomial-divisor mirror map'' of Aspinwall, Greene, and Morrison.

\Date{12/94}

\newsec{Introduction}

\nref{\rnonren}{M. T. Grisaru, W. Siegel, and M. Ro\v{c}ek, ``Improved methods
for supergraphs,'' Nucl.\ Phys.\
{\bf B159} (1979) 429--450; 
for a recent version with a simpler proof see
N. Seiberg, ``Naturalness versus supersymmetric non-renormalization
theorems,'' Phys.\ Lett.\ {\bf 318B} (1993) 469--475, hep-ph/9309335.}

\nref\rmir{See {\sl Essays on mirror manifolds\/}, S.-T. Yau, ed.
International Press, Hong Kong, 1992; 
{\sl Essays on mirror manifolds II\/}, B. R. Greene and S.-T. Yau, eds.
International Press, Hong Kong, to appear.}

Supersymmetric nonlinear sigma models with
K\"ahler target space form an interesting class of nontrivial field
theories in two spacetime dimensions. One reason for this interest is
that geometrical intuitions can be fruitfully applied in studying
them.  The nonrenormalization theorems \rnonren\ following from the $N=2$
supersymmetry show that a subset of the correlation
functions in these theories are not corrected from their classical
values at any order in perturbation theory. The classical values can
be interpreted in terms of geometrical properties of the target space
and thus some of these are directly manifest in the physical
theory. When the target space is a \CY\ manifold the theory is in fact
conformally invariant (for an appropriate choice of metric); such
models comprise a large class of consistent string vacua for which the
geometric interpretation yields useful insights. One striking example
of this is mirror symmetry \rmir\ which relates pairs of topologically
distinct manifolds leading to isomorphic conformal field theories (but
with a sign change in a certain quantum number).

\nref\rtopsim{E. Witten, ``Topological sigma models,'' Commun.\ Math.\
Phys.\ {\bf 118} (1988) 411--449.}

The correlation functions in the distinguished set referred to above
are in fact independent of the worldsheet metric. (We will follow the
conventions of string theory and call the two-dimensional manifold
$\Sigma$ on which our field theory is defined the worldsheet.) There
is a simpler version of the theory \rtopsim\ which isolates these
correlators, obtained by modifying the spins of the
fields and performing a projection on the space of states. (This is
called ``twisting'' the theory.)  The result is a topological field
theory. The correlation functions in this theory are topological
invariants of the target space $X$ corrected by instanton
contributions, which themselves have a more elaborate geometrical
interpretation as intersection numbers in spaces of maps $\Sigma\to
X$. Thus, in principle the model is completely solved in terms of some
invariants of the target space $X$.  In practice, however, computing
these rather complicated invariants of mapping spaces is a daunting
task. The first few terms in the instanton expansion have been
computed for a number of examples, and some progress has been made in
developing more powerful tools, but in general the problem is far from
solved.

\nref\rWitmir{E. Witten, ``Mirror manifolds and topological field
theory,'' in {\sl Essays on Mirror Manifolds\/}, S.-T. Yau, ed.
International Press, Hong Kong, 1992, pp.~120--159, hep-th/9112056.}
\nref\rDG{J. Distler and B. Greene, ``Exact results on the
superpotential from Calabi--Yau compactifications,'' Nucl.\ Phys.\
{\bf B309} (1988) 295--316.}

When the target space is \CY\ there is a choice of metric for which
the sigma model is conformally invariant; in fact one then obtains
invariance under the $N=2$ superconformal algebra. Since supersymmetry
transformations are generated by chiral currents, which can be either
left- or right-moving on the worldsheet, there are two sets of
operators whose correlators are protected by the $N=2$
nonrenormalization theorems.  Correspondingly, there are two ways to
twist the superconformal field theory to obtain a topological
model. These were called the {\bf A} and {\bf B} model in \rWitmir ;
the {\bf A} twist, discussed in the previous paragraph, is defined for
any almost complex target space.  The two twists are related by mirror
symmetry so that if $M$ and $W$ are a mirror pair of \CY\ manifolds
then the {\bf A} model with target $M$ is isomorphic to the {\bf B}
model with target $W$. Correlation functions in the {\bf A} model
receive instanton corrections as mentioned, while correlators in the
{\bf B} model are given exactly by their semiclassical values
\rDG. This has been used to replace the difficult instanton
computations on $M$ with tractable computations on $W$. In these
applications one rewrites the sum over all instanton sectors in one
model as a classical computation in the other; in this sense one has
exactly summed the instanton series and produced exact correlation
functions in the conformal field theory. Interpreting the coefficients
in the instanton expansion in terms of the mapping spaces has thus far
been the most powerful method for computing the invariants involved.

\nref{\rphases}{E. Witten, ``Phases of $N{=}2$ theories in two
dimensions,'' Nucl. Phys. {\bf B403} (1993) 159--222, hep-th/9301042.}

An important development in the study of these models was the
discovery by Witten \rphases\ that for a particular class of target
spaces they can be obtained as the low-energy approximation to certain
two dimensional $N=2$ supersymmetric models with Abelian gauge
symmetry~--~we will call these ``gauged linear sigma models''
(GLSM). This formulation led immediately to several important insights
into the structure of the moduli space of these models \rphases . In
particular, the moduli space of the conformal field theory extends
beyond the radius of convergence of the instanton expansion. Since the
parameters are expressed as couplings in the GLSM Lagrangian we can
directly study the model in regions beyond the radius of convergence.
One finds that the low-energy dynamics in these regions does not
necessarily correspond to nonlinear sigma models but can be described
by other, less geometrical constructions. The different regimes in
which one description or another is valid were termed ``phases'' in
\rphases , and we will use this terminology. In some of these phases
the low-energy theory is of a type familiar from previous studies (such
as Landau--Ginzburg orbifold models) while in others we find
theories which are not as well understood.  An important observation
is that the gauged linear model can also be twisted to obtain a
topological version.  (We focus on the {\bf A} twist which is always
defined; the {\bf B} twist is only defined when certain
criteria, equivalent to the requirement that the low-energy nonlinear
model is conformal, are met.) Since correlation functions in this
topological field theory are completely independent of the worldsheet
metric they are in particular independent of scale and quantities
relevant to the low-energy effective theory can be computed directly
in the high-energy limit.

In this paper we will continue the program initiated in \rphases\ of
studying the topological nonlinear sigma models using the topological
gauged linear model. In particular, we will study the correlation
functions in the {\bf A} twisted model. We find that as in the
nonlinear case these are described in terms of a semiclassical result
corrected by instanton effects. In the present case the instantons are
in fact a subset of the familiar $U(1)$ instantons. A new feature of this
model is that the instanton corrections can be computed exactly, and
in fact the entire instanton series can be explicitly summed. The
crucial difference between the GLSM computation and the nonlinear
sigma model computation lies in the fact that the relevant instanton
moduli spaces in the GLSM (for gauge and matter degrees of freedom)
are {\it compact}\/ and relatively simple spaces. As mentioned above,
however, the GLSM correlation functions we compute are in fact
scale-independent. Thus our computations
necessarily agree with the corresponding
correlators in the nonlinear sigma model which is obtained at low
energies. So, in summing the series we have computed the correlation
functions in the nonlinear {\bf A} model directly. We stress that this
computation does {\it not}\/ involve mirror symmetry. The instanton
contributions are instead computed directly, then summed. Where the
mirror partner is known (or conjectured) we can compare our results to
those obtained from mirror symmetry and find complete agreement.

\nref\rdoit{D. R. Morrison and M. R. Plesser, work in progress.}%

The fields that survive the projection to the {\bf A} twisted GLSM are
twisted chiral fields. We find a rather simple description of the
dynamics of these fields which in some cases allows us to obtain the
exact correlation functions directly. In effect, we compute an exact
expression for the effective twisted superpotential interaction in the
low-energy theory. This computation is very suggestive of
mirror symmetry itself, in that we obtain the exact correlators directly. In a
forthcoming paper \rdoit\ we hope to make this connection explicit,
obtaining a direct argument for mirror symmetry in these models. (We
present some of the preliminary results of \rdoit\ at appropriate
points in the present paper.)
Even in cases where a direct computation of the exact result exists,
we find it instructive to also derive it as an explicit sum of instanton
contributions. This latter approach is at present more general, and it more
directly illustrates the modification to the nonlinear sigma model instanton
sum which has rendered it computable.

The topological GLSM correlation functions we compute are rational
functions of the parameters.  In particular, their only singularities
are poles and there is no difficulty in performing analytic
continuations of the series beyond their radius of convergence to the
entire {\bf A} model moduli space. Of course, this statement assumes a
particular choice of the coordinates on parameter space and of a basis
for the Hilbert space over every point in this space. The choice we
have made is familiar from studies of {\bf B} models; in some sense we
will have found the mirror, {\bf A} model interpretation of this
choice. We note that this is {\it not}\/ the choice usually assumed in
studies of the nonlinear {\bf A} model. In particular, therefore, the
coefficients in our series (which give the contributions from specific
instanton sectors) are not the invariants of mapping
spaces mentioned above. The two sets of computations are certainly
related, as we will discuss further below, but our current methods do
not suffice to extract one directly from the other.

In fact, we will find that the correlation functions can be determined
from a purely algebraic computation, which is probably as near to a
closed formula as one can expect to obtain. Using this formula we can
check the agreement with mirror constructions.
Namely, when the mirror manifold is conjecturally given by
a certain construction, we can use our computation of {\bf A} model
correlation functions on one member of the pair and compare to the
{\bf B} model correlators on the other. Where we can demonstrate
equality we will have gone a long way towards proving that the two are
related by mirror symmetry. We will succeed in doing this for a class
of models for which mirror constructions have been conjectured but not
previously proved, and indicate how the extension to a much broader
class of conjectured constructions could be performed.

\nref\rbator{V. V. Batyrev, ``Quantum cohomology rings of toric
manifolds,'' in {\sl Journ\'ees de G\'eom\'etrie Alg\'ebrique d'Orsay
(Juillet 1992)}, Ast\'erisque, vol. 218, Soci\'et\'e Math\-\'ematique
de France, 1993, pp.~9--34, alg-geom/9310004.}

In more detail, the structure of the paper is as follows. In section
two we will quickly review the construction of the nonlinear sigma
model and the twisted {\bf A} model, and their properties. This brief
review will be unnecessary for many readers; it is included to
establish notation and make the work more self-contained.  In section
three we use the linear model to study the nonlinear topological model
with target space a toric variety $V$. (This section begins with a
brief introduction to toric geometry.) These models are interesting
as solvable nonlinear sigma models. Our principal interest in this
work, however, is in superconformal models related to them; in
this section we will make a detailed study of the model which will lay
the groundwork for subsequent developments.
We find the structure of (part
of) the parameter space, construct the moduli spaces of instantons, and
sum the instanton series to compute the exact correlation
functions. We use this formulation of the model to reproduce and
clarify an algebraic construction of the solution for smooth $V$ first
proposed by Batyrev in \rbator. We show how the equations obtained by
Batyrev follow directly from the twisted superpotential, and also give
a derivation based upon the combinatorial structure of the instanton
moduli spaces (closer to the argument of \rbator ).
In section four we modify the model, following
\rphases , to obtain a theory whose low-energy approximation is the
nonlinear sigma model with target space $M$ a \CY\ hypersurface in
$V$. We use this modified model
to describe the structure of the moduli space.
Repeating the analysis of section three we are once more able
to compute the instanton corrections, sum the series, and compute the
exact correlation functions. Finally, we derive an algebraic formula
for the solution of the $M$ model in terms of the $V$ model. In
section five we present a different approach to the same model,
related to the nonlinear sigma model with target space $V^+$ the
(noncompact \CY ) total space of the canonical line bundle of $V$. We
show that in this model instanton expansions can be computed about the
semiclassical limit point in any phase. Further, we rewrite the
correlation functions in terms of ``expectation functions'' in an
algebra which we construct. We then apply this formulation to proving
mirror conjectures, by comparing our algebraic formula for the
solution to existing algebraic formulae for {\bf B} model
correlators. For smooth $V$ (where our algebraic solution is valid)
and in the most general case for which an analogous solution to the
{\bf B} model on the conjectured mirror manifold $W$ is known, we
prove that the correlation functions in the two theories are equal.
In section six we discuss some questions raised by these results and
discuss open problems and directions for future work.

In most of the paper we restrict attention to a worldsheet
of genus zero. The methods can be generalized to higher genus
worldsheets, and we point out the generalization where appropriate. We
note, however, that at higher genus the correlation functions of the
twisted model will no longer coincide with those of the superconformal
model.

Throughout the paper we illustrate our results by studying two
particular examples, the simplest known examples of one- and
two-parameter families of three-dimensional \CY\ hypersurfaces in
toric varieties. Each development is explicitly illustrated for these
examples immediately after its introduction. We hope this format,
while cumbersome, will help to clarify the discussion. We stress that
the restriction to one- and two-parameter families is here one of
convenience only.

This work is in a certain sense a sequel to Witten's paper \rphases,
and in fact some of the
computations for the first example discussed in this paper
were performed by Witten. We thank him
for sharing these with us, and encouraging us to study the generalization.

\newsec{Pr\'{e}cis of Nonlinear Sigma Models}

\nref{\rAMii}{P. S. Aspinwall and D. R. Morrison,
``Chiral rings do not suffice: {$N$=(2,2)}
  theories with nonzero fundamental group,'' Phys. Lett.  {\bf 334B} (1994)
  79--86, hep-th/9406032.}

Given a manifold $X$, equipped with a Riemannian metric $g$ as well as
a closed\foot{However, see \rAMii\ for a generalization of this.}
two-form $B$, the nonlinear sigma model is written in terms of maps
$\Phi:\Sigma\to X$ where $\Sigma$ is a Riemann surface. (We work with
a metric of Euclidean signature on $\Sigma$ throughout.)
Choosing local
coordinates $z,\zbar $ on $\Sigma$ and $\phi^I$ on $X$, we write the
map in terms of functions $\phi^I(z,\zbar )$.  In addition the model
includes the superpartners of $\phi^I$, Grassman-valued sections
$\psi^I_{\pm}$ of $K^{\pm 1/2}\otimes\Phi^*(TX)$, where $TX$ is the
complexified tangent bundle to $X$ and $K$ is the canonical line
bundle of $\Sigma$. In terms of these we write a component action
\eqn\eSnonl{\eqalign{
S = \int_{\Sigma} d^2 z \Bigl(
\half (g_{IJ}+iB_{IJ})  \p_z\phi^I\p_{\zbar}\phi^J +
&{i\over 2} g_{IJ} \psi^I_-D_z\psi^J_- +
{i\over 2} g_{IJ} \psi^I_+D_{\zbar}\psi^J_+ + \cr
&{1\over 4} R_{IJKL}\psi^I_+\psi^J_+\psi^K_-\psi^L_-\Bigr)
\ ,\cr
}}
where the covariant derivatives $D$ are constructed by pulling back
the Christoffel connection on $TX$. When $X$ is complex and $g$ is
K\"ahler, the model has $N=2$ supersymmetry. We assume this
henceforth.

As pointed out in \rtopsim\ (see \rWitmir\ for the notation used here)
there is a sector of this theory for which correlation functions can
be computed using a greatly simplified version~--~the topologically
twisted sigma model. This differs in the assignment of worldsheet
spin to the fields. In particular, choosing local complex
coordinates $\phi^i, \phi^{\ibar}$ on $X$ we consider $\psi^i_-$ to be
a section of $\Phi^*(T^{(1,0)}X)$ (a scalar on $\Sigma$) and
$\psi^{\ibar}_+$ a section of $\Phi^*(T^{(0,1)}X)$~--~together they
form a scalar section $\chi$ of $\Phi^*(TX)$. On the other hand,
$\psi^{\ibar}_-$ becomes a $(1,0)$ form with values in
$\Phi^*(T^{(0,1)}X)$ called $\psi^{\ibar}_z$ and likewise we have
$\psi^i_{\zbar}$. In this theory one of the two supercharges under
which \eSnonl\ is invariant becomes a nilpotent global fermionic
symmetry $Q$. It is natural to employ this as a BRST-like projection,
restricting attention to $Q$-closed observables, i.e., those
annihilated by $Q$. Then the nilpotency means operators of the form
$\{Q,\CO\}$ will decouple. The space of operators is thus the
quotient, or the cohomology of $Q$. In the case at hand one finds that
correlation functions of $Q$-closed observables will depend on the
parameters $g$ and $B$ of the model only through the complexified
K\"ahler class they define in $H^2(X)$. The parameter space of the
model is thus naturally the space $H^2(X,\IC)/H^2(X,\IZ)$; the
quotient reflects the fact that under shifts of $B$ by integral
classes \eSnonl\ changes by an integer so the quantum mechanical
measure $e^{2\pi i S}$ is unchanged.

Explicitly, the local operators are of the form
\eqn\eobs{
{\cal O}_\xi  = \xi_{i_1\ldots i_s}(\Phi) \chi^{i_1}\cdots\chi^{i_s}
}
where $\xi$ is an $s$-form on $X$. One finds
\eqn\eQ{
\{Q ,{\cal O}_\xi\} = -{\cal O}_{d\xi}
}
with $d$ the exterior derivative on $X$. Thus the space of operators
in the topological nonlinear sigma model is the total cohomology
$H^*_{\rm DR}(X)$. We can choose representatives of these classes dual
to homology cycles $H$ on $X$ as having delta-function support on $H$,
and denote them by ${\cal O}_H$.

\nref\rNmat{E. Witten, ``The $N$-matrix model and gauged WZW models,''
Nucl.\ Phys.\ {\bf B371} (1992) 191--245.}

The path integral computing a correlation function of
$Q$-closed observables localizes on field configurations invariant under
an odd symmetry \rWitmir ; in the case at hand this requires the map
$\Phi$ to be holomorphic. The classical action of such a configuration
is given by
$S_{cl} = \int_\Sigma \Phi^*(J)$ where $J$ is the (complexified) K\" ahler
form on $X$. Further, the $N=2$ supersymmetry ensures that the determinants
of nonzero modes cancel between bosons and fermions.
Thus the path integral reduces to a sum over homotopy classes
of maps, weighted by $e^{2\pi iS_{cl}}$, of integrals over the
finite-dimensional moduli space of holomorphic maps (instantons)
in a given homotopy class. When $H_1(X,\IZ) = 0$ we can label each instanton
sector by the homology class $h$ in $H_2(X,\IZ)$ of the image of
any of the corresponding maps.  The virtual
dimension of the ``sector $h$''
moduli space $\CM_h$ is $d_h = d - K\cdot h$ where $K$ is
the canonical class of $X$ and $d=\dim_{\IC}X$.
Let us consider a given correlation function
\eqn\ecor{
\la \CO_{H_1}(z_1)\ldots \CO_{H_s}(z_s) \ra
}
for $z_1\ldots z_s$ generic points on $\Sigma$ and $H_i$ homology
cycles of codimension $q_i$. The contributions to this correlator from
the various instanton sectors are referred to as {\it Gromov--Witten
invariants}; the contribution from
$\CM_h$ vanishes unless $d_h = \sum q_i$. When nonzero the
contribution is given by the integral over $\CM_h$ of a density
constructed as follows. Our choice of representatives $\CO_H$ means
the integrand has delta-function support on maps $\Phi$ such that
$\Phi(z_i) \in H_i$.  If ${\rm dim}_{\IC} \CM_h = d_h$ then maps
satisfying this constraint will be isolated and the contribution is
simply the number of such maps. In general, of course, $\CM_h$ may
have a dimension larger than $d_h$. In this case there are also zero
modes of the fermions $\psi^i_\zbar$.  These vary as sections of a
vector bundle $\cal V$ over $\CM_h$ of rank ${\rm dim}_{\IC} \CM_h -
d_h$ (by the index theorem). The contribution to \ecor\ is given by
inserting into the integral the Euler class of $\cal V$ \rNmat.
Eqn.~\ecor\
does not depend on the choice of the points $z_i$ at which we insert
the operators; we will often drop these.  Summing over all possible
$h$ we obtain a (formal) series expansion
for \ecor. This is expected to converge for ${\rm Im}(J)$ deep in the
interior of the K\"ahler cone. Note that for such $J$,
${\rm Im}(S_{cl})$ can be made
arbitrarily large for any nontrivial class $h$, so the series
collapses in the limit onto its first term, determined by {\it
constant}\/ maps. The moduli space here is simply $X$, and the
nonvanishing correlators are the intersection numbers $\la
\CO_{H_1}(z_1)\ldots \CO_{H_s}(z_s) \ra _0 = H_1\cdot H_2\cdots H_s$.
The arguments of \rWitmir\ ensure that for $s=2$ this is in fact the
full answer; nontrivial instantons do not contribute. In particular,
this guarantees that the bilinear pairing given by the two-point
function is nondegenerate and a deformation invariant.

\nref\rCR{C. W. Curtis and I. Reiner, {\sl Representation Theory of
Finite Groups and Associative Algebras}, Interscience Publishers,
New York, 1962.}%
\nref\rKarp{G. Karpilovsky, {\sl  Symmetric and $G$-algebras: With
Applications to Group Representations},  Kluwer Academic Publishers, 1990.}%
\nref\rDubrov{B. Dubrovin, ``Integrable systems in topological field theory,''
Nucl. Phys. {\bf B379} (1992) 627--689.}%

We can use the three-point functions to define a ``quantum product''
turning the vector space $H^*(X,\IC)$ into an associative,
super-commutative
algebra. We define the product $\CO_1 *\CO_2$ in this algebra by using the
nondegenerate pairing: it is the unique element satisfying
\eqn\eQprod{
\la \left(\CO_1*\CO_2\right) (z) \CO_3(w)\ra =
\la \CO_1(z_1)\CO_2(z_2)\CO_3(z_3)\ra \quad \forall \CO_3
\ .}
(Notice that the $z_i$ in \eQprod\ are inserted for clarification
only; nothing depends upon these; dropping them as we will often do
exhibits the ring structure on $H^*$ directly.)
This  algebra has some additional
 structure which makes it into a {\it Frobenius algebra}.\foot{We
follow standard mathematical usage \refs{\rCR,\rKarp} and do not require
a Frobenius algebra to be commutative; our definition therefore differs
slightly from that in \rDubrov.
However, we will primarily be interested in the even part
$H^{ev}(X)$ of the cohomology of $X$, on which the quantum product will
in fact be commutative.}
This means that the algebra~--~call it $A$~--~has a
multiplicative identity element $\boldone$, and that there is
a linear functional
$\varepsilon:A\to\IC$ such that the induced bilinear
pairing $(x,y)\mapsto\varepsilon(x*y)$ is nondegenerate.  There does
not seem to be a standard name for such a functional; we will call it
an {\it expectation function}.  If an expectation function exists at all,
then most linear functionals on $A$ can serve as expectation functions.
If $A$ is $\IZ$-graded, we call $\varepsilon$ a {\it graded expectation
function}\/ when $\ker(\varepsilon)$ is a graded subalgebra of $A$ (and we
call $A$ a {\it graded Frobenius algebra}\/ when such a function exists). There
is much less freedom to choose graded expectation functions.

The cohomology of a compact manifold $X$ has the structure of a graded
Frobenius algebra, with multiplication given by cup product, $\boldone$ given
by the standard generator of $H^0(X)$, and a graded
expectation function given by ``evaluation on the fundamental class.''
The ``quantum cohomology'' algebra\foot{This
is often referred to in the literature
as the quantum cohomology {\it ring}, and indeed we will also use that
terminology.  However, here we use the term ``algebra'' in order to
stress that in addition to the ring structure, this object has the
structure of a complex vector space.}
of $X$ is also a Frobenius algebra, more or less by definition:  the
correlator gives an expectation function, and the product \eQprod\ is
defined with the aid of this function; the multiplicative identity is
again provided by the standard generator of $H^0(X)$.
The arguments above about the
limit for large $J$ show that this quantum cohomology algebra
is a deformation of the
classical cohomology algebra of $X$, reducing to the latter for
appropriate limiting values of $J$.  This deformation however is
nontrivial. In general the nilpotent, $\IZ$-graded
cohomology algebra $H^*(X)$ is
deformed into an algebra which is not nilpotent and which has
at most a $\IZ_p$-grading for some
integer $p$ (when the canonical class $K$ of $X$ satisfies $-K = pe$ for some
$e\in H^2(X,\IZ)$). In the Calabi--Yau case, however, where $K=0$,
 the quantum cohomology algebra is actually $\IZ$-graded and nilpotent.

\nref\rNakayama{T. Nakayama, ``On Frobeniusean algebras I,''
Annals of Math. {\bf 40} (1939) 611--633.}%

Eqn.~\eQprod\ shows that the correlation functions determine the ring
structure. In fact they can also be used to specify a presentation
of the ring in terms of generators and relations.
For example, if we have a set of generators $\CO_1,\ldots ,\CO_N$ for
the even cohomology of the
space $X$ then nondegeneracy of the two-point function implies that
the (even part of the) quantum cohomology ring takes the form
\eqn\ering{
\CR = \IC[\CO_1,\ldots ,\CO_N ]/{\cal J}
}
where
\eqn\eJ{
{\cal J} = \{ {\cal P}\in \IC[\CO_1,\ldots ,\CO_N ]\ |\
\la \CO {\cal P} \ra = 0
\quad\forall \CO\}
\ .}
More generally, given any associative
algebra $A$ with multiplicative identity,
and any linear functional $\varphi$ on $A$,
the kernel of the bilinear form $(x,y)\mapsto\varphi(x*y)$
is an ideal ${\cal J_\varphi}$, and the quotient ring
$A/{\cal J_\varphi}$ is a Frobenius algebra
with expectation function induced by $\varphi$.
If $A$ is itself a Frobenius algebra with an expectation function
$\varepsilon$,
then by a theorem of Nakayama \rNakayama\ (see \rKarp\ for a modern
discussion),
$\varphi$ takes the form $\varphi(x)=\varepsilon(\alpha*x)$ for some fixed
element $\alpha\in A$, and $\cal J_\varphi$ coincides with
 the annihilator of $\alpha$.

Although the correlation functions determine the ring structure, the
opposite does not hold in general~--~there can be many
expectation functions on a given algebra. However,
if $A$ is a graded Frobenius algebra of finite length (i.e., of finite
dimension as a complex vector space) and all elements of $A$ have non-negative
degree, then the graded expectation functions
on $A$ are in one-to-one correspondence with degree $0$ elements of $A$
which are not zero-divisors.  (This is because they must all be of the
form $\varphi(x)=\varepsilon(\alpha*x)$ for some $\alpha$ which is not
a zero-divisor, but every element of degree ${}>0$ must be a zero-divisor.)
In particular, in the case of the quantum cohomology algebra of a \CY\
manifold $X$, we have a graded Frobenius algebra of finite length in which the
degree $0$ elements are just the one-dimensional vector space $H^0(X)$.
This means that the graded expectation function
is unique up to a scalar multiple~--~a single normalization constant~--~and
that the ring structure determines the
correlation functions up to this overall factor.
(It is not hard to see that the graded
expectation function is nonzero precisely
on the top degree piece $H^{2d}(X)$, where $d=\dim_{\IC}X$, and that
$H^{2d}(X)$ must also be one-dimensional.)
When we have a family
of rings depending on parameters, of course, the normalization can
depend on these. These facts will be important in section five.

The topological field theory is obtained from the original, untwisted
theory by modifying the spins of some of the fields and
projecting out some of the states. As mentioned above, the fields eliminated
decouple from correlators of those we keep, so their elimination has no
effect on these. Furthermore, if we choose $\Sigma$ so that $K$ is
trivial, the twisting itself alters nothing; this allows us to relate
correlation functions computed in the topological field theory
directly to correlators in the original theory, for $\Sigma$ of genus
zero. The mapping is described very clearly in \rWitmir .

\nref\rshel{S. Katz, ``Rational curves on Calabi--Yau threefolds,''
(revised version), alg-geom/9312009.}%
\nref\rBCOV{M.~Bershadsky, S.~Cecotti, H.~Ooguri, and C.~Vafa,
 ``Holomorphic anomalies in topological field theories,''
with an appendix by S.~Katz,
 Nucl. Phys. {\bf B405} (1993) 298--304, hep-th/9302103.}%
\nref\rKatz{S. Katz, ``Rational curves on Calabi--Yau manifolds:
verifying predictions of mirror symmetry,'' in {\it Projective
Geometry with Applications}, (E.~Ballico, editor), Marcel Dekker, 1994,
pp.~231--239, alg-geom/9301006.}%
\nref\rRuan{Y. Ruan,
``Topological sigma model and {D}onaldson type invariants in {G}romov theory,''
 preprint, 1993.}%
\nref\rRT{Y. Ruan and G. Tian, ``A mathematical theory of quantum
cohomology,'' Math. Res. Lett.
{\bf 1} (1994)  269--278.}%
\nref\rMcDS{D. McDuff and D. Salamon,
{\sl $J$-holomorphic Curves and Quantum Cohomology,}
American Mathematical Society,  1994.}%
\nref\rKM{M. Kontsevich and Yu. Manin,
``{G}romov--{W}itten classes, quantum cohomology, and enumerative geometry,''
Commun. Math. Phys. {\bf 164} (1994) 525--562,
hep-th/9402147.}%
\nref\rKont{M. Kontsevich, ``Enumeration of rational curves via
torus actions,'' hep-th/9405035.}%
\nref\rManin{Yu. I. Manin, ``Generating functions in algebraic geometry and
sums over trees,'' alg-geom/9407005.}%

\CY\ manifolds are a special case of the above. In particular, for these
we have $K=0$ and $d_h=d$ independent of $h$. Thus the sum defining
nonzero correlators is over all classes $h\in H_2(X,\IZ)$.  This makes
the model extremely hard to study, and indeed the relevant
computations have thus far been essentially impossible to carry out
for this class of models.%
\foot{In any event, computations of even the contribution of a given
instanton sector are difficult to make at any level of mathematical
rigor; one example of the difficulties encountered is the fact that
the instanton moduli spaces are noncompact, and the integrals~--~or
the intersection problem they represent~--~must be treated with great
care for boundary contributions. See \refs{\rshel{--}\rManin} for some
recent progress in this direction.}

The nonlinear sigma model is conformally invariant when $g$ and $B$
satisfy some nonlinear differential equations (which are not actually
known explicitly).  These can have a solution only if the target space
is a \CY\ manifold. When this holds, these equations should have at
most one solution for each given choice of complex structure on $X$
and a class in $H^{1,1}(X)$ to which the (complexified) K\"ahler class
determined by $g$ and $B$ should belong, and there should be a
solution corresponding to $(g,B)$ whenever the K\"ahler class is
sufficiently large.  The resulting theory is then invariant under the
$N=2$ superconformal algebra. The important features of this
structure, for present purposes, are the existence of chiral left- and
right-moving U(1) $R$-symmetries, and the fact that the supercharges
themselves break into left- and right-moving components. The
nonrenormalization theorems protect correlators of fields annihilated
by one-half of the supercharges. In the conformal case these can be
chosen independently for the left- and right-moving sectors.  There
are thus two distinct topological sectors, and correspondingly two
distinct versions of the twisted model.

The discussion up to this point has referred to the {\bf A} model of
\rWitmir . Correlation functions in this model are independent of the
complex structure; the moduli space is thus the (complexified) space
of K\"ahler forms. The other possible twist leads to the {\bf B}
model, which is consistent only for a \CY\ target. The correlation
functions of the operators which survive the {\bf B} model projection
do not receive instanton corrections, and are given exactly by their
semiclassical values \rDG . Correlation functions in this theory are
independent of the K\"ahler structure and the parameter space is thus
the space of complex structures on $X$.  The superconformal models
exhibit mirror symmetry. This relates a \CY\ manifold $M$ to a
topologically distinct ``mirror manifold'' $W$ such that the nonlinear
sigma models defined by the two are isomorphic. The mirror isomorphism
is such that the two twists are exchanged. Thus, if $M$ and $W$ are
mirror \CY\ manifolds then the {\bf A} model with target $M$ is
isomorphic to the {\bf B} model with target $W$. The converse need not
hold, i.e., the isomorphism of topological models does not guarantee
an isomorphism of conformal field theories.

This property has been used to compute the quantum cohomology rings of
a number of \CY\ examples, as correlation functions in the {\bf B}
model are simpler objects geometrically and often computable.
Translated using mirror symmetry they are interpreted as the exact sum
of the instanton series for {\bf A} model correlators. Using this
interpretation, the coefficients in the expansion express geometrical
properties of the moduli spaces of holomorphic maps, bypassing the
technical difficulties of a direct calculation to which we alluded
above. This has led to a wealth of new predictions regarding these
mapping spaces, some of which have been verified. (See \rshel\ for a
review.) In principle, making these predictions requires two
ingredients: a construction of the mirror manifold $W$ given $M$, and
an understanding of the ``mirror map'' between the moduli spaces of
the two (isomorphic) topological field theories.

\nref\rGP{B. R. Greene and M. R. Plesser, ``Duality in Calabi--Yau
moduli space,'' Nucl.\ Phys.\ {\bf B338} (1990) 15--37.}%
\nref\rpertris{P. Berglund and T. H\"ubsch, ``A generalized
construction of mirror manifolds,'' in {\sl Essays on Mirror
Manifolds\/}, S.-T. Yau, ed. International Press, Hong Kong, 1992,
pp.~388--407, hep-th/9201014.}
\nref\rbatmir{V. V. Batyrev, ``Dual polyhedra and mirror symmetry for
Calabi--Yau hypersurfaces in toric varieties,'' J.\ Alg.\ Geom. {\bf 3}
(1994) 493--535, alg-geom/9310003.}
\nref{\rborisov}{L. A. Borisov, ``Towards the mirror symmetry for Calabi--Yau
complete intersections
in  Gorenstein toric Fano varieties,'' alg-geom/9310001.}
\nref{\rbatbor}{V. V. Batyrev and L. A. Borisov, ``Dual cones and mirror
symmetry for
generalized Calabi--Yau manifolds,'' alg-geom/9402002.}
\nref\rpershel{P. Berglund and S. Katz, ``Mirror symmetry
constructions: a review,'' preprint IASSNS-HEP-94/38, OSU-M-94/2,
to appear in {\sl Essays in mirror manifolds II},
B. R. Greene and S.-T. Yau, eds. International Press, Hong Kong,
hep-th/9406008.}
\nref\rAGM{P. S. Aspinwall, B. R. Greene, and D. R. Morrison, ``Calabi--Yau
moduli space, mirror manifolds and spacetime topology change in string
theory,'' Nucl.\ Phys.\ {\bf B416}  (1994) 414--480, hep-th/9309097.}

The existence of mirror manifolds, even in this weak sense of
equivalence of the topological field theories, is proved only for an
extremely restricted class of examples \rGP. The proof uses a
construction of the mirror manifold which is demonstrated to yield an
isomorphic conformal field theory using the fact that at a point in
the parameter space the superconformal theory is solvable.
There are many
conjectured generalizations of this construction
\refs{\rpertris{--}\rbatbor} (for a review see \rpershel). However,
no {\it direct}\/ proof of mirror symmetry has been given for any of
these, and since the models are not exactly solvable, an indirect
proof is difficult to construct. One of the broadest classes for which
there is a conjectured mirror construction \rbatmir\ is the class of all
\CY\ manifolds realized as hypersurfaces in toric varieties (see
section three for a brief description of these objects, or \rAGM\ for
an elementary introduction for physicists).

\nref\rdV{B. de Wit and A. Van Proeyen, ``Potentials and symmetries
of general gauged $N{=}2$ supergravity-Yang--Mills models,''
Nucl.\ Phys.\ {\bf B245}
(1985) 89--117.}
\nref\rdLV{B. de Wit, P. Lauwers, and A. Van Proeyen, ``Lagrangians of $N{=}2$
supergravity-matter systems,'' Nucl.\ Phys.\ {\bf B255}
(1985) 569--608.}
\nref\rCKVDdG{E. Cremmer, C. Kounnas, A. Van Proeyen,
J. P. Derendinger, B. de Wit, and
L. Girardello, ``Vector multiplets coupled to $N{=}2$ supergravity: super
Higgs effect, flat potentials and geometric structure,''
Nucl.\ Phys.\ {\bf B250} (1985) 385--426.}
\nref\rStrom{A. Strominger, ``Special geometry,''
Commun. Math. Phys. {\bf 133} (1990) 163--180.}%
\nref\rttstar{S. Cecotti and C. Vafa, ``Topological--anti-topological
fusion,'' Nucl. Phys. {\bf B367} (1991) 359--461.}
\nref\rBCOVtwo{M. Bershadsky, S. Cecotti, H. Ooguri, and C. Vafa,
``Kodaira--Spencer theory of gravity and exact results for quantum
string amplitudes,'' Commun. Math. Phys. {\bf 165} (1994) 311--427,
hep-th/9309140.}
\nref\rhid{B. R. Greene, D. R. Morrison, and M. R. Plesser, ``Mirror
symmetry in higher dimension,'' preprint CLNS-93/1253,
IASSNS-HEP-94/2,  YCTP-P31-92, hep-th/9402119.}

The mirror construction relates two families of conformal field
theories. The members in each family are obtained by marginal
deformations (of any fixed initial theory) preserving the
superconformal structure.  The theories are labeled by points in a
``moduli space'' of such deformations. Explicitly constructing the
mirror isomorphism requires both a map between the two moduli spaces
and a related map between the Hilbert spaces at a given point (a
frame). In general, when considering families of conformal field
theories there is no canonical choice of coordinates on the moduli
space or of bases for the Hilbert spaces. The ambiguity is related
physically to the ambiguous choice of contact terms in operator
products, and geometrically to the fact that the moduli space is a
nontrivial manifold and the Hilbert spaces of the theories form a
nontrivial bundle over this. In twisted $N=2$ superconformal theories,
however, there is in fact a canonical choice. The reason is that the
operator products of the operators which survive the projection to the
topological theory are nonsingular as operators approach one
another. This leads to a canonical choice of contact terms in which
the insertion of multiple operators at a point is defined by
point-splitting~--~it is the limit (as the points coincide) of
insertions at distinct points. Geometrically, the existence of these
coordinates means that the moduli space carries an extra
structure. This structure is known as ``special geometry''
\refs{\rdV{--}\rStrom} when the superconformal theory has central charge 9
(corresponding to a \CY\ threefold); there are natural generalizations
to any dimension \refs{\rttstar{--}\rhid}.

\nref\rCdGP{P. Candelas, X. C. de la Ossa, P. S. Green, and L. Parkes,
``A pair of Calabi--Yau manifolds as an exactly soluble superconformal
theory,'' Nucl.\ Phys.\ {\bf B359} (1991) 21--74.}
\nref\rmondiv{P. S. Aspinwall, B. R. Greene, and D. R. Morrison, ``The
monomial-divisor mirror map,'' Int.\ Math.\ Res.\ Notices (1993)
319--337, alg-geom/9309007.}

In the {\bf A} model this canonical choice has been shown to
correspond to a natural coordinate on the moduli space given by the
coefficients in an expansion of the K\"ahler form in a fixed
(topological) basis for $H^2(M)$; the preferred choice of frame is
similarly given by a topological basis for $H^*(M)$. The mirror image
of this in the moduli space of the {\bf B} model on $W$ is constructed
using the periods of the holomorphic form of top degree on $W$ using
an {\it Ansatz}\/ proposed in \rCdGP\ and justified in \rBCOVtwo . When $W$
is a hypersurface in a toric variety there is also a (geometrically)
natural choice of coordinates on the moduli space of the {\bf B} model
on $W$, related to the coefficients of the defining polynomial. The
mirror maps computed in specific examples were obtained by computing
the period integrals in terms of these ``algebraic'' coordinates (the
reason for the terminology will become clear) and using the {\it
Ansatz}\/ referred to above. A conjecture for the asymptotic form of
this map, termed the ``monomial-divisor mirror map'' was given in
\rmondiv . (Some control over asymptotics
is needed because the {\it Ansatz}\/ leaves a finite
number of parameters undetermined.) The conjecture relates, for
hypersurfaces in toric varieties, the coefficients of monomials in the
defining equation directly to the expansion of the K\"ahler form on
the mirror manifold. We will discuss this map in section five, in
which we find that the GLSM naturally yields an interpretation of the
algebraic coordinates in the interior of parameter space.

\newsec{The Linear Sigma Model for $V$}

In this section we will present a GLSM whose low-energy limit is the
nonlinear sigma model with target space a toric variety $V$. We will
then use this to completely solve the {\bf A} model computing the
exact correlation functions. For smooth $V$ we verify an algebraic
computation of these, first proposed by Batyrev \rbator . This section
is the longest in the paper, and by the end we will not only have
solved the models in question but also developed most of the
techniques we will need for the models studied in the following
sections.

\subsec{Toric Varieties on One Leg}

\nref\rGIT{D.~Mumford, {\sl Geometric Invariant Theory},
{E}rgeb. {M}ath. {G}renzgeb.
  (N.F.), vol.~34, Springer-Verlag, Berlin, Heidelberg, New York, 1965.}%
\nref\rKirwan{F.~C. Kirwan,
  {\sl Cohomology of Quotients in Symplectic and Algebraic Geometry},
  Mathematical Notes, no.~31,  Princeton University Press, 1984.}%
\nref\rNess{L.~Ness,
``A stratification of the null cone via the moment map,''
 Amer. J. Math. {\bf 106} (1984) 1281--1329.}%
\nref\rFulton{W. Fulton, {\sl Introduction to Toric Varieties}, Annals
of Math. Studies, vol. 131, Princeton University Press, 1993.}%
\nref\rhomog{D. A. Cox, ``The homogeneous coordinate ring of a toric
variety,'' Amherst preprint, 1992, alg-geom/9210008.}

A toric variety is a natural generalization of projective space.
Just as $\CP d$ can be described in the form
$\CP d = (\IC^{d+1}-\{0\})/\IC^*$,
a general $d$-dimensional toric variety $V$ is best thought
of for our present purposes as a quotient space\foot{For smooth toric
varieties or ones with mild singularities (so-called ``simplicial''
toric varieties), this is an ordinary quotient; however, in general
we must take the quotient in the sense of Geometric Invariant Theory
\rGIT\ (cf.\ also \refs{\rKirwan,\rNess}).}
\eqn\ehol{
V_{\Delta} = (Y -  F_{\Delta } ) /T_{\Delta}
}
with $Y = \IC ^n$, $T_{\Delta}\sim {\IC ^*}^{(n-d)}$
acting diagonally on
the coordinates of $Y$ as
\eqn\eFact{
g_a(\lambda) : x_i\rightarrow \lambda^{Q_i^a} x_i\qquad
a = 1,\ldots ,(n-d), \quad i = 1,\ldots ,n
\ ,}
and $F_{\Delta}$ a subset of $Y -  {\IC^*}^n$ which is a
union of certain intersections of
coordinate hyperplanes.  We often refer to \ehol\ as a ``holomorphic
quotient'' construction of $V_\Delta$, to emphasize the holomorphic
nature of the group $T_\Delta$ and its representation on $Y$.

The precise set of intersections of
coordinate hyperplanes which constitute $F_\Delta$,
as well as the integers $Q_i^a$ which specify the representation,
are determined by the combinatorial data $\Delta$ defining $V$ as we
will describe below. The
name ``toric variety'' refers to the algebraic torus
${\IC^*}^d$ contained in the interior of $V$; the
combinatorial data are in effect
telling us how to (partially) compactify this torus. The fact that the torus
itself is trivial means that topological (and in fact
algebro-geometrical) information about $V$ is encoded in $\Delta$,
making it easily accessible as we shall see.  For details of this
construction see \refs{\rFulton,\rhomog} or \rAGM.
In what follows we will consider the case in
which $V$ is smooth, compact, and irreducible. (Noncompact toric
varieties will appear in the following sections.)
We give two examples
which will serve to illustrate our results in the sequel.

\line{\hrulefill}\nobreak
\noindent{\it Example 1.}\par\nobreak
This is perhaps the simplest possible example, complex projective space
$\CP4$. Here $Y = \IC ^5$, $F = \{0\}$, and $T = \IC ^*$ acts on the
standard coordinates in $Y$ as \eqn\eac{ g(\lambda) :
(x_1,x_2,x_3,x_4,x_5)\mapsto (\lambda x_1,\lambda x_2,\lambda x_3,\lambda
x_4,\lambda x_5) \ .}

\nref{\rCdFKM}{P. Candelas, X. de la Ossa, A. Font, S. Katz, and
D. R. Morrison, ``Mirror symmetry for two parameter models (I),''
Nucl. Phys. {\bf B416} (1994) 481--538, hep-th/9308083.}%
\nref\rHKTY{S. Hosono, A. Klemm, S. Theisen, and S.-T. Yau, ``Mirror
symmetry, mirror map and applications to Calabi--Yau hypersurfaces,''
Harvard preprint  HUTMP-93-0801, hep-th/9308122.}

\noindent{\it Example 2.}\par\nobreak
The second example is chosen among other reasons because it has been
studied by Candelas, de la Ossa, Font, Katz, and Morrison \rCdFKM\
(see also \rHKTY).
The toric variety in question is obtained by resolving the curve of
$\IZ_2$ singularities in the weighted projective space
\WCP{4}{1,1,2,2,2}. Each point on the curve is blown up to a \CP1.  We
have $Y = \IC ^6$,
\eqn\eF{
F =
\{x_1=x_2=0\}\cup\{x_3=x_4=x_5=x_6=0\}
\ ,  }
and $T = (\IC ^*)^2$ with the action on the coordinates
\eqn\egact{
\eqalign{
g_1(\lambda) &: (x_1,x_2,x_3,x_4,x_5,x_6)\mapsto
(x_1,x_2,\lambda x_3,\lambda x_4,\lambda x_5,\lambda x_6) \cr
g_2(\lambda) &: (x_1,x_2,x_3,x_4,x_5,x_6)\mapsto (\lambda x_1,\lambda
x_2,x_3,x_4,x_5,\lambda ^{-2}x_6) \ .\cr
}}
Note that the group element
\eqn\egelt{
g_1^2 g_2(\lambda) : (x_1,x_2,x_3,x_4,x_5,x_6)\mapsto
(\lambda x_1,\lambda x_2,\lambda^2 x_3,\lambda^2 x_4,\lambda^2 x_5, x_6)
}
reproduces the familiar weighted projective space action on the
first five variables.
\par\nobreak\line{\hrulefill}

We now give a somewhat more detailed version of the construction
described above, explicitly describing the combinatorial structures
used. This level of detail is necessary for some of the developments
in section five.
Let $\bN\sim \IZ^d$ be a lattice in $\bN_{\IR} = \bN\ztim \IR$.
To define
$V$ we need a {\it fan of strongly convex rational polyhedral cones}\/
$\Delta$ in $\bN_{\IR}$. This is a collection of cones with apex at the
origin, each of which is
spanned by a finite collection (``polyhedral'')
of elements in $\bN$ (``rational'') and such
that the angle subtended by any two of these at the apex is less than
$\pi$ (``strongly convex''). To be a fan the collection must have the
property that (i) any two members of the collection intersect in a
common face (i.e.~a cone of lower dimension bounding each)~--~note
that this includes the origin as a possible intersection~--~and (ii)
for each member of $\Delta$ all its faces are also in $\Delta$. This data
determines the holomorphic quotient described above as follows.

\item{1.} Let $\{v_1,\ldots,v_n\}$ be the integral generators of the
one-dimensional cones in $\Delta$, then in \ehol\ we set $Y=\IC^n$.

\nref\rbatoh{V. V. Batyrev, ``On the classification of smooth projective
toric varieties,'' T\^ohoku Math. J. {\bf 43} (1991) 569--585.}%

\item{2.} The set $F_{\Delta}$ is the union of intersections of
coordinate hyperplanes $x_{i_1} = \cdots = x_{i_p} = 0$ for each set
$1\leq i_1\leq\cdots i_p\leq n$ such that $v_{i_1},\ldots v_{i_p}$ are
not contained in any cone of $\Delta$. The irreducible components of
$F$ are seen to be determined by collections as above such that any
subset of $k<p$ of the vectors in the collection spans a
$k$-dimensional cone in $\Delta$. Following \rbatoh,
these are called {\it primitive collections}.

\item{3.} To get $T_{\Delta}$ let $D\subset\IZ^n$ be the
sublattice of vectors $d = (d_1,\ldots,d_n)$ such that $\sum_i d_i v_i =
0$. Choosing a basis $\{Q^1,\ldots,Q^{n-d}\}$ for $D$ we obtain the
$T$ action of \eFact .

Our interest in toric varieties here stems from their relation to
GLSM. In mathematics, these are useful examples of nontrivial
varieties whose topological and algebro-geometric properties are rather
directly encoded in the combinatorics of
$\Delta$. Some examples of this will be useful in what follows.
A good first example is the question whether $V$ is compact. This will
be true precisely when the fan $\Delta$ is {\it complete}, i.e., when
the cones in $\Delta$ cover $\bN_{\IR}$. To see this, let $n^*\in \bN$. To
this point we can associate a $\IC^*$ action on $V$ as follows. Write
$n = \sum_{i=1}^n k_i v_i$, where $k$ is determined up to adding an
element of the lattice $D$ defined above. Then
\eqn\elact{
g(\lambda): z_i\to \lambda^{k_i} z_i
}
defines an action on $V$ which suffers from no such ambiguity. Now
consider the limit point $\lambda\to 0$. This is contained in $V$
precisely when (i) we can choose $k$ such that all $k_i\ge 0$; let
$I\subset\{1,\ldots ,n\}$ be the set for which $k_i>0$. This
means the point $n^*$ lies in the cone spanned by the $v_i,\ i\in I$.
The limit is then the point $z_i=0,\ i\in I$ which is contained in
$V$ precisely when (ii) this cone is a member of our collection
$\Delta$. In all, $V$ will contain all such limit points precisely
when any $n^*$ is contained in some cone of $\Delta$.

Another property encoded in $\Delta$ is the intersection theory on
$V$ (at least when $V$ is ``simplicial'', which we now assume).
In general, the cohomology of $V$ is nonzero only in even
dimensions; further, when $V$ is compact $H^*(V)$ is generated by
$H^2(V)$ under the intersection product. Thus the complete
intersection ring is determined by relations on the elements of
$H^2(V)$. These are easily read off from $\Delta$ as follows. The group
$H^2(V)$ itself is generated by classes $\xi_i\quad i=1,\ldots,n$ dual to
the divisors $\{x_i=0\}$, subject to linear relations. These
essentially express the fact that $T$-invariant monomials in the
homogeneous coordinates $x_i$ (and their inverses) are meromorphic
functions on $V$ and hence correspond to trivial divisors.  These
monomials are parameterized by the lattice $\bM\sim \IZ^d$ dual to $\bN$,
since 3.\ above guarantees that for $m\in \bM$ the monomial $\chi_m =
\prod_{i=1}^n x_i^{\la m,v_i\ra}$ is $T$-invariant. Thus we have
\eqn\eok{
\sum_{i=1}^n \la m,v_i\ra \xi_i = 0
}
for every $m\in \bM$.
There are $d$ independent relations of this type, which
reduce the dimension of $H^2(V)$ to $n{-}d$. This determines a
basis $\eta_a$ of $H^2(V)$ in which we can write
\eqn\elinr{
\xi_i = \sum_{a=1}^{n-d} Q_i^a \eta_a\ .
}

There are also nonlinear relations in the ring $H^*(V)$. These are read
off most easily in the dual picture as excluded intersections of the
coordinate hyperplanes.  That is, for each irreducible component of
$F$, described as $\{ x_a = 0\ |\ a\in A\}$ for some set
$A\subset\{1,\ldots,n\}$,
we get a relation
\eqn\enonl{
\prod_{a\in A} \xi_a = 0\ .
}
(These relations comprise what is known as the
{\it Stanley--Reisner ideal}.)

The relations \elinr\ and \enonl\
determine the ring structure of $H^*(V)$ completely; the one thing
left undetermined is the normalization of the expectation function
$\la\ \ra_V$
given by evaluation on the fundamental class of $V$.  This can also be
determined by the toric data, as follows.  Given a collection of
$d$ distinct coordinate hyperplanes $\{x_{i_1}=0\}$, \dots, $\{x_{i_d}=0\}$
which {\it do}\/ intersect on $V$, we have
\eqn\enorm{
\la \xi_{i_1} \dots \xi_{i_d} \ra_V =
{1\over \mult(v_{i_1},\ldots,v_{i_d})}
\ ,}
where $\mult(v_{i_1},\ldots,v_{i_d})$ denotes the index in ${\bf N}$
of the lattice spanned by these vectors.  (This index is always $1$
if $V$ is smooth.)

\line{\hrulefill}\nobreak
\noindent{\it Example 1.}\par\nobreak
Here there is precisely one $\eta$, and \elinr\ becomes $\xi_i =
\eta\quad\forall i$. From the form of $F$ we have $\eta^5 = 0$.
The one-dimensional cones in $\Delta$ are generated by
\eqn\evi{\eqalign{
v_1 &= (-1,-1,-1,-1)\cr
v_2 &= (1,0,0,0)\cr
v_3 &= (0,1,0,0)\cr
v_4 &= (0,0,1,0)\cr
v_5 &= (0,0,0,1)\ ,\cr
}}
and \enorm\ yields $\la \eta^4\ra_V = 1$.

\noindent{\it Example 2.}\par\nobreak
Here the linear relations read
\eqn\elinrel{
\eqalign{
\xi_1 &= \xi_2 = \eta_2\cr
\xi_3 &=\xi_4 = \xi_5 = \eta_1\cr
\xi_6 &= \eta_1 - 2 \eta_2 \cr
}}
and the nonlinear relations are
\eqn\enlrel{
\eqalign{
\xi_1 \xi_2 & = \eta_2^2 = 0\cr
\xi_3 \xi_4 \xi_5 \xi_6 &= \eta_1^3 (\eta_1-2 \eta_2 ) = 0\cr
}}
(compare \eF). The one-dimensional cones in $\Delta$ are spanned by
\eqn\evii{\eqalign{
v_1 &= (-1,-2,-2,-2)\cr
v_2 &= (1,0,0,0)\cr
v_3 &= (0,1,0,0)\cr
v_4 &= (0,0,1,0)\cr
v_5 &= (0,0,0,1)\cr
v_6 &= (0,-1,-1,-1)\ ,\cr
}}
and \enorm\ yields $\la\eta_1^4\ra_V = 2\la\eta_1^3\eta_2\ra_V = 2$.
\par\nobreak\line{\hrulefill}

\nref{\rGS}{V.~Guillemin and S.~Sternberg,
``Birational equivalence in the symplectic category,''
Invent. Math. {\bf 97} (1989) 485--522.}%
\nref{\rDH}{J.~J. Duistermaat and G.~J. Heckman,
``On the variation in the cohomology of the symplectic form of the reduced
phase  space,''
Invent. Math. {\bf 69} (1982)  259--269;  ibid. {\bf 72} (1983) 153--158.}%
\nref\rOP{T. Oda and H. S. Park,
``Linear Gale transforms and Gelfand-Kapranov-Zelevinskij
decompositions,''
T\^ohoku Math. J. {\bf 43} (1991) 375--399.}%

There is an equivalent construction of $V$, in which we consider $Y$
as a symplectic manifold (with the standard symplectic form $\omega =
i \sum_{i=1}^n d z^i\wedge d \zbar ^{\ibar}$), with a symplectic
action by the maximal compact subgroup $G\subset T$.  In our case $G =
U(1)^{(n-d)}$ acts as in \eFact\ with $|\lambda|=1$. The {\it symplectic
reduction}\/ of $Y$ by $G$ depends on a choice of
``moment map'' $\mu:Y\to \left( {\rm
Lie}(G)\right)^{\vee}$. In coordinates, the components of
$\mu:Y\to \IR ^{(n-d)}$ are
simply generators (by Poisson brackets) of the $G$ action, and can
be described by
\eqn\emu{
\mu_a =
\sum_{i=1}^n Q_i^a |x_i|^2 - r_a \ ,
}
where $r_a$ are undetermined additive constants.  The symplectic
reduction is then defined as
\eqn\esymp{
V(r)
\equiv \mu^{-1}(0) / G \ .
}
The structure of $V(r)$ depends on $r$, of course. Its topology and even its
dimension will change as $r$ varies. From the form of \emu\ it is
clear that every $T$-orbit in $Y$ will contribute at most one point
to $V(r)$ (or one $G$-orbit to $\mu^{-1}(0)$). Which orbits will
contribute is determined by the value of
$r$~--~these are the $T$-orbits which intersect $\mu^{-1}(0)$.
There is a cone in $r$-space for which the set of $T$-orbits
which do not contribute is precisely $F_\Delta$. For these values of
$r$ we obtain a manifold topologically identical to $V$ as defined by
\ehol. The quotient space $V(r)$ inherits a symplectic form $\omega_r$ by
reducing $\omega$.  (That is, the restriction of $\omega$ to
$\mu^{-1}(0)$ is $G$-invariant, and so induces a symplectic form on
$V$.)  The cohomology class of $\omega_r$ in $H^2(V,\IR)$ depends
linearly on $r$ (for $r$ in the cone corresponding to $V$).\foot{This
was shown in \rGS\ to follow from \rDH.}  The coefficients of $r_a$
are essentially the $\eta_a$ of \elinr .  The symplectic reduction
carries a natural complex structure, in which the reduced symplectic
form becomes a K\" ahler form. The range of allowed $r$ is in fact
precisely the K\" ahler cone of $V$ \rOP; for $r$ in this cone the two
constructions are identical. The symplectic construction
makes sense for {\it any}\/ value of $r$. However, it does not
necessarily produce a smooth space or one of dimension $d$; there
will in general be values of $r$ for which the space $V(r)$ is altogether
empty. We denote $\CK_c$ the $(n{-}d)$-dimensional cone of values of $r$
for which $V(r)$ is nonempty. This is the cone spanned by the
vectors $Q_i$, and in general is larger than the K\"ahler cone
$\CK_V$. However, for a compact $V$, the cone $\CK_c$ is always
convex, since otherwise we would have a combination $\sum_i a_i Q_i^a
= 0$ for some non-negative integers $a_i$ not all zero; then
$\prod_i x_i^{a_i}$ would be a nonconstant holomorphic function on $V$.

\subsec{The Lagrangian}

We are interested in studying the nonlinear sigma model
with target space $V$.  As is the case for the well-known example of
$\CP n$, this is related to an $N=2$ supersymmetric gauged linear
sigma model with target space $Y$ and gauge group $G=U(1)^{(n-d)}$
such that $G_{\IC}=T$.

This model contains vector superfields $V_a$ with component
expansions%
\foot{The conventions are those of \rphases ; we work in WZ gauge.}
\eqn\ecompv{
V = -\sqrt{2}(\theta^-\thetabar^- v_\zbar + \theta^+\thetabar^+ v_z -
\theta^-\thetabar^+\sigma - \theta^+\thetabar^-\sigmabar) +
i(\theta^2 \thetabar^\alphad{\bar\lambda }_\alphad -
\thetabar^2\theta^\alpha\lambda_\alpha ) +
\half \theta^2\thetabar^2D\ ,
}
and chiral superfields $\Phi_i$ with component expansions
\eqn\ecomphi{
\Phi = \phi +
\sqrt{2}(\theta^+\psi_+ + \theta^-\psi_- ) + \theta^2 F + \cdots\ ,
}
where $\cdots$ are total derivative terms, as well as their complex
conjugates $\Phibar$. We couple $n$ chiral matter multiplets
$\Phi_i$ with charges $Q_i^a$ under $G$ to the $n{-}d$ abelian gauge
superfields $V_a$, and introduce Fayet-Illiopoulos terms for the
abelian gauge symmetry.
In the limit of infinite coupling the gauge fields
act as constraints, and the action may be written in superspace as
\eqn\eLV{
S = \int_{\Sigma} d^2 z\, d^4\theta \left[ \sum_{i=1}^n
\Phibar_i {\rm exp}\left( 2 \sum_{a=1}^{n-d} Q_i^a V_a \right)
\Phi_i - \sum_{a=1}^{n-d} r_a V_a \right] \ .
}
The $V_a$ can be
eliminated by their equations of motion leading to a nontrivial kinetic
term for the matter fields corresponding to the analog of the
Fubini--Study metric on $V$. (Expanding in components we reproduce
\eSnonl .)

\nref\rpower{N. Seiberg, ``The power of holomorphy: exact results in
4D SUSY field theories,'' preprint RU-94-64, IASSNS-HEP-94/67,
hep-th/9408013, to appear in proceedings of PASCOS 94.}%

For the present application it is perhaps better to start off with
nonzero kinetic terms for the gauge fields (these will be generated
dynamically in any case, so this is more a shift of perspective than a
change in the model).  We thus add to \eLV\ a kinetic term for the
gauge fields, and explicitly allow for the inclusion of $\theta$ angles,
so the total action is
\eqn\eSVii{
\eqalign{
S = \int_{\Sigma} d^2z\,d^4\theta \left[ \sum_{i=1}^n
\Phibar_i {\rm exp}\left( 2\sum_{a=1}^{n-d} Q_i^a V_a \right)
\Phi_i - \right.
&\left.\sum_{a=1}^{n-d} {1\over 4e_a^2} {\overline\Sigma}_a\Sigma_a -
\sum_{a=1}^{n-d} r_a V_a \right] +\cr
&\int_{\Sigma} d^2 z
\sum_{a=1}^{n-d} {\theta_a\over 2\pi i} f_a\ ,\cr
}}
where $f_a$ is the curvature of the gauge connection, and
$\Sigma_a = {1\over\sqrt 2} \overline {D}_+ D_- V_a$ is the
(twisted chiral) gauge-invariant field
strength associated to the gauge field $V_a$ with component expansion
\eqn\ecompsig{
\Sigma = \sigma -
i\sqrt{2}(\theta^+ {\bar\lambda}_+ + \thetabar^-\lambda_-)
+\sqrt{2}\theta^+\thetabar^-(D - f) + \cdots\ .
}
The last two terms in
\eSVii\ can be rewritten as
\eqn\eSdt{
S_{D,\theta} =
\sum_{a = 1}^{n-d}\int d^2 z (-r_a D_a + {\theta_a\over 2\pi i} f_a)
= \int d^2 z d\theta^+ d\thetabar^-
\widetilde W(\Sigma)|_{\theta^-=\thetabar^+=0} + {\rm c.c.}\ ,
}
where
\eqn\ewtildei{
\widetilde W(s) = {i\over 2\sqrt 2}\sum_{a=1}^{n-d} \tau_a s_a
}
with
\eqn\xiu{
\tau_a = i r_a + {\theta_a\over 2\pi}\ .
}
This interaction is a twisted superpotential for the twisted chiral
fields $\Sigma_a$. As we shall see, there are regions in which these
are the low-energy degrees of freedom. Then integrating out the
massive chiral fields we will obtain an effective action for
$\Sigma_a$. As is familiar with chiral superpotentials, this effective
(twisted) superpotential is constrained by the requirements of holomorphy
in the fields $\Sigma$ and in the couplings $\tau$.\foot{A
holomorphic dependence on the
couplings in a two-dimensional theory is familiar. For a recent, more general
application in four dimensions see \rpower .} In
some cases this
property, together with computable limiting properties,
will suffice to yield an exact expression.

The action \eSVii\ is invariant under a large group of global
symmetries, in
addition to the local $G$ symmetry. Of special importance is the
existence of left- and right-moving $U(1)$ $R$-symmetries. We can
choose these so that under  the left-moving $R$-symmetry
$U(1)_L$ the charged fields are $(\psi^i_-,F^i,\sigma_a,\lambda^-_a)$
with charges $(-1,-1,-1,1)$
(and
of course their complex conjugates with opposite charges),
while under the right-moving $R$-symmetry
$U(1)_R$ the charged fields are
$(\psi^i_+,F^i,\sigma_a,\lambda^+_a)$ with charge $(-1,-1,1,1)$.
These symmetries suffer from gauge
anomalies, given by
\eqn\eanom{
\Delta (Q_R) = -\Delta (Q_L) = -{1\over 2\pi}
\sum_{i=1}^n \left( \sum Q_i^a \int_\Sigma d^2 z\,f_a \right)\ .
}
The nonchiral combination $Q_L + Q_R$ is conserved. In addition there
is a group $H = U(1)^{n-d}$ of nonanomalous chiral global symmetries,
acting by phases on $\Phi_i$. This is the ungauged subgroup of the
full $U(1)^n$ phase symmetry.\foot{This is the full symmetry group in
generic cases. When the charges $Q_i^a$ are degenerate or satisfy
appropriate divisibility conditions the group is actually larger. In
our two examples the groups are $U(5)$ and $U(2)\times U(3)\times
U(1)$ respectively.}

\subsec{The Low-Energy Limit}

As mentioned above, this model is expected to reduce to the
nonlinear sigma model with target space $V$ in the infrared (at strong
coupling). We can see this explicitly by solving for the auxiliary
fields $D_a$ (setting all the gauge couplings equal)
\eqn\eD{
D_a = -e^2 ( \sum_{i=1}^n Q_i^a |\phi_i|^2 - r_a)\ .
}
Comparing to eqn.~\emu\ we see that $D_a = -e^2\mu_a$ are just the
components of the moment map to within an irrelevant constant.

We are interested in finding the space of classical ground states of
the theory. To this end we consider the potential for the bosonic
fields
\eqn\eU{
U = \sum_{a=1}^{n-d} {{(D_a)}^2\over 2 e^2} +
2\sum_{a,b=1}^{n-d} {\overline\sigma}_a \sigma_b
\sum_{i=1}^n Q_i^a Q_i^b |\phi_i|^2 \ .
}
Setting $U =0$ we obtain the restriction to $D^{-1}(0)$.
For values of $r$ in the appropriate range (see below) solving $D=0$
will lead to expectation values for the $\phi_i$ which break $G$ by
the Higgs mechanism\foot{The attentive reader will wonder at the
appearance of a Higgs phase, since the model is usually thought to
exhibit confinement. In fact the question is moot. Gauge-invariant
correlation functions cannot settle the confinement
issue. Certainly the Higgs
description is valid at weak coupling. We
thank E.~Rabinovici and N.~Seiberg for discussions of this point.}
to a discrete subgroup, and
lead to a nondegenerate mass matrix for the complex scalars $\sigma$
whose expectation values thus vanish. When this obtains we see that
the manifold of gauge-inequivalent vacua is indeed
\eqn\evac{
D^{-1}(0)/G = V(r) \ .
}
The massless modes (classically) will be oscillations of $\phi,\,\psi$
tangent to $V$. Thus, when $r$ lies in the K\"ahler cone of $V$ we
obtain the nonlinear sigma model with target space $V$ as the
low-energy limit. The discussion in the previous subsection then
identifies the $\tau$ coordinates with the canonical coordinates on the
moduli space of K\"ahler structures on $V$, since the class of the
K\"ahler form $\omega_r$ on
the symplectic reduction $V(r)$ will depend linearly on $r$.

What happens when $r$ lies outside this cone? For $r$ outside the cone
$\CK_c$ defined above, the equation $D=0$ has no solutions and \eU\
suggests that supersymmetry is broken. We will find later that this is
not the case (in general the theory has a nonzero Witten index so this
is not too surprising). For now, we restrict attention to $r\in\CK_c$.
This is still larger than the K\"ahler cone, so there are
values of $r$ for which we cannot expect the model to reduce to the
nonlinear sigma model with target space $V$ at low energies. In
general, the classical analysis predicts the existence of a set of
codimension-one cones on which the model is singular. The
singularities occur whenever there are solutions to $D=0$ which leave
a continuous subgroup of $G$ unbroken. With $\phi$ set to one of these
solutions, the $\sigma$ field associated to this subgroup has a flat
potential and the space of classical ground states is noncompact.  The
form of \eD\ shows that the subgroup generated by $g_a$ can be
unbroken when $D=0$ is consistent with $\phi_i=0$ for all $i$ such
that $Q_i^a\neq 0$. This in turn constrains $r$ to lie in a
codimension-one cone.

The singular loci will divide $\CK_c$ into components, one of
which will be the K\"ahler cone $\CK_V$. The low-energy theory in
other components will have a different
interpretation. An equivalent characterization of the singularities is
as those values of $r$ at which some component of $F$ disappears. On
the two sides of the singular locus the sets $F$ differ, leading to
topologically distinct quotients $V(r)$.
Several possible types of models can be encountered
here. The first possibility is that the quotient space
is a smooth manifold.
In this case the considerations of the previous paragraph hold and the
manifold in question is a toric variety. There can be any number of
such topologically distinct quotients in the parameter space of a
model; they are related by birational transformations.
Another possibility is that singular quotient spaces arise. In
this case the holomorphic interpretation is not as clear. In
particular, the interpretation of $r$ as coefficients in an expansion
of the K\"ahler class requires some generalization here (see \rmondiv).
In summary, the classical analysis predicts the existence of several
``phases'' \rphases\ of the model, separated in $r$-space by
singularities. We will refine this picture in what follows.

\line{\hrulefill}\nobreak
\noindent{\it Example 1.}\par\nobreak
For  \CP 4, we have
$D = -e^2(\sum_{i=1}^5 |\phi^i|^2-r)$. Thus we see that
$\CK_c = \{r>0\} = \CK_V$; in this region the quotient space is just $V$.

\noindent{\it Example 2.}\par\nobreak

In the second example we have
\eqn\edd{
\eqalign{
D_1 &= -e^2(|\phi_3|^2+|\phi_4|^2+|\phi_5|^2+|\phi_6|^2 - r_1)\cr
D_2 &= -e^2(|\phi_1|^2+|\phi_2|^2-2|\phi_6|^2 - r_2)\ .\cr
}}
We see that $\CK_c$ is given by the inequalities
\eqn\eK{
\eqalign{
r_1 & > 0 \cr
2r_1 + r_2 &> 0\ .\cr
}}
In this cone there is a ray (codimension-one cone) $r_2 = 0,\ r_1>0$
on which we can have $\phi_1=\phi_2=\phi_6=0$ solving $D=0$. When this
obtains we see from \egact\ that $g_2$ is unbroken; hence $\sigma_2$
is free. The existence of this noncompact component of field space
means that the theory is singular. This singularity separates the
first quadrant from the cone $r_2<0,\ 2r_1 + r_2 > 0$. In
the first quadrant \edd\ shows that the excluded set is precisely \eF .
This region is thus the K\"ahler cone $\CK_V$.
For $r$ in the other cone in $\CK_c$, we see from \edd\ that the
excluded set is
$F = \{\phi_6=0\}\cup\{\phi_1 =\phi_2 = \phi_3 = \phi_4 = \phi_5 = 0\}$.
Thus the quotient space is topologically distinct from $V$.  In some sense
this phase corresponds to the original (unresolved) weighted
projective space. (Following \rAGM, we refer to this as the ``orbifold''
phase since the space of classical vacua has orbifold singularities.)
In the holomorphic version this means that since
$\phi_6$ is nonzero we can use $g_2$ to fix it (say at $\phi_6=1$); the
remaining symmetry ($g_1^2 g_2$) is the expected $\IC ^*$ action on
the homogeneous coordinates.  Figure~1 shows this structure in $r$
space.
\par\nobreak\line{\hrulefill}

\iffigs
\midinsert
\centerline{\epsfxsize=3.1in\epsfbox{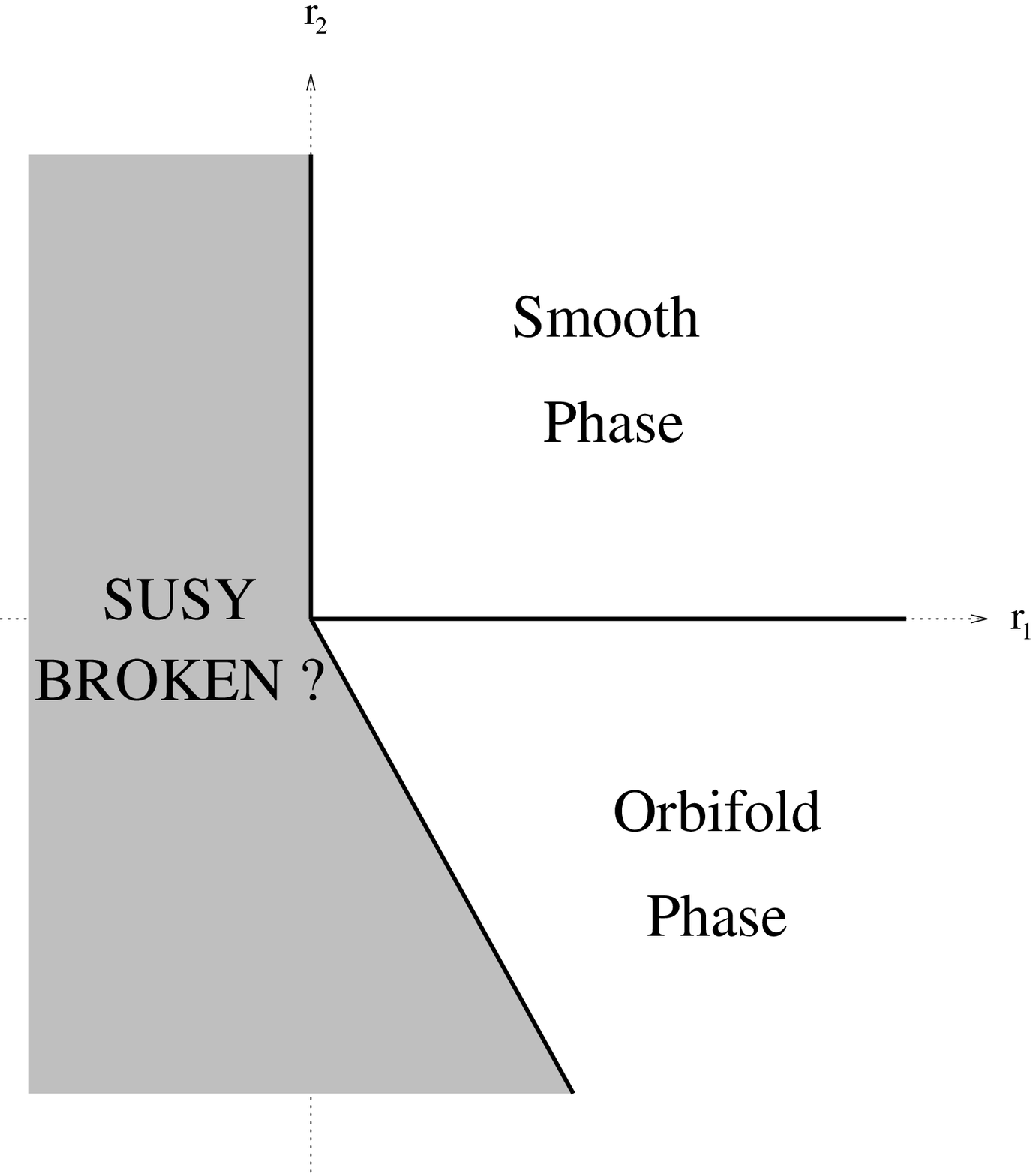}}
\centerline{Figure 1. Classical phase diagram.}
\endinsert
\fi

\subsec{Singularities and Quantum Corrections}

The classical analysis of the preceding subsections led us to the
conclusion that for $r\in \CK_V$ the GLSM is equivalent in the
low-energy limit to a nonlinear sigma model with target space $V$,
determined by the K\"ahler form obtained via symplectic reduction. In
terms of a fixed topological basis for $H^2(V)$ this is proportional to
$\omega = \sum_{a=1}^{n-d} r_a \eta_a$, and the cone $\CK_V$ is the
K\"ahler cone. The $B$-field on $V$ is likewise linearly related to
the $\theta$ angles. In terms of the canonical coordinates introduced
in section two this classical result is
\eqn\eclast{
t_a = \tau_a\ .
}
This is expected to approximate the exact
answer for values of $r$ deep in the interior of the cone $\CK_V$,
because there the model is weakly coupled (from our normalization of
the gauge multiplet it follows that $1/r_a$ is the coupling constant
for the $a$th factor of $G$)
and the classical analysis should be approximately valid.

As we decrease the value of
$r$ we will find quantum corrections to these results. There is a
natural deformation of the low-energy model (given by deforming the
K\"ahler class) which one expects to correspond to the deformation of
the ultraviolet theory by changing $r$. Thus we can expect that
at least for sufficiently large $r$ (we use this language somewhat
loosely; what we mean is $r$ deep in $\CK_V$, so that all of the gauge
couplings are small) the low-energy theory is given by the nonlinear
sigma model with target space $V$ and some K\"ahler form. The quantum
corrections then take the form of corrections to the
simple relation between the parameters of the low-energy theory and
those of the original model. The corrections are constrained by $N=2$
nonrenormalization theorems as follows.
Correlation functions which are holomorphic in $\tau$ in the
microscopic theory (correlators of twisted chiral fields~--~see the
next subsection) will map to correlation functions holomorphic in
$t$. This implies that $t(\tau)$ is holomorphic. Thus the perturbation
series for $t = \tau + a_1 + a_2 r^{-1} + \cdots$ must
terminate at $a_1$, since higher-order terms are not analytic in $\tau$.
Thus we expect that to all orders in $1/r$, \eclast\ is modified at most
by the addition of a constant term.
In addition there can be nonperturbative corrections which drop off
exponentially at large $r$. These corrections are of course important
in our application of the GLSM to a study of the low-energy nonlinear
model. We will find it convenient to compute them rather indirectly,
using the properties of instanton corrections to the correlation
functions. In the models of this section, the corrections will be
found to vanish and in fact \eclast\ holds {\it exactly}.

As $r$ approaches the boundaries of $\CK_V$ the classical theory is
singular signaling a possible breakdown of the approximation. The
singularity arises from the region in field space in which the scalars
$\sigma$ are large. We now turn to a more careful study of this
region. We will find that some of the singularities predicted by the
classical analysis are true singularities while others are removed by
quantum corrections. Further, we will find as expected that
supersymmetry breaking does not occur and explicitly find the vacuum
structure outside the cone $\CK_c$.

\nref{\rsidney}{S. Coleman, ``More on the massive Schwinger model,''
Ann.\ Phys.\ {\bf 101} (1976) 239--267.}

Let us then consider the large-$\sigma$ region in field space. In this
region the scalars $\ph_i$ are massive and their expectation values
vanish because of the second term in \eU . Then $G$ is unbroken and
the theory is approximated by free $N=2$ gauge theory, weakly coupled to
heavy matter fields. In this theory, as is well-known, the
Fayet-Illiopoulos term breaks supersymmetry at the classical level
(eqn.~\eU\ gives $U\ge \sum_a r_a^2/2e^2$). In fact, as one would expect by
holomorphy, the $\theta$ angles, which we have ignored thus far, also
break supersymmetry. These are equivalent to a constant background
electric field \rsidney\ and the ground-state energy is
\eqn\eUfree{
U(r,\theta) = {e^2\over 2}
\sum_{a=1}^{n-d}
\left( r_a^2 + \left({{\hat \theta_a}\over2\pi}\right) ^2\right) \ ,
}
where $\hat{\theta}_a = \theta_a + 2\pi m_a$ for an integer $m_a$ such that
\eUfree\ is minimized. Physically, this shift reflects the fact
that a field corresponding to $|\theta|>\pi$ will be
screened by pair creation.

We have neglected the chiral fields. Integrating them out
will lead to small corrections suppressed by their large mass (of
order $|\sigma|$), unless they appear in the loops of divergent
graphs. In fact, there is precisely one relevant divergence \rphases ~--~the
one-loop correction to the expectation value of
the auxiliary field $D$. This arises through boson loops; the
$\sigma$-dependence enters through the effective mass of these
bosons. Thus\foot{We thank E. Witten for pointing out to us that
the corresponding formula in \rphases\ needed a correction, as given here.}
\eqn\edeld{
{1\over e^2}\langle D_a\rangle_{\rm 1-loop} =
\sum_{i=1}^n Q_i^a \int {d^2 k\over (2\pi)^2} \,
{1\over k^2 + 2Q_i^aQ_i^b\sigmabar_a\sigma_b}\ .
}
Cutting off the divergent integral we have
\eqn\edeldii{
{1\over e^2}\langle D_a\rangle_{\rm 1-loop} =
{1\over 4\pi} \sum_{i=1}^n Q_i^a
\log \left( {\Lambda^2\over 2Q_i^aQ_i^b\sigmabar_a\sigma_b}\right)\ .
}
The correction can be interpreted as a $\sigma$-dependent shift in
$r$. There is also a corresponding shift in $\theta$, verified by
computing the one-loop correction to $\la \lambda_+\bar\lambda_-\ra$.
We incorporate both of these in a perturbative correction to the twisted
chiral superpotential for $\Sigma$ \eSdt , valid for large $\sigma$
\eqn\ewtilde{
\widetilde W(\Sigma) = {1\over2\sqrt 2} \sum_{a=1}^{n-d}\Sigma_a
\left( i\hat\tau_a-{1\over 2\pi}
\sum_{i=1}^n Q_i^a
\log (\sqrt{2}\sum_{b=1}^{n-d}Q_i^b\Sigma_b/\Lambda)\right)\ .
}
This corrected twisted superpotential leads to a modification in the
potential energy
\eqn\eUcor{
U(\sigma) = {e^2\over 2} \sum_{a=1}^{n-d}\left\vert i\hat\tau_a -
{\sum_{i=1}^n Q_i^a\over 2\pi}
(\log (\sqrt{2}\sum_{b=1}^{n-d}Q_i^b\sigma_b /\Lambda )+ 1)
\right\vert^2\ .
}

The computation we have performed is valid at large values of
$\sigma$. One observes that deep in the interior of $\CK_V$ \eUcor\
will vanish (indicating new vacua missed by the classical analysis)
for small $\sigma$. Here the approximation fails; more precisely,
these classical vacua are unstable. The true vacuum states
are found at large $\phi$ and $\sigma=0$ as described in the previous
subsection. This can change at the boundaries of this cone,
where singularities were predicted classically. These singularities
arise precisely from the existence of low-energy states at large
$\sigma$. We first consider values of $r$ in the vicinity of the
classical singularity, and very far from the origin in $r$-space. In
this region the classical description is a good approximation for most
of the theory, the exception being the strongly-coupled dynamics of
the unbroken gauge symmetry, which we  relabel $V_1$.
The fields $\sigma_a$ for $a\neq 1$ are massive; the low-energy
degrees of freedom are the chiral fields neutral under $g_1$ and the
gauge multiplet $V_1$. The two sectors are weakly coupled by the
massive chiral fields charged under $g_1$. The effective twisted
superpotential for $\Sigma_1$ after integrating out these massive
chiral fields can be computed by setting $\sigma_a=0$ for $a>1$ in
\ewtilde , in which case the potential
\eUcor\ reduces to
\eqn\eUbdry{
U(\sigma_1) = {e^2\over 2}
\left\vert i\hat\tau_1 - {\sum_{i=1}^n Q_i\over 2\pi}
(\log (\sqrt{2}Q_i\sigma_1 /\Lambda )+ 1)\right\vert ^2\ ,
}
where $Q_i\equiv Q_i^1$.
The effect of
this correction on the physics differs dramatically, depending on
whether or not the equality
\eqn\esumq{
\sum_{i=1}^n Q_i = 0
}
holds.

Let us first address the situation in which \esumq\ is satisfied. Then
\eUbdry\ reduces to
\eqn\eUdiv{
U(\sigma_1) = {e^2\over 2} \, \left\vert i\hat\tau_1 - {1\over 2\pi}
\sum_{i=1}^n Q_i\log (Q_i)\right\vert ^2\ .
}
Indeed in this case
the integral \edeld\ is convergent, and the correction to $\tau$ is
in fact finite and $\sigma$-independent
\eqn\eref{
\tau_{\rm eff} = \tau + {i\over 2\pi} \sum_{i=1}^n Q_i \log (Q_i)\ .
}
This is in accord with the fact that instantons in this factor of $G$
do not contribute to the anomaly. The linear twisted superpotential is
the unique form that does not break the symmetry explicitly. Since the
correction in $\sigma$-independent we will assume that it holds away
from the region (large-$\sigma_1$) in which the computation is valid.
For a particular value of $\theta_1$ (0 or $\pi$)
\eUdiv\ approaches zero as
$r_1\to 0$, leading to the singularity in the low-energy theory
discussed earlier. For other $\theta_1$, however, there is no
singularity; the minimum energy at which
the large-$\sigma_1$ region of field space becomes accessible is
nonzero, the space of supersymmetric ground states is compact and the
theory nonsingular. Thus the singularities occur in codimension-two
subspaces of parameter space. This will be very important in what
follows. In particular, it means that we can continuously deform the
theory ``around'' the singular loci to connect models in different
``phases'' \refs{\rphases,\rAGM}.
We stress that \eref\ is valid {\it only}\/ in the vicinity
of one of the asymptotic components of the singular locus (i.e.~$r_1$
small and all other $r_a$ large). It is useful for predicting the
location of the singularities far from the origin in $r$-space. It
should not, however, be confused with perturbative corrections to
\eclast . We will compute these in the sequel.

Things are very different if \esumq\ is not satisfied.
In this case the integral \edeld\ is
divergent, quantum corrections are large, and we can expect
qualitative modifications to our classical conclusions.
Indeed, in this case \eUbdry\ grows as $|\log (\sigma_1)|^2$ at large
$\sigma_1$, so the field space accessible to
very low-lying states is effectively compact and there is no
singularity for any value of $\tau_1$.\foot{There will always be one
such symmetry in $G$ if $V$ is compact. This corresponds to the fact
that the semipositive anticanonical class is nonzero for a compact toric
variety
$V$. It is also in accord with the existence of an anomalous
$R$-symmetry, which means that one $\theta$ angle could be
eliminated by field redefinitions.} On the other hand, continuing past
the nonexistent ``singularity'' to a region in which $r_1$ differs in
sign from  $\sum_i Q_i$, we now find at large $|r_1|$ new vacuum states.
In these states the expectation values of $\phi_i$ are restricted to
those values invariant under $g_1$ (hence leaving $\sigma_1$
massless), and the expectation value of $\sigma_1$ is determined by
the requirement that \eUbdry\ vanish. These values are given by
\eqn\enuvac{
\sigma_1 = {\Lambda\over e}
\exp \left[ {2\pi i\hat\tau_1\over\sum_i Q_i}\right]\ .
}
Note that for large $|r_1|$ these values lie at large $|\sigma_1|$, justifying
the use of \eUbdry .
There will be $\sum_i Q_i$ of these because of the freedom to shift
$\hat\tau_1$ by an integer. The boundaries of $\CK_c$ are necessarily
of this type, and the analysis above shows that supersymmetry is not
broken for values of $r$ outside this cone, correcting the classical
prediction.

In all, we have been led to modify our classical
analysis as follows. First, the regions outside the cone $\CK_c$
are in fact continuously connected to the interior; there is no
supersymmetry breaking. However, in these regions the physical theory
has additional vacuum states involving the gauge multiplet in a
nontrivial way and cannot be expected to reduce to a nonlinear sigma
model at low energies. Furthermore, some of the regions inside
$\CK_c$ may be of this type. This will be the case if in reaching
them from the K\"ahler cone we have crossed a boundary for
which \esumq\ does not hold. The models for which a (standard) geometrical
interpretation is expected will occupy a smaller cone $\CK_q$; this is
the cone of $r$'s for which the corresponding
symplectic reduction $V(r)$ is a space of
complex dimension $d$. Finally, our na\"{\i}ve predictions for the
singularities of the model are corrected. The true singular locus is,
as we shall see, quite complicated in general. What we have found is
that far from the origin it asymptotically approaches the cones
in $r$ predicted classically, shifted as in \eref .
At each of these the singularity is restricted to a
particular value of $\theta$.
The fact that the singularity occurs in complex codimension one
(consistent with the holomorphic properties of these models) is
important; it means one can connect two
points lying in different regions in $r$-space without
encountering a singularity. Despite this fact, we will follow \rphases\
and refer to the theories in regions of $r$-space
separated by singularities as different phases of the model.

\line{\hrulefill}\nobreak
\noindent{\it Example 1.}\par\nobreak
\nref\rcp{S. Cecotti and C. Vafa, ``Exact results for supersymmetric
sigma models,'' Phys.\ Rev.\ Lett.\ {\bf 68} (1992) 903--906,
hep-th/9111016.}%
The singularity at $r=0$ does not satisfy \esumq. We conclude that
there is in fact no singularity, and that the $r<0$ theory exists but is not
described by a nonlinear sigma model. A different interpretation of
the analytic continuation of this model was proposed in \rcp.

\noindent{\it Example 2.}\par\nobreak
The singularity at $r_2=0$ does satisfy \esumq. There is thus a
true singularity (at $\theta=0$; we can of course choose a path in
parameter space connecting the two ``phases'' that does not meet this
singularity) and it occurs at $\tau_2 = i{\log 2\over \pi}$. In this
example $\CK_q = \CK_c$. The singularities at the boundary of
$\CK_q$ do not satisfy \esumq\ and are thus not true singularities.
\par\nobreak\line{\hrulefill}

\subsec{Determining the Singular Locus}

\nref\rILS{K. Intriligator, R. Leigh, and N. Seiberg, ``Exact
superpotentials in four dimensions,'' Phys. Rev. {\bf D50} (1994)
1092--1104, hep-th/9403198.}%

The analysis in the previous
subsection has corrected the classical predictions for the
singularities of the model far from the origin. The exact singular
locus of the model diverges from the semiclassical predictions due to
the effects of instantons in the Higgs sector of the theory, and
interpolates between the asymptotic pieces computed above. To find
this locus would apparently require computations in the strongly-coupled
interior of $r$-space near the origin. This appearance is misleading, however,
because in fact the twisted superpotential \ewtilde\ is an exact
expression. Perturbative corrections at higher order are of course
precluded by holomorphy in $\tau$. We claim, however, that
nonperturbative corrections are absent as well. This will be borne out
in the following subsections by explicit computations of the exact
correlation functions and a study of their singularities.\foot{Similar
methods were used in \rILS\ to compute exact superpotentials in four
dimensions. N. Seiberg has recently presented an argument for the
assumption made in this work (and justified by its results as we do
here) that the exact effective superpotential is linear in the
tree-level coupling.}  We defer a complete argument for the vanishing
of these corrections to \rdoit .

The equations of motion for $\sigma$ which follow from the twisted
superpotential interaction
\eqn\eomsig{
{\partial \widetilde W\over\partial\sigma_a} = 0\ ,
}
can be rewritten using \ewtilde\ as
\eqn\ealsol{
\prod_{i=1}^n \left( {\sqrt{2} e\over\Lambda}
\sum_{b=1}^{n-d} Q_i^b\sigma_b\right)^{Q_i^a} = e^{2\pi i\tau_a}\ ,
a = 1,\ldots, n{-}d\ .
}
These equations will lead to some exact properties of the correlation
functions of the $\sigma_a$ in what follows, but we first show
how they lead directly to the exact singular locus. This happens
because the singularity, as observed above, arises from the existence
of solutions to \ealsol\ (nontrivial vacuum states) at large $\sigma$.
Let us choose a
basis for  $G$ such that $\sum_{i=1}^n Q_i^a = Q\delta^{a1}$
for some constant $Q$, so that
\esumq\ is satisfied for $a>1$. Then we have seen that the
large-$\sigma_1$ region of field space does not lead to a singularity;
the ``geometric'' phases of the theory are then in the region (say) $r_1>0$,
and for the moment we concentrate on this region. Then the field
$\sigma_1$ is massive and we integrate it out to obtain the low-energy
effective action. We will in fact simply set $\sigma_1=0$ and drop it,
an approximation which should be valid for sufficiently large $r_1$.
At generic large values of the other $\sigma_a$ the chiral fields
which remain massless are the set $\{\phi_i\}_{i\in I^c}$ of fields
satisfying $Q_i^a=0$ for $a>1$. Expectation values for these do not
break the subgroup $H\subset G$ generated by $g_a$ for $a>1$. The
remaining chiral fields are massive and will
be integrated out as well. This will lead to an effective twisted
superpotential for the massless $\sigma_a$, yielding as equations of motion
\eqn\ealsolii{
\prod_{i\in I} \left(
\sum_{b=2}^{n-d} Q_i^b\sigma_b\right)^{Q_i^a} = e^{2\pi i\tau_a}\ ,
a = 2,\ldots, n{-}d\ ,
}
where we have used \esumq\ to eliminate the $\Lambda$-dependent constant.

We now note that the equations \ealsolii\ are all
homogeneous (of degree zero) in $\sigma$ because of \esumq . This
homogeneity means two things. First, if the equations have a solution
at all there are solutions at arbitrarily large $\sigma$ and the model
is indeed singular. Further, because \ealsolii\ can be
interpreted as $n{-}d{-}1$
equations for the $n{-}d{-}2$ {\it ratios\/} of the nonzero $\sigma_a$'s,
the equations are overdetermined. They can be considered the
parametric equations of some subvariety of $\tau$-space of dimension
$n{-}d{-}1$
(recall $\tau_1$ is unconstrained), along which the models are
singular.

This subvariety (a part of the ``singular locus'' in the parameter
space) will have several components at
large-$r$, and by inspection these can be seen to coincide with some
of the singularities in this region predicted semiclassically above.
In these
limits the solutions indeed tend to the limiting direction in
$\sigma$-space which was predicted from the classical action. In the
interior of $r$-space, the true singular locus interpolates smoothly
between these various limits. At a generic point on it the $\sigma$
vacua are in a generic direction in $\sigma$-space (generic, that is,
among $\sigma$'s constrained by
$\sigma_1=0$), leading to nonzero masses for all of the chiral fields
$\phi_i,\ i\in I$. We have made a computation at large
$r_1$ to describe this part of the singular locus,
but note that the result is completely
independent of $\tau_1$. This is actually to be expected. The $\tau_1$
dependence of correlation functions is determined by the anomalous
$R$-symmetry. Thus any correlator of the $\sigma_a$ is given
by $q_1^m f(q)$ for some $m$, with $f$ independent of $q_1$. The
singularities of such an object cannot depend upon $q_1$.

The discussion in this subsection has until now neglected one subtlety.
We have computed the singular locus as the locus at which \ealsol\
have solutions. These equations were obtained by integrating out all
of the chiral fields charged under any of the symmetries involved. Of
course, there can be intermediate situations in which some factors of
$G$ are Higgsed by massless chiral fields and others are in the
Coulomb phase. (This simply generalizes the situation in the previous
paragraph in which $g_1$ was Higgsed.)
In this case one should integrate out only the massive
chiral fields, and seek solutions to the equations of motion for the
massless twisted chiral $\sigma_a$ fields. This will lead to extra
components of the singular locus.
Indeed the divisor computed above will reproduce only some of the
asymptotic singularities predicted in the semiclassical approximation.
The extra components will interpolate smoothly between the remaining
limits. To extract a coherent picture from this idea we need to
understand which limits are joined, or in other words which subsets
of the chiral fields can be integrated out to yield a consistent
low-energy theory for the remaining light fields. The structure we are
looking for is the following. Consider a linear subspace $h$ of
$S = \{s\in \IR^{n-d} | \sum_{a=1}^{n-d} s_a \sum_{i=1}^n Q_i^a=
0\}\sim \IR^{n-d-1}$, of dimension $k$, and let $H\subset G$ be the
corresponding subgroup (of rank $k$). We
divide the chiral fields into two subsets, the set $\{\phi_i\}_{i\in
I}$ of fields charged under some element of $H$ and its complement,
the set $\{\phi_i\}_{i\in I^c}$ of $H$-invariant fields.
We then ask whether a component of the singular locus exists such that
at a generic point in it the $\sigma$ field is in a generic direction
within the subspace
$h$. This means the chiral fields $\phi_i$ are massless for $i\in
I^c$ and that their expectation values will lead to a mass matrix for
$\sigma$ whose zero eigenspace is $h$. We then compute a
twisted superpotential $\widetilde W$ for the massless components (in
$h$) of $\sigma$. We can choose coordinates such that these massless
components are $\sigma_{n-d-k+1},\ldots,\sigma_{n-d}$, and write
\eqn\ewtildeh{
\widetilde W(\Sigma) = {1\over2\sqrt 2} \sum_{a=n-d-k+1}^{n-d}\Sigma_a
\left( i\hat\tau_a-{1\over 2\pi}
\sum_{i\in I} Q_i^a
\log (\sqrt{2}\sum_{b=n-d-k+1}^{n-d}Q_i^b\Sigma_b/\Lambda)\right)\ .
}
As above we will then find solutions to the equations of motion
derived from this for $\tau$ in a codimension-one subvariety, of the
form
\eqn\ealsolbis{
\prod_{i\in I} \left(
\sum_{b=n-d-k+1}^{n-d} Q_i^b\sigma_b\right)^{Q_i^a} = e^{2\pi i\tau_a}\ ,
a = n{-}d{-}k{+}1,\ldots, n{-}d\ .
}

When will this construction lead to a component of the singular locus?
The
special status of the origin in $r$-space as the unique point at which
all of the gauge symmetries can be unbroken has been removed (there is
in fact no such point). Instead, along the singular divisor, $\sigma$
rotates smoothly from one asymptotic direction to another. Thus we
expect the components of the true singular locus not to end in the
strong coupling region (as those corresponding to convex cones do
classically), but to have as boundaries only the semiclassical limits.
To check this we consider the classical limit and compute the
$D$-terms using only the massless fields $\{\phi_i\}_{i\in I^c}$.
Setting these to zero will restrict $r$ to a cone, which is precisely
the cone spanned by the vectors $\{Q_i\}_{i\in I^c}$. If this cone is
in fact all of $\IR^{n-d-k-1}$, then the singular locus computed above
can extend to the classical boundaries of $\tau$-space and there is a
component of the singular locus given by the above data. Otherwise, we
will find that $h$ is contained in a larger subspace (possibly all of
$S$) and that in the strong coupling regime the singular
locus interpolates between the asymptotic limits we have used and
others so that in the interior $\sigma$ rotates out of $h$. This
condition can be naturally expressed in terms of the combinatorial
presentation of $V$. Given a subset $I\subset\{1,\ldots ,n\}$ as
above, the condition on $Q_i$ is precisely the condition that the
set $\{v_i\}_{i\in I}$ is the set of lattice points in a
face of the polyhedron $\cal P$, the convex hull of the $v_i$.
This construction of the singular
locus which contains an irreducible component for each face of
$\cal P$ is familiar from studies of {\bf B} model moduli spaces, as
we will discuss in detail in section five.

\line{\hrulefill}\nobreak
\noindent{\it Example 1.}\par\nobreak
In this example the discussion is completely trivial, since there {\it
is\/} no singular locus.

\noindent{\it Example 2.}\par\nobreak
The one true singularity in this model, at $\tau_2 = i{\log 2\over
\pi}$, is once more a rather trivial example of the above with $H$
generated by $g_2$.
We will see a better example of these phenomena in the models of
section four.
\par\nobreak\line{\hrulefill}

We have already managed to extract some exact results from our
one-loop computation of \ewtilde . But we can do more; the correlation
functions of the twisted chiral fields $\sigma_a$ are holomorphic in
$\tau$ (and do not depend on the location in the worldsheet at which
the operators are inserted). We can use \ewtilde\ to obtain exact
relations among these which in some cases suffice to compute them all
exactly.\foot{We thank N. Seiberg and E. Witten for
urging us to solve the model in this fashion.}
In fact, for values of $r$ outside the cone $\CK_c$ (recall such
values always exist for compact $V$) the theory is far simpler to
analyze. The massless fields are just $\sigma_a$ and these are
governed by the interaction \ewtilde . The correlation functions are
then simply products of the expectation values for $\sigma_a$ derived
from \ealsol . Note that here we do {\it not\/} drop the first
equation (indeed there will be solutions at large $\sigma_1$ as in
\enuvac ); the equations are then not homogeneous and determine
isolated vacua for generic $\tau_a$ in the complement of $\CK_c$. The
equations \ealsol\ are then interpreted as constraints on the
correlation functions. Since the latter are holomorphic in $\tau$, the
relations must continue to hold when we analytically continue to other
regions in parameter space. Of course the $\sigma_a$ should then be
considered as operators and not as numbers. Since some of the $Q_i^a$
are negative, this equation may not make
sense as written; however, we can take combinations of \ealsol\ such
that the powers on the left-hand side are always positive. These then
give a set of nonlinear relations on the $\sigma_a$, as we
discuss in the next subsection.
In some cases
these are sufficient to determine the correlators completely.

\subsec{The Topological Model}

The above discussion has been somewhat imprecise. In particular, the
description we have given is a semiclassical one relevant at small
$e$. We have used this reasoning to understand the low-energy
(large-$e$) behavior. Ultimately, the justification for this is that
we will limit ourselves here to properties of the supersymmetric
theory which are in fact {\it independent}\/ of $e$. These
quasitopological properties are shared by the topological field theory
obtained by ``twisting". We now turn to discuss this theory. We will
initially base our discussion in the K\"ahler cone; later we shall see
how computations can be done in other phases.

Like any $N=2$ supersymmetric theory, the model described in the
previous subsections can be twisted to obtain a topological field
theory \rtopsim . We will study the {\bf A} twist. In this model,
the supercharges $Q_-$ and ${\overline Q}_+$ become worldsheet
scalars. These generate a $(0|2)$ dimensional supergroup $\cal F$ of
symmetries. The spin-$\half$ fermions $\lambda ,\ \psi^\pm$ and their
conjugates become either one-forms on $\Sigma$ or scalars.
Setting $Q = Q_- + {\overline Q}_+$ we have $Q^2=0$. If we restrict
attention to correlation functions of operators that are $Q$-closed,
then all $Q$-exact operators decouple (vanish in correlators). We can
thus consider a theory in which the operators are cohomology classes
of $Q$.
Correlation functions in this theory are independent of the
worldsheet metric because the twisted
energy-momentum stress tensor, which
couples to the metric, is itself $Q$-exact. In particular, all
correlators are scale-invariant (and the gauge kinetic term, coupling
to $e$, is also exact). Thus we can compute directly in the weak-coupling
(high energy) limit quantities which will
be relevant to the strongly-coupled theory in the infrared.

The twisting procedure will change the anomaly equation \eanom. By
modifying the spins of the fermions we have added to the gauge anomaly
\eanom\ a gravitational anomaly from the coupling of the fermions to
the spin connection on $\Sigma$. The modified equation is now
\eqn\eanomt{
\Delta (Q_R) = -\Delta (Q_L) = {d\over 2}\chi(\Sigma) -{1\over 2\pi}
\sum_{i=1}^n \left( \sum Q_i^a \int_\Sigma d^2 z\,f_a \right)\ ,
}
the additive constant being simply the difference between the number
of left- and right-moving fermions to which the current couples. The
difference $Q_L-Q_R$ is called ghost number.

In the nonlinear sigma model the cohomology of $Q$ is
precisely the de~Rham cohomology $H^*_{\rm DR}(V)$, discussed in
subsection {\it 3.1\/}. By the arguments of
the previous paragraph we expect to find a similar structure for the
local observables in the
linear model. The $Q$ variations of the fields are:
\eqn\eqvarg{\eqalign{
[Q, v_z] &= -i \lambda^+_z\cr
[Q, v_\zbar ] &= -i {\bar\lambda}^-_\zbar\cr
[Q,\sigma ]  &= 0 \cr
[Q,\sigmabar ] &= i\sqrt{2}({\bar\lambda}^+ - \lambda^-)\cr}
\qquad
\eqalign{
\{Q,\lambda^+_z\} &= 2\sqrt{2}\p_z\sigma\cr
\{ Q,\lambda^-\} &= i(D+f)\cr
\{ Q,{\bar\lambda}^+\} &=i(D+f)\cr
\{ Q,{\bar\lambda}^-_\zbar\} &= 2\sqrt{2}\p_\zbar\sigma\cr
}}
for each of the gauge multiplets, and
\eqn\eqvarc{\eqalign{
[Q,\phi^i] &= \sqrt{2} \chi^i\cr
[Q,\phibar^\ibar ] &= -\sqrt{2}{\bar\chi}^\ibar\cr}
\qquad\eqalign{
\{ Q,\chi^i \}&= 2 \sum_a Q_i^a \sigmabar_a\phi^i\cr
\{ Q,{\bar\chi}^\ibar \}&= -2\sum_a Q_i^a \sigmabar_a\phibar^\ibar\cr
\{ Q,\psi^i_\zbar \}&= 2\sqrt{2}iD_\zbar\phi^i + \sqrt{2}F_\zbar^i\cr
\{ Q,{\bar\psi}^\ibar_z \}&= 2\sqrt{2}iD_z\phibar^\ibar +
\sqrt{2}{\overline F}_z^\ibar\cr
}}
for the chiral multiplets.

\nref\rgreg{S. Cordes, G. Moore, and S. Ramgoolam, ``Lectures on 2D
Yang-Mills theory, equivariant cohomology and topological field
theories,'' preprint YCTP-P11-94, hep-th/9411210.}%

The supersymmetry variations above lead us to
identify representatives for the cohomology of $Q$ in the space
of local operators:  every class can be represented by a
function of the $\sigma_a$'s. That is, the
$Q$-variations realize the Cartan model for $\cal G$-equivariant
cohomology of field space, where $\cal G$ is the group of gauge
transformations. A recent review of this procedure can be found in
\rgreg.
The cohomology is graded by ghost number, with $Q$ carrying ghost
number one; $\sigma_a$ carries ghost number two.
Note that the nonlinear relations \enonl\ are not satisfied by the
operators. We will identify the correct nonlinear relations using the
correlation functions as discussed in section two. We wish to compute
correlation functions in the low-energy model, and hence look for the
map between the operators in the nonlinear sigma model and the space
obtained in the GLSM. Restricting to the zero modes, the GLSM
operators will be given by $G$-equivariant cohomology. This in turn is
naturally identified with the cohomology of $V$, leading to the
identification $\sigma_a \sim \eta_a$ to within a normalization.

In a topological field theory, given a local $Q$-closed operator we
can form other $Q$-closed operators by the descent equations. These
lead to $Q$-cohomology classes expressed as the integrals over
cycles on the worldsheet of appropriate forms. The
two-form operators thus obtained yield deformations of the action that
preserve the topological invariance. Applying this method to
$\sigma_a$ leads to  $\int_{\Sigma} (D_a - f_a)$, the deformation
coupling to $\tau$. This is in line with the identification made
above, of course. Further, one can show that the operator coupling
to $\taubar$ is $Q$-exact. The correlation functions
of $\sigma_a$ will thus be holomorphic functions of $\tau$.

The nonlinear relations among the $\sigma_a$ will be determined by the
correlation functions we compute in the sequel. However, a subset of
these relations can be computed directly from the twisted
superpotential \ewtilde , as mentioned in the previous subsection.
If for each $i$ we fix an operator $\delta_i$
which in $Q$-cohomology is given by
\eqn\elinrd{
\delta_i \sim {\sqrt{2} e\over\Lambda}\sum_{a=1}^{n-d} Q_i^a \sigma_a\ ,
}
then we can write the relations \ealsol\ in the form
\eqn\ealbor{
\prod_{i=1}^n
\delta_i^{Q_i^a} =  e^{2\pi i\tau_a}\ .
}
Relations of this type had already been proposed by Batyrev \rbator ,
in the form
\eqn\ealbat{
\prod_{i=1}^n \xi_i^{Q_i^a} = e^{2\pi i t_a}\ .
}
In fact, for smooth $V$ Batyrev showed that
such relations would suffice to compute the quantum cohomology of $V$ (see
a fuller discussion in subsection {\it 3.9\/}). We will recover these
relations by a direct study of the moduli spaces of instantons
(refining the analysis of Batyrev) in the
sequel. Our direct approach has the advantage of not being limited to
smooth $V$, and will prove very useful in the next sections. At this
point, we can use \ealbor\ and \ealbat\ to find the normalization
constant in our identification of operators with divisors and to
verify that indeed \eclast\ is not corrected. Notice that
for values of $a$ satisfying \esumq\ the constants in \elinrd\ drop out of
\ealbor\ (in particular, $\Lambda$ does not appear as expected since
the instantons in this factor do not contribute to the anomaly).

\subsec{Reduction to Moduli Space}

The computation of correlation functions in the topological field theory
is greatly simplified by the existence of the $(0|2)$ dimensional
symmetry group $\cal F$. The path integral computing correlation
functions of $Q$-closed operators (for which $\cal F$-invariant
representatives can be chosen) reduces to an integral over the fixed
point set of $\cal F$. Fluctuations about this can be treated exactly
in the Gaussian approximation. Furthermore, exact cancellation between
the bosonic and fermionic determinants means they can be
altogether dropped. (Note that in general the ratio of determinants can
be $\pm 1$; in the present case it is in fact just 1, as we discuss
more fully in section four.)
In the nonlinear sigma model, the fixed point set of $\cal F$ is the
space of holomorphic maps
$\Sigma\to V$ and its irreducible components are classified by the
degree of the map. The contribution of each component can be expressed
as an intersection computation in that space, but the moduli spaces
are highly nontrivial and noncompact and explicit computations are
extremely difficult.
We will find a qualitatively similar picture in the linear model. The
essential difference will be that the moduli spaces will be simple
and the computation tractable.

The fixed point set of $\cal F$ is found by considering field
configurations annihilated by ${\overline Q}_+$ and $Q_-$. As in
\rphases\ this leads to
\eqna\einst
$$\eqalignno{
d\sigma_a &= 0 &\einst a\cr
\sum_{a=1}^{n-d} Q_i^a \sigma_a \phi_i &=0 &\einst b\cr
D_a + f_a &=0 &\einst c\cr
D_\zbar \phi_i &= 0 \ ,&\einst d\cr
}$$
where $D_\zbar$ is a covariant derivative constructed from the
gauge connection. The space of solutions to these will split into
``instanton sectors'' labeled by
\eqn\enins{
n_a = -{1\over 2\pi} \int d^2 z\, f_a\ .
}
The action for a solution of \einst{}\ is
\eqn\elins{
L = -2\pi i \sum_{a=1}^{n-d} \tau_a n_a\ .
}

\nref\rBD{
S. B. Bradlow and G. D. Daskalopoulos, ``Moduli of stable pairs for
holomorphic bundles over Riemann surfaces,'' Int. J. Math. {\bf 2} (1991)
477--513.}%
\nref\rGarcia{O. Garc\'\i a-Prado, ``Dimensional reduction of stable bundles,
vortices and stable pairs,'' Int. J. Math. {\bf 5} (1994) 1--52.}%

For $r$ in the interior of the cone, \einst{a{-}b}\ lead to $\sigma=0$,
leaving \einst{c{-}d}\ to determine $\phi$ and $f$. The equations are
of course gauge-invariant; the moduli space $\CM_\vn$ of solutions at
instanton number $\vn$ is a quotient of the space of solutions satisfying
\enins\ by the group of gauge transformations
\eqn\egauge{
\phi_i\to e^{i\Sigma_a Q_i^a \epsilon_a}\phi_i\ ,
\qquad v_a\to v_a - d\epsilon_a\ .
}
What makes the computation
tractable is the fact that this quotient can be recast as a toric variety.
The first equation \einst c\ is in fact invariant under \egauge\ with
$\epsilon$ a {\it complex}\/ function on $\Sigma$, while the second
(with $D$ expressed in terms of $\phi$
using \eD ) is not. The essential observation \refs{\rphases,\rBD,\rGarcia}
is that for each solution of \einst c\ there is at most one value
of $|\epsilon|$ transforming it to a solution of \einst d\ as well.
Thus $\CM_\vn$ can be expressed as the set of solutions of \einst c\
which can be transformed this way modulo complex gauge transformations.

\nref\rfunctor{D. A. Cox, ``The functor of a smooth toric variety,''
Amherst preprint, 1993, alg-geom/9312001.}%

By an appropriate complex gauge transformation we can make $v$ a
holomorphic connection. Then \einst c\ states simply that $\phi_i$ is
a holomorphic section of a line bundle over $\Sigma$ of degree
$d_i(\vn) = \sum_a Q_i^a n_a$. Restricting attention to $\Sigma$ of
genus zero, this is the line bundle $\CO (d_i)$. For $d_i<0$ such sections
do not exist and we have $\phi_i =0$. For $d_i\geq 0$ there is a
$(d_i+1)$-dimensional vector space of sections. These can be expressed as
homogeneous polynomials of degree $d_i$ in the homogeneous coordinates
$(s,t)$ on $\Sigma$
\eqn\epoly{
\phi_i = \phi_{i0} s^{d_i} + \phi_{i1} s^{d_i-1} t +
\cdots + \phi_{id_i} t^{d_i}\ ,
}
i.e., we can think of $\{\phi_i\}$ as a $G_{\IC}$-equivariant map
$\IC^2\to Y$. It is now easy to see which sets of sections do not lead
to a solution of \einst d\ for any $|\epsilon |$. These are those
solutions for which the image lies completely in $F$, determined by
$r$ as in subsection {\it 3.1}.
We denote this set of solutions $F_\vn$. We have not completely fixed the
complex gauge invariance. Complex gauge transformations with constant
$\epsilon$ will not affect our choice of connection, so we must still
quotient our space by the action of these. In summary, we have the following
expression for the moduli spaces (compare \rfunctor):
\eqn\eMn{
\CM_\vn = (Y_\vn -  F_\vn)/T
}
where $Y_\vn = \bigoplus_{i=1}^n H^0 (\CO(d_i) )$ and $T = G_{\IC}$ acts by
$\phi_{ij}\to (\prod_a \lambda_a^{Q_i^a}) \phi_{ij}$.
The components $\phi_{ij}$ of the $\phi_i$'s give
coordinates on $Y_{\vec n}$.  Since $\phi_i$ has $\max\{0,d_i+1\}$
components, we find
\eqn\edim{
\dim Y_{\vn}=\sum_{i:d_i\ge0}(d_i+1)
}
and hence
\eqn\edimm{
\eqalign{
\dim \CM_{\vn}=\dim Y_{\vn}&-(n-d)=\cr
\left(d+\sum_{i=1}^nd_i\right)
&+\sum_{i:d_i\le-1}(-d_i-1)\geq d+\sum_{i=1}^n d_i\ .\cr
}}

A similar analysis can be carried out for worldsheets $\Sigma$
of higher genus: each $\phi_i$ is a section of a line bundle ${\cal L}_i$
on $\Sigma$, and these line bundles must have compatible degrees.
(In fact, a bit more compatibility is required: there must be
isomorphisms among certain
tensor powers of these bundles~--~see \rfunctor\ for details of this
construction.)

The toric variety $\CM_{\vn}$ can also be described via symplectic reduction.
If we start with a moment map $D$ (depending on a choice of $r$
in $\CK_q$) which defines $V$ in the form
$D^{-1}(0)/G$, then $F$ is characterized as the set of points in $Y$
whose $G_{\IC}$-orbit does not meet $D^{-1}(0)$.  If we replace each
term $|\phi_i|^2$ in $D$ with a corresponding term
$\sum_{j=0}^{d_i} |\phi_{ij}|^2$ to form a map $D_{\vec n}$
\eqn\eDinst{
D_{\vn, a}=\sum_{i=1}^n Q_i^a
\Bigg(\sum_{j=1}^{d_i}|\phi_{ij}|^2\Bigg)-r_a
}
then we see that $D_{\vec n}$ serves as a
moment map in this context with $F_{\vec n}$ being precisely the
set of points in $Y_{\vec n}$ whose $G_{\IC}$-orbit does not meet
$D_{\vec n}^{-1}(0)$.  Note that the constants $r_a$ which determine
the moment map are the same as those used for the moment map of $V$
itself.

We observe that the moduli space $\CM_{\vec n}$ must be empty unless
$\vec n$ belongs to the dual of the cone of allowed $r$'s for this
model. Let $\vec n$ be such
that $\sum n_ar_a<0$.  We consider  ${1\over e^2}\sum n_a D_a$, which
takes the form
\eqn\enada{
\sum_i (\sum_a n_a Q^a_i)|x_i|^2 - (\sum n_a r_a)
.}
{}From this form, we see that if $I=\{i\ |\ \sum_a n_a Q^a_i<0\}$,
then $F$ contains the set $\{x_i=0\quad\forall i\in I\}$.  But now for
polynomials $\phi_i$ in this sector, $d_i<0$ for $i\in I$
which means that all such $\phi_i$ ($i\in I$) must vanish identically.
Thus, the corresponding collection $(\phi_i)$ lies in $F_{\vec n}$,
i.e., $Y_{\vec n}=F_{\vec n}$ so $\CM_{\vec n}$ is empty.
Since this holds for any $r$ which determines the same toric variety,
the moduli spaces must be empty for any $\vec n$ not in the (closed)
cone $\CK^\vee$ dual to the cone $\CK$ in $r$-space determining the
phase in which we compute.

The cohomology of the moduli space $\CM_{\vec n}$ is generated by
classes of toric divisors.  The coordinates in the space $Y_{\vec n}$
are given by the set of coefficients $\phi_{ij}$ of the sections
$\phi_i$.  Each can be set to zero, giving divisor classes
$\xi_{ij}$.  Among the linear relations we find that for a fixed $i$,
all $\xi_{ij}$'s are linearly equivalent to each other.  (This
essentially says that we can vary the insertion point of the operator
without affecting the cohomology class on the moduli space and hence
with no effect on correlation functions as expected.)  Let us identify
all of these with a fixed class $\xi_i$ (more precisely,
$(\xi_i)_{\vec n}$).

These classes can be written in terms of some classes $(\eta_a)_{\vec n}$
as follows:
\eqn\eetan{
(\xi_{ij})_{\vec n}
=(\xi_i)_{\vec n}=\sum_{a=1}^{n-d}Q_i^a(\eta_a)_{\vec n}\ .
}
The $(\eta_a)_{\vec n}$'s generate the cohomology in this case.

It is useful at this point to compare the rather simple instanton
moduli spaces \eMn\ to the intractable ones found in the nonlinear
sigma model. In fact, the equivariant maps $\phi_i$ are closely
related to maps $\Sigma\to V$. In particular, setting $\vn=0$ we
immediately recover $\CM_0 = V$. However, not every instanton
represents an actual map from the worldsheet to $V$.  A collection
$\phi_i$ which sends {\it any}\/ point from $\IC^2-\{(0,0)\}$ to $F$
is disqualified from being a true map.  Following Witten, we call
these the {\it pointlike instantons}, and note that they comprise a
subset of $\CM_\vn$ of positive codimension.
The moduli space $\CM_\vn$ is obtained from the
corresponding moduli space in the nonlinear model by adding these
configurations. We can interpret them physically by returning to
\einst c. For points in $\CM_\vn$ which correspond to true maps we see
that the curvature $f$ is small. For pointlike solutions, on the other
hand, there are points on the worldsheet at which the curvature is
large (of order $r$). The strong fluctuations of the gauge field
around these points extend to a distance of order the Compton
wavelength of the massive gauge boson. From the point of view of the
low-energy theory these field configurations are singular at a point
on $\Sigma$. Mathematically, the pointlike instantons lead to a natural
compactification of the space of holomorphic maps from $\CP1$ to $V$.
It is the natural occurrence of this compactification, and the extreme
simplicity of the compactified space, that make the instanton
computation tractable.

This comparison of the instanton spaces will also allow us to
show that the relation \eclast\ between the GLSM parameters and those
of the low-energy nonlinear model is exact.
Since the instanton spaces for the GLSM differ from those for the
nonlinear model only in positive codimension, and since in either case
the contributions to correlation functions arise by intersection
calculations, or equivalently by integrating a closed form of top
degree, the contribution of a given instanton sector is expected to be
the same in both theories. Thus, for the models studied here, the
relation $t_a=\tau_a$ holds {\it exactly}, and the computations of the
next subsection yield the Gromov--Witten invariants of $V$ directly.
(The analogous statement for the linear models related to Calabi--Yau
hypersurfaces $M\subset V$ is false, as we shall see later.)

\line{\hrulefill}\nobreak
\noindent{\it Example 1.}\par\nobreak
In the first example, working in the K\"ahler cone $r>0$, we find
$\CK^\vee$ given by $n\geq 0$. Thus $\phi_i$ is a section of $\CO (n)$.
Pointlike instantons are configurations in which the maps $\phi_i$
have a common zero. The set $F_n$ contains the zero map, and
$Y_n = \IC^{5(n+1)}$. Thus $\CM_n = (Y_n -  F_n)/\IC^* =
\CP{5n+4}$. This is a compactification of the space of degree $n$ maps
to $V=\CP4$. Note that ${\rm dim}_{\IC}\CM_n = 4 + 5n$ as expected from
\edimm.  The cohomology of $\CM_n$ is generated by the hyperplane class
$\xi_{ij} = \eta\quad \forall i,j$ subject to the relation (analog of
\enonl ) $\eta^{5n+5} = 0$.

\noindent{\it Example 2.}\par\nobreak
In the second example, working once more in the K\"ahler cone
$\CK_V = \{r_1,r_2>0\}$, we have $\CK^\vee = \{n_1,n_2\geq 0\}$ and
$d =(n_2,n_2,n_1,n_1,n_1,n_1-2n_2)$. Here there are two cases to
consider. The first is $n_1-2n_2\geq 0$. In this case we find
$Y_\vn = \IC^{4n_1+6}$. Comparing $d$ and \eF\ we find the nonlinear
relations defining $F_\vn$ are
\eqn\enlrelm{
\eqalign{
\xi_1^{n_2+1} \xi_2^{n_2+1} &= \eta_2^{2n_2+2} = 0\cr
\xi_3^{n_1+1}\xi_4^{n_1+1}\xi_5^{n_1+1}\xi_6^{n_1-2n_2+1} &=
\eta_1^{3(n_1+1)}(\eta_1-2\eta_2)^{n_1-2n_2+1} =0\ .\cr
}}
So $\CM_\vn = (Y_\vn -  F_\vn)/{\IC^*}^2$ is defined as a toric
variety of dimension $4n_1+4$. On the other hand, if $n_1-2n_2<0$ we
see that $d_6$ is negative, hence $\phi_6\equiv 0$. When this obtains
the two $U(1)$'s in $G=U(1)^2$ act on disjoint sets of fields, hence
$\CM_\vn = \CP{2n_2+1}\times\CP{3n_1+2}$. The hyperplane class of the
first factor is $\eta_2$, and of the second $\eta_1$. The total
dimension is $3n_1+2n_2+3$.
\par\nobreak\line{\hrulefill}

\subsec{Computing Correlators}

We are now almost ready to compute the correlation functions
\eqn\eY{
Y_{a_1\ldots a_s} =
\langle \sigma_{a_1}(z_1) \cdots \sigma_{a_s}(z_s)\rangle
}
where $z_i\in\Sigma$ and of course the result does not depend upon
the choice of the $z_i$'s. We compute $Y$ in an instanton expansion as
\eqn\eYex{
Y_{a_1\ldots a_s} =
\sum_{\vn\in \CK^\vee} Y_{a_1\ldots a_s}^\vn
\prod_{a=1}^{n-d} q_a^{n_a}\ ,
}
where $q_a = e^{2\pi i \tau_a}$.
(This series is expected to converge for $|q_a|$ sufficiently small.)
The contributions $Y_{a_1\ldots a_s}^\vn$ are given by the restriction
of the path integral to this component of the moduli space. Clearly,
this will vanish unless the anomalous ghost number conservation law
\eanomt\ is satisfied, i.e., $s = d + \sum_{i=1}^n d_i$. This is in
complete accord with the discussion in subsection {\it 3.4}. The sum
\eYex\ can diverge at small $r_a$ only if it is an infinite series in
$q_a$; this is turn can only occur if \esumq\ is
satisfied.

\nref\rgras{E. Witten, ``The Verlinde algebra and the
cohomology of the Grassmannian,'' preprint IASSNS-HEP-93/41,
hep-th/9312004.}

The $Y_{a_1\ldots a_s}^\vn$ will be given by an intersection
computation on $\CM_\vn$. To each insertion of $\sigma_a$ we
associate a class in $H^*(\CM_\vn)$. This can be determined using the
methods of \rgreg . In our context, we can make the identification
more directly, in a similar manner to our treatment of the zero modes.
In this case, restricting attention to field configurations of the
form \epoly\ we see that \eqvarc\ leads to $G$-equivariant
cohomology of $Y_\vn$. To within a normalization, the class
corresponding to $\sigma_a$ is thus $\eta_a$. We will assume this
holds exactly (the normalization factor is independent of $\vn$),
though we do not have a
complete justification for this choice. We can consider a change
of this normalization as a change of the contact terms between the
inserted class and the perturbing operator, so our choice is simply a
choice of some natural contact terms.\foot{E. Witten has suggested to us
that the computation of quantum cohomology could be carried out
by the methods of ref.\
\rgras, which would also determine the normalization.} We note that
this normalization is suggested by the correspondence between \ealbor\ and
\ealbat .
Making this identification we find, for the case $d_i\geq 0$
that
\eqn\eYvn{
Y_{a_1\ldots a_s}^\vn =
\la (\eta_{a_1})_\vn\cdot(\eta_{a_2})_\vn\cdots\cdot(\eta_{a_s})_\vn\ra_\vn
}
when \eanomt\ is satisfied, zero otherwise. Here $\la\ \ra_\vn$
denotes the intersection form on $\CM_\vn$, and when no confusion is
likely we will simply write $\eta_a$, the lift to moduli space being
understood. We have ignored the constant normalization factors in
\eYvn\ and will continue to do so.

When some $d_i<0$ the dimension of moduli space is too large. This
means that the section whose zero set is $\CM_\vn$ is not generic. In this
case the solution is well-known \refs{\rNmat}. The contribution
is obtained by inserting in \eYvn\ the Euler class $\chi_\vn$ of the
obstruction bundle. Physically, in this case there are also
$\psi^i_\zbar$ zero modes; $\chi_\vn$ is obtained by integrating over
these.
Since the obstruction bundle has a $G$-action, it is determined by the
corresponding representation of $G$.  It follows that, up to a normalization,
$\chi_\vn$ takes the form
\eqn\eexcess{
\chi_\vn = \prod_{d_i<0} (\xi_i)_\vn^{-d_i-1}\ ,
}
where the product is an intersection as above, and $\xi_i$ is defined
by \eetan .
We assume that \eexcess\ in fact gives the correct normalization.
It may appear strange to insert $\xi_i$, since the maps
take values in the subset $\phi_i = 0$ to which this is certainly not
transverse. The point is that we use \eetan\ to define the insertion,
and the $\eta_a$ do have representatives meeting $\CM_\vn$
transversely. The general formula is thus
\eqn\eYvnchi{
Y_{a_1\ldots a_s}^\vn =
\la (\eta_{a_1})_\vn\cdot(\eta_{a_2})_\vn\cdots\cdot(\eta_{a_s})_\vn
\chi_\vn\ra_\vn
\ .}

\nref{\rsmall}{P. S. Aspinwall, B. R. Greene, and D. R. Morrison,
``Measuring small distances in
  {$N{=}2$} sigma models,'' Nucl.\ Phys.\  {\bf B420} (1994) 184--242,
hep-th/9311042.}

\line{\hrulefill}\nobreak
\noindent{\it Example 1.}\par\nobreak
Since the index $a$ takes but one value in this example, we will
simplify notation by dropping it and introducing $Y_s$ to stand for
the previous $Y_{a\dots a}$ ($s$ factors of $a$). Here $\chi_n$ is
trivial and we find $Y_s^n = \delta_{s,5n+4}$, so that
\eqn\eexi{
Y_{5m+4} = \sum_{n\geq 0} q^n Y_{5m+4}^n = q^m
\ ,}
all others vanishing. All correlation
functions are analytic in $q$ and there is no singularity, as predicted
in subsection {\it 3.4} above.

\noindent{\it Example 2.}\par\nobreak
In the second example we will have to work a little harder. The
correlation functions we compute will be
$Y_{m_1,m_2} = \langle \sigma_1^{m_1}\sigma_2^{m_2}\rangle$.
Ghost number conservation implies $Y_{m_1,m_2}^\vn = 0$ unless
$m_1+m_2 = 4n_1 + 4$. We thus need $Y_{4n_1+4-m_2,m_2}^\vn$, and to
get it we will need to study the intersection form on $\CM_\vn$. Let
us start with the case $n_1-2n_2\geq 0$. In this case we will use the
relations in the algebra to determine the expectation function to within a
normalization as discussed in section two. To determine this we need
to compute one intersection and toric methods yield easily
\eqn\erln{
\la \eta_2^{2n_2+1}\eta_1^{4n_1-2n_2+3}\ra_\vn = 1\ ,}
(cf.\ \enorm).
Using the
relations we can reduce any nonzero intersection to these.  Let
\eqn\efa{
 \varphi_a = 2^{-a} \la \eta_2^{2n_2+1-a} \eta_1^{4n_1-2n_2+3-a}\ra _\vn
}
for $0\leq a \leq 2n_2+1$, and set $ \varphi_a=0$ for $a<0$.
Then $ \varphi_0=1$ is
the normalization condition, and we have the following recursion relation
by expanding the second of \enlrelm
\eqn\erecf{
 \varphi_a + \sum_{j=1}^{n_1-2n_2+1} (-1)^j
\left({n_1-2n_2+1\atop j}\right)  \varphi_{a-j} =0
}
for $1\leq a\leq 2n_2+1$ determining $ \varphi_a$. This is solved by
\eqn\esolf{
 \varphi_a = \left( {n_1-2n_2+a\atop a}\right)\ ,
}
(see appendix A) which yields finally
\eqn\esoly{
Y_{4n_1+4-m_2,m_2}^\vn
= 2^{2n_2+1-m_2} \left( {n_1+1-m_2\atop 2n_2+1-m_2}\right) \ .
}

Let us now consider the case $n_1-2n_2<0$. The moduli space was
described above. Since $d_6<0$ the contribution is
\eqn\eynex{
Y_{4n_1+4-m_2,m_2}^\vn = \la\eta_2^{m_2}\eta_1^{4n_1+4-m_2}\chi\ra_\vn\ ,
}
where
$\chi = \xi_6^{2n_2-n_1 - 1} = (\eta_1-2\eta_2)^{2n_2-n_1 - 1}$.
The only nonzero contribution to \eynex\ is simply the coefficient of
$\eta_1^{3n_1+2}\eta_2^{2n_2+1}$, hence expanding $\chi$
\eqn\ethesame{
Y_{4n_1+4-m_2,m_2}^\vn =
(-2)^{2n_2+1-m_2}\left( {2n_2-n_1-1\atop 2n_2+1-m_2}\right) =
2^{2n_2+1-m_2}\left( {n_1+1-m_2\atop 2n_2+1-m_2}\right)\ .
}
Remarkably, this is identical to \esoly. We can now sum the instanton
series to obtain
\eqn\esumit{
Y_{4n_1+4-m_2,m_2} = q_1^{n_1} \sum_{n_2\geq 0}
2^{2n_2+1-m_2}\left( {n_1+1-m_2\atop 2n_2+1-m_2}\right) q_2^{n_2}\ .
}
One observes that for $m_2>n_1+1$ the series is infinite, and its sum has
a pole at $q_2=1/4$, in complete agreement with the discussion of
subsection {\it 3.5\/} above.
\par\nobreak\line{\hrulefill}

\subsec{Quantum Cohomology}

In the preceding subsection we have completely solved the model and
computed all the correlators. As observed in section two, these
determine a deformation of the cohomology algebra of $V$ to an object
known as the {\it quantum cohomology algebra}. This is essentially the
ring of local operators in the topological field theory. This ring can
be viewed as
a deformation (with parameters $q_a$) of the cohomology ring $H^*(V)$,
which approximates the latter in the limit $q_a\to 0$, i.e.\ as we
move in $r$-space to a point deep in the interior of the K\"ahler
cone. Conversely, the
structure of the algebra determines the correlation functions to within
an overall (possibly $q$-dependent) normalization. In
subsection {\it 3.6\/} we computed some of the relations in this
algebra \ealsol . In
this subsection we will show how the direct analysis reproduces these
relations. Our treatment closely follows the work of \rbator ;
the method of proof will be useful in what follows. If $V$ is smooth,
these relations determine the algebra completely as we discuss.
We stress however, that the computations given
above are valid without this assumption. In fact, in appendix B we
consider a model in which $V$ is not smooth and show that the details of
the solution for such a model differ somewhat from the smooth case.

\nref\rST{B. Siebert and G. Tian, ``On quantum cohomology rings of
Fano manifolds and a formula of Vafa
 and Intriligator,'' alg-geom/9403010.}%

The correlation functions \eY , since they do not depend on the
points $z_i$ at which we insert the local operators, can be
interpreted as defining a linear function, the ``expectation function,'' on
${\cal Y} = \IC [\sigma_1,\ldots,\sigma_{n-d} ]$, the ring of
polynomials in
these formal variables. This, as pointed out earlier, is essentially the
ring of local observables in the theory. This statement is somewhat
misleading, however.  There is an ideal $\cal J$ in $\cal Y$ annihilated by all
correlation functions, i.e.\ for ${\cal P}\in\cal J$ we have
$\la \CO {\cal P} \ra = 0$
for all $\CO\in{\cal Y}$. The ideal $\cal J$ of course depends on $q$. The true
space of local operators is the quotient space ${\cal Y}/{\cal J}(q)$.
The problem of computing the quantum cohomology of $V$ is thus
equivalent to computing the ideal ${\cal J}(q)$.\foot{This approach
to studying quantum cohomology by finding generators $\sigma$ and
analyzing the ideal $\cal J$ of relations among them (as determined
by the correlation functions) has been systematized by Siebert
and Tian \rST.}
It will prove
convenient, in what follows, to work instead with the presentation
${\cal Y} = \IC [\delta_1,\ldots,\delta_n ]/{\cal L}$ where the $\delta$ are
determined by \elinrd\ and $\cal L$ is the ideal generated by the
linear relations among these following from \elinrd.  (Note that
just as $\sigma_a$ induces cohomology classes $(\eta_a)_\vn$ in
the moduli spaces, $\delta_i$ induces the corresponding cohomology
classes $(\xi_i)_\vn$ given by \eetan.)

As an example, we can set $q_a=0$. In this limit we expect to
reproduce the classical cohomology ring of $V$. Physically, the
contributions of nontrivial instanton sectors are suppressed in
this limit and we obtain correlators as intersection computations on
$\CM_0 = V$ as observed above. The ideal ${\cal J}(0)$ is then simply
related to the set $F$ of excluded intersections of hypersurfaces.
Consider a component of $F$ given by $\phi_{i_1} = \cdots =\phi_{i_p}=0$
(where $\phi_{i_1}, \ldots, \phi_{i_p}$ is a ``primitive
collection'' as before). Then of course any correlation function
$\la \CO \delta_{i_1}\cdots\delta_{i_p}\ra_0 =
\la \CO \xi_{i_1}\cdots\xi_{i_p}\ra$ vanishes in this limit because the
corresponding hypersurfaces do not meet in $V$. This
is precisely the presentation of the cohomology ring of $V$ which
appeared above in subsection {\it 3.1}.

Let us now move away from the locus $q_a=0$. When $q_a\ne0$, there can be
nonzero contributions to correlation functions containing operators
from ${\cal J}(0)$. In the nonlinear language, these come from
nontrivial rational curves in $V$ which intersect the $p$ divisors,
even though these do not intersect in $V$. We can readily study these
by taking advantage of the fact that correlators in the topological
field theory do not depend on $z_i$. Essentially we will use this
freedom to insert the operators at the {\it same}\/ point on $\Sigma$.
As always, the local operator corresponding to this situation suffers
from ambiguities related to contact terms. We will essentially be
making the analog of the canonical choice, i.e., we will use the
point-splitting definition and use computations with all the
operators inserted at a point to learn about the generic situation.

The ideal $\cal J$ can
be constructed as follows (we are here following closely the argument
of \rbator ). We will start by writing one relation for every set
$\vn^*$ of $n{-}d$ integers lying in the $(n{-}d)$-dimensional
cone $\CK^+\subset \CK^\vee$
determined by the inequalities $d_i(\vn) \geq 0$. Given such a vector
$\vn^*$ we will show that for any operator $\CO$ in the ring and any
$\vn\in \CK^\vee$
\eqn\etosho{
Y_\CO^\vn = Y_{\CO^*}^{\vn +\vn^*}
}
where $\CO^* = \CO\prod_{i=1}^n \delta_i^{d_i^*}$.

To see this, let us first assume that $\vn\in \CK^+$. In this case
\etosho\ is
\eqn\etoshoi{
\la \CO \ra_\vn = \la \CO \prod_{i=1}^n (\xi_i)_{\vn+\vn^*}^{d_i^*}\ra_{\vn
+\vn^*}\ .
}
We can choose explicit representatives for the $\xi_i$ on the right-hand
side such that in the space of equivariant maps $\phi_i$ we impose all
of the constraints at $s=0$ on $\Sigma$. This means that the
expression \epoly\ for $\phi_i$ as a homogeneous polynomial of degree
$d_i+d_i^*$ takes the form
\eqn\epolyi{
\phi_i = s^{d_i^*} P_i(s,t)
}
where $P_i$ is a homogeneous polynomial of degree $d_i\geq 0$. But the
space of such polynomials is precisely $Y_\vn$. Further, the image
lies in $F$ precisely when the maps $P_i$ lie in $F_\vn$. Thus
choosing these representatives on the right-hand side of \etoshoi\
leads precisely to the left-hand side.

Next, note that if for some $i$ we have $d_i+d_i^* <0$, the discussion
of the previous paragraph is modified as follows. On both sides of
\etosho\ the moduli space in question is restricted to maps for which
$\phi_i=0$ identically. However, on both sides we must modify
\etoshoi\ by inserting the appropriate classes $\chi$. On the
left-hand side this contains a factor of $\xi_i^{-d_i-1}$, while on
the right-hand side the factor that appears is
$\xi_i^{-d_i-d_i^*-1}$. Comparing the two we see that \etosho\ holds
in this case as well. Finally, we must also address the case
$0<-d_i<d_i^*$. In this case, on the left-hand side of \etosho\ we
have set $\phi_i=0$ identically and inserted $\xi_i^{-d_i-1}$. On the
right-hand side, we have $\phi_i$ of degree $d_i+d_i^*$ initially;
however, we can use $d_i+d_i^*+1$ of the explicit insertions of
$\xi_i$ to restrict to the set $\phi_i=0$ identically; this leaves us
with $-d_i-1$ insertions to be imposed in this subset, once more in
agreement with \etosho . There is one final case to study, $\vn\not\in
\CK^\vee$ but $\vn+\vn^*\in \CK^\vee$. It is easy to see that in this
case both sides of \etosho\ vanish.

In conclusion, what we have shown is that for each $\vn^*\in \CK^+$ we
have a relation following from \etosho
\eqn\etherel{
\prod_{i=1}^n \delta_i^{d_i^*} = \prod_{a=1}^{n-d} q_a^{n_a^*}\ ,
}
in agreement with \ealbor\ (the cone $\CK^+$ of course contains
precisely the combinations leading to non-negative powers of $\xi_i$
referred to there).

If $V$ is smooth these relations generate the ideal ${\cal J}(q)$. In
fact, the
argument to this point has demonstrated only that the ideal generated
by our relations is contained in $\CJ(q)$. The converse, for the case
of smooth compact $V$, essentially follows from the work of Batyrev.
Under these hypotheses,
he showed that the ring (more precisely algebra) defined by these
relations approximates $H^*(V)$ in the limit $q\to 0$. Note that our
choice of $n^*$ makes this limit well-defined; thus the dimension of
the algebra is independent of $q$ for values away from the singular
loci and in particular does not change as $q\to 0$. Now, imposing any
additional relations would change the dimension of the ring (at least
for nonzero $q$, the new relation could be ill-defined at $q=0$). But
this would be in disagreement with the fact that the space of
operators at any $q$ is isomorphic as a graded vector space to
$H^*(V)$. The restriction to smooth $V$ enters the argument precisely
at this point. When $V$ is singular, the relations \etosho\ are
certainly contained in the ideal $\CJ$, but they do not suffice to
generate it. In appendix B we describe an example of a singular toric
variety for which this occurs. Of course, since we solve the model for
all values of the parameters, the algebraic solution is valid whenever
there is {\it some\/} toric model for $V$ which is smooth.

Notice that our choice of contact terms is seen to agree with natural
choices made from other points of view for which the nonlinear model
was the point of departure. In particular, the $\tau_a$ are indeed the
canonical coordinates as expected. To construct the canonical basis
one begins with the classes $\eta_a$. Their products (modulo the
relations) are a basis for the algebra; the change of basis to a
canonical set is determined by the requirement that all two-point
functions be constant.

\line{\hrulefill}\nobreak
\noindent{\it Example 1.}\par\nobreak
\nref\rtopphase{E. Witten, ``On the structure of the topological phase
of two-dimensional gravity,'' Nucl. Phys. {\bf B340} (1990) 281--332.}

Here we have seen that
$\CK^+=\CK^\vee$ is generated by $n=1$, from which we obtain the
relation $\prod_{i=1}^5 \delta_i = \sigma^5 = q$, in agreement with the
well-known results for the nonlinear $\CP n$ model \rtopphase.

\noindent{\it Example 2.}\par\nobreak
In the second example $\CK^+$ is
generated by $\vn = (1,0),\ (2,1)$, leading to the relations
\eqn\enlrelf{
\eqalign{
\delta_3\delta_4\delta_5\delta_6 &=
\sigma_1^3(\sigma_1-2\sigma_2) = q_1 \cr
\delta_1\delta_2\delta_3^2\delta_4^2\delta_5^2 &=
\sigma_1^6\sigma_2^2 = q_1^2 q_2\
.\cr
}}
Dividing the second of these by the square of the first we obtain
\eqn\enlrelh{
\sigma_2^2 = q_2(\sigma_1-2\sigma_2)^2\ ,
}
which together with the first of \enlrelf\ form a deformation of
\enlrel .
\par\nobreak\line{\hrulefill}

\newsec{The Linear Sigma Model for $M$}

Topological sigma models with toric target spaces are interesting as
solvable nontrivial field theories. However, the recent interest in
toric geometry among physicists has in fact focused on different
though related theories. These are the superconformal nonlinear sigma
models with Calabi--Yau
target spaces that can be embedded as hypersurfaces
$M\subset V$ in toric varieties. For these models there exists a
conjecture by Batyrev \rbatmir\ which if true allows one to construct
the mirror manifold $W$ as a hypersurface in a toric variety $\varLambda$
related to $V$ by a combinatorial duality. In \rbatmir\ the most basic
implication of mirror symmetry~--~the relation between the Hodge
numbers of $M$ and $W$~--~was verified. In this section we will use
the results of our careful study of the model with target space $V$ to
study the model with target space $M\subset V$ a Calabi--Yau
hypersurface. We will find that the {\bf A} model with target space
$M$ is solvable in the same sense (and using the same methods) as that
for $V$. We will return to discuss mirror symmetry and Batyrev's
construction in greater detail in section five. The modification to
the model of section three which leads to the nonlinear sigma model
with target space $M$ was given in \rphases . We note that the methods
of \rphases\ and probably the extension given here are not restricted
to hypersurfaces; however this is the simplest case to study and we
restrict ourselves to that case in this paper.

\subsec{Toric Geometry on the Other Leg}

A toric variety $V$ as described in subsection {\it 3.1\/} naturally
determines a family of \CY\ manifolds, provided that $V$ satisfies a
certain combinatorial condition which we shall describe in section five.
A hypersurface (holomorphically
embedded codimension-one submanifold) $M\subset V$ inherits a K\"ahler
metric by restriction from the embedding in $V$. $M$ determines a
homology class in $H_2(V)$. $M$ is \CY\ precisely when this class is
the anticanonical class $-K$. The family of \CY\ manifolds mentioned
above is the (finite-dimensional) family of hypersurfaces in this
homology class. These are related of course by deformations. They thus
represent different complex structures on one underlying
differentiable manifold. The significance of this is that the
topological {\bf A} models on all of these are isomorphic, this model
being independent of complex structure. A submanifold of this type is
locally the vanishing locus of a holomorphic function. In a toric
variety the homogeneous coordinates $z_i$ allow us to represent $M$
{\it globally\/} as the vanishing locus of a homogeneous polynomial
$P(z):Y\to \IC$. The \CY\ condition requires that $P$ have degree
$\sum_{i=1}^n Q_i^a$ under the $a$th factor of $T=G_{\IC}$. The deformations
of $M$ (or at least a subset of the deformations)
are described by varying the
coefficients of the polynomial $P$. When $V$ satisfies the combinatorial
condition alluded to above, $M$ will be a quasi-smooth hypersurface
for generic choices of these coefficients, meaning that $M$ is smooth
away from possible singularities of $V$. If $V$ is smooth, so is $M$.
This condition means, in practice, that the solutions of $dP=P=0$ (in
$Y$) are contained in $F$.

We can obtain cohomology classes on $M$ by
restriction from classes on $V$. In general, this will not yield the
full cohomology of $M$ but only a subspace, which we will call
the ``toric'' subspace $H_V^*(M)$.
In practice, this leads to an extremely simple relation between the
intersection form on $H_V^*(M)$ and that on $H^*(V)$. The
hypersurface $M$ determines some divisor, and the \CY\ condition on
$M$ is equivalent to the requirement that this be the anticanonical
divisor
\eqn\ek{
-K = \sum_{i=1}^n \xi_i\ .
}
The ring $H_V^*(M)$ is thus generated by the $\eta_a$ of the previous
section (or rather their intersections with $M$ for which we will use
the same notation). The linear relations \elinr\ of
course still hold. The new intersection form in this presentation is
given simply by the standard
{\it restriction formula}\/ for intersections in hypersurfaces:
\eqn\eclass{
\la\eta_{a_1}\cdots\eta_{a_p}\ra_M =
\la \eta_{a_1}\cdots\eta_{a_s}(-K)\ra_0\ .
}
This leads to a graded ring of length (highest degree in which
ring is nontrivial) $d{-}1=\dim_{\IC}M$. Since we have solved the
problem of finding the quantum cohomology ring of $V$, solving the
{\bf A} model with target space $M$ can be thought of as finding the
quantum analog of this relation.

\line{\hrulefill}\nobreak
\noindent{\it Example 1.}\par\nobreak
A Calabi--Yau hypersurface of \CP4 is determined by a generic quintic
polynomial $P(z)$. In terms of the homology of $V$, this determines
the class dual to $5\eta$. Thus $H^*_V(M)$ is generated by $\eta$
subject to the relation $\eta^4=0$.

\noindent{\it Example 2.}\par\nobreak
A Calabi--Yau hypersurface is determined by a polynomial of
multidegree $(4,0)$, e.g.
\eqn\ehyp{
P(x) = x_1^8x_6^4+x_2^8x_6^4+x_3^4+x_4^4+x_5^4+
x_1x_2x_3x_4x_5x_6=0\ .
}
The class determined by this is $4\eta_1$, so that $H_V^*(M)$ is
determined by the nonlinear relations (compare \enlrel )
\eqn\enlrelmm{
\eqalign{
\eta_2^2 &=0\cr
\eta_1^2 (\eta_1 - 2\eta_2) &= 0\ .\cr
}}
\par\nobreak\line{\hrulefill}

\subsec{The Model}

We now modify the GLSM so that the low-energy theory is the nonlinear
sigma model with target space $M$, following \rphases . To this end,
we add an additional chiral multiplet $\Phi_0$ with charges
\eqn\eqzero{
Q_0^a = -\sum_{i=1}^n Q_i^a\ .
}

This modification has the effect of canceling the gauge anomalies in
the $R$-symmetries $Q_{L,R}$ \eanom . This is encouraging; we expect
to be obtaining in the infrared a nontrivial fixed point of the
renormalization group flow. This should then enjoy $N=2$
superconformal symmetry, which contains left- and right-moving $U(1)$
$R$-symmetries.

Analyzing this modified model as in section three, we find that
the space of supersymmetric ground states is now
\eqn\evplus{
V^+ = (Y^+ -  F^+)/T = (D^+)^{-1}(0)/G\ ,
}
where $Y^+=\IC^{n+1}$, and $F^+$ can be read off as in section three
from the modified $D$-terms
\eqn\eDplus{
D_a^+ = -e^2 ( \sum_{i=0}^n Q_i^a |\phi_i|^2 - r_a)\ .
}
This determines $V^+$ as a (noncompact \CY ) toric variety of
dimension $d{+}1$. In fact, \eqzero\ implies that for $r$ in the
K\"ahler cone $\CK_V$ this is the total
space of the canonical line bundle over $V$. This space is the natural
setting for Batyrev's mirror construction and will play a central r\^ole
in section five. At this point, however, its salient feature is that
it is not $M$.

To rectify this we make one more modification to the model. The
introduction of $\Phi_0$ makes it possible to add a superpotential
interaction
\eqn\esup{
L_W = - \int d^2 z d\theta ^+ d\theta ^- W(\Phi)|_{\thetabar^+=\thetabar^-=0}
- {\rm h.c.}
}
with $W$ a holomorphic, $G$-invariant function. Note that before
introducing the extra field such a function does not exist; it would
correspond to a global holomorphic function on $V$. By contrast, the
coefficient of $\Phi_0^k$ corresponds to a section of the $k$-th power
of the anticanonical line bundle. A generic choice of $W$ will in fact
break the $R$-symmetry, but if $W$ is homogeneous of some degree in
$\Phi_0$ we can recover invariance by accompanying the action of $Q_R$
by a rotation of this superfield. We will choose $W = \Phi_0 P(\Phi)$
where $P$ does not depend on $\Phi_0$. Then \eqzero\ shows that
this is gauge invariant precisely when $P=0$ determines a Calabi--Yau
hypersurface in $V$. The nonanomalous $R$-symmetry is thus directly
related to the Calabi--Yau condition. This is natural, as the latter
is expected to be the condition for the existence of a nontrivial
conformal theory in the low-energy limit.

\nref\rorbifold{P. S. Aspinwall, ``Resolution of orbifold singularities
in string theory,''
to appear in {\sl Essays on mirror manifolds II},
B. R. Greene and S.-T. Yau, eds. International Press, Hong Kong,
hep-th/9403123.}%

We can now study the model following the same steps as in section
three. We begin with the space of classical ground states.
The bosonic potential \eU\ is modified by the addition of
\eqn\euw{
U_W = \sum_{i=0}^n |{\partial W\over\partial\phi_i}|^2 =
|P|^2 + |\phi_0|^2 \sum_{i=1}^n |{\partial P\over\partial\phi_i}|^2
\ ,}
and by the extension of the $D$-terms as in \eDplus . This affects
the space of classical supersymmetric vacua as follows. The first
modification to the discussion in section three is that the apparent
supersymmetry breaking for $r$ outside the cone $\CK_c$ does not
arise. The classical moduli space is the entire complexified
K\"ahler space $\IC^{n-d}/\IZ^{n-d} = \IR^{n-d}\times U(1)^{n-d}$.
The classical theory is singular along certain cones in $r$-space
found as in section three (without the restriction to $\CK_c$), for
the same reason, dividing (real) $r$-space into
regions corresponding to different ``phases.''

Restricting attention once more to the K\"ahler cone we find that
requiring the vanishing of \euw\ in addition to the $D$-terms
will set $\phi_0 = 0$ (restricting to $V$) and then \euw\ requires
that the remaining fields satisfy $P=0$, in other words that the
(point) image of the worldsheet lie in $M$. This will in fact hold for
any $r\in\CK_q$. One can study the model more closely and see that the
massless modes are precisely the variations of $\phi$ tangent to
this and their superpartners, so we have as the low-energy limit
precisely the nonlinear sigma model on the \CY\ hypersurface.
The metric on $M$ is classically just the restriction of the metric on
$V$. Notice that with this metric the nonlinear model is not
conformally invariant; quantum corrections will presumably shift this
to the conformally invariant solution. As is usual, these corrections
will not change the K\"ahler class of the metric, parameterized by
$t_a$. The classical solution determines these in terms of $\tau$ as
in \eclast ; there can be corrections to this as discussed in subsection
{\it 3.4}.
In other regions of $r$-space we obtain other types of models. In
general these include hypersurfaces in various birational models of
$V$, including models with unresolved orbifold singularities, as well
as phases in which the space of vacua is of dimension less than $d$.
In these cases there are massless excitations about these vacua,
governed by the superpotential interaction. When the space of vacua is
a point the model is what is commonly known as a Landau--Ginzburg
theory, intermediate cases in which there are massless fluctuations
about a nontrivial space of vacua were termed ``hybrid'' models
\rphases . In many vacua there are discrete subgroups of $G$ unbroken
by the expectation values; the low-energy theory is then a quotient by
this subgroup. In the Landau--Ginzburg case this leads to the familiar
Landau--Ginzburg orbifold models. The physics of hybrid phases is not
very well understood. The various theories that arise
are classified very naturally by the combinatorics of $\Delta$;
equivalently they are obtained from the excluded set $F$ determined by
$r$ and the implications of the expectation values required by this
when inserted in \euw . For more details on these, see
\refs{\rphases,\rAGM,\rorbifold}.

\line{\hrulefill}\nobreak
\noindent{\it Example 1.}\par\nobreak
This is treated in detail in
\rphases . There are two phases separated by a singularity which by \eref\
is located at
$r = r_c = {5\log 5\over 2\pi}$ and $\theta=\pi$.
For $r\gg r_c$ we find that setting
$U=0$ leads to $\phi_0 = 0$ and $P(\phi =0)$ as well as
$\sum_{i=1}^5 |\phi_i|^2 = r$. In this region the low-energy limit is
the conformal nonlinear sigma model with target space a quintic
hypersurface of \CP4. For $r\ll r_c$ we see that the vacuum solutions
satisfy $\phi_i=0$, and $|\phi_0|^2=r$; the vacuum is point-like. Here
however the fluctuations of $\phi_i$ are massless and governed by the
superpotential $W = \phi_0 P(\phi)$. This type of model is known as a
Landau-Ginzburg theory. Note as well that the choice of expectation value
for $\phi_0$ does not completely break the gauge symmetry, leaving
rather an unbroken $\IZ_5$ subgroup. The low-energy theory in this
region is thus an orbifold quotient of the LG model by this symmetry.

\noindent{\it Example 2.}\par\nobreak
In the second example, we find the modified $D$ terms
\eqn\eDex{
\eqalign{
D_1^+ &= -e^2 (|\phi_3|^2+|\phi_4|^2+|\phi_5|^2+|\phi_6|^2-
4|\phi_0|^2 - r_1)\cr
D_2^+ &= -e^2 (|\phi_1|^2+|\phi_2|^2-2|\phi_6|^2 - r_2)\ .\cr
}}
Let us first find the phase boundaries, by seeking those $r$ values
for which an unbroken continuous symmetry is consistent with
$D_a=0$ using \eU. One finds that $g_1$ is unbroken if
$\phi_3=\phi_4=\phi_5=\phi_6=\phi_0=0$, which from \eDex\ can
happen at zero energy if $r_1 = 0,\ r_2\geq 0$. Similarly, $g_2$ is
unbroken if $\phi_1=\phi_2=\phi_6=0$ which implies $r_2=0$ but
leads in fact to two cones (rays) because both signs of $r_1$ are possible.
Finally, if $\phi_6$ is the only nonvanishing coordinate, then we
see that $g_1 g_2^2$ is unbroken. This implies $r_1\geq 0,\ 2r_1+r_2=0$.
Figure~2 shows the structure in $r$-space, taking account of the shift
\eref . There are four phases, labeled I--IV.

\iffigs
\midinsert
\centerline{\epsfxsize=3.5in\epsfbox{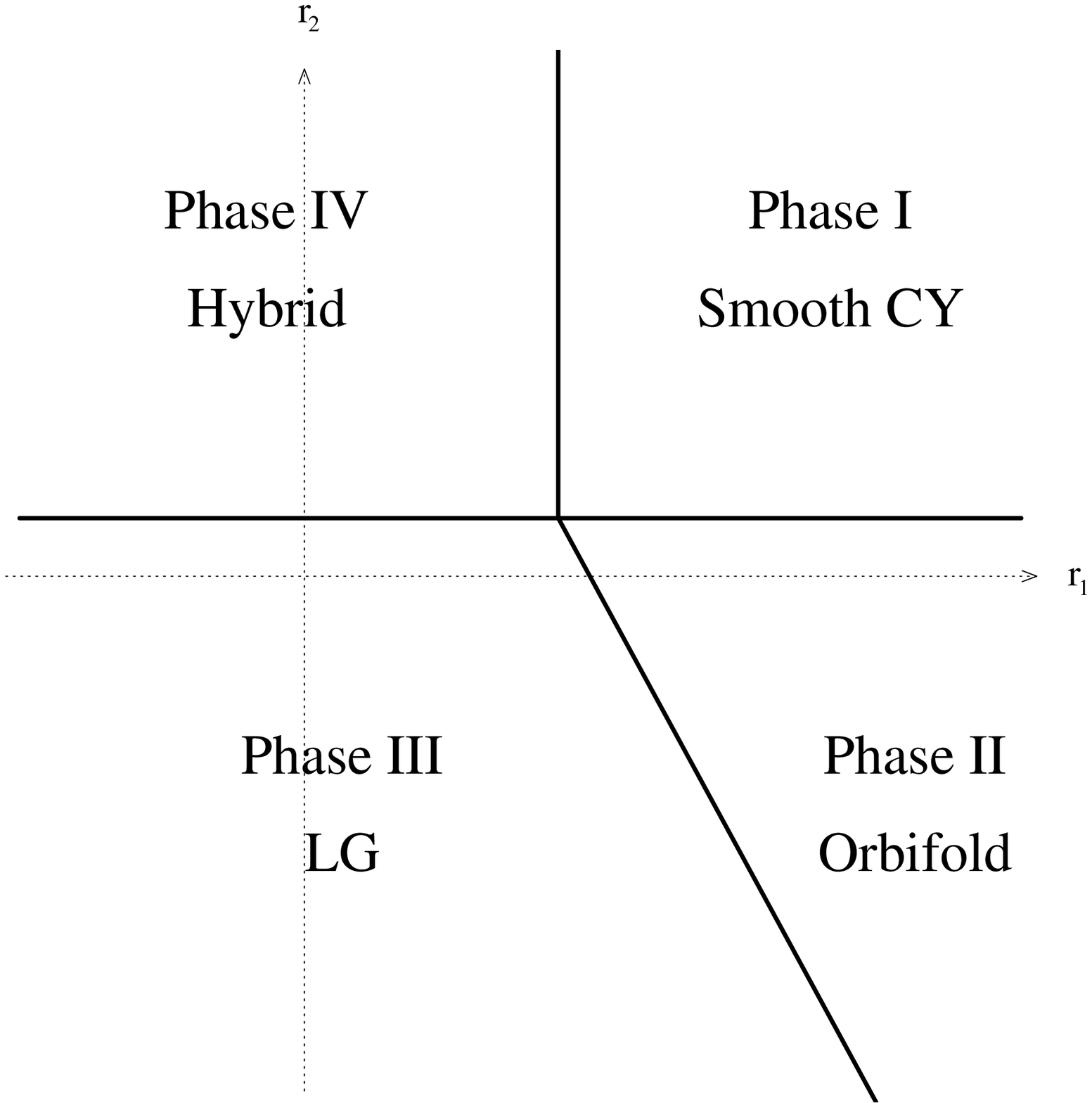}}
\centerline{Figure 2. Phase diagram for $M$.}
\endinsert
\fi

In phase I we see from \eDex\ that the excluded regions are
precisely those of \eF . Requiring the vanishing of $U_W$
then implies $\phi_0=P=0$ so the low-energy modes describe the
nonlinear sigma model on the \CY\ hypersurface in the smooth toric
variety $V$.

In phase II the excluded regions are
$\{\phi_6=0\}\cup\{\phi_1=\phi_2=\phi_3=\phi_4=\phi_5=0\}$.
Once more $U_W=0$ implies $\phi_0=0$. This corresponds to the
original (unresolved) projective space; the low-energy limit is the
nonlinear sigma model with target space a hypersurface in this space.
More precisely we have a deformed
version of this because $\phi_6$ while nonzero is not fixed to
a constant; however for the computations we are interested in
this is equivalent. This is the ``orbifold'' phase.

In phase III the excluded region is $\{\phi_0=0\}\cup\{\phi_6=0\}$.
Then $U_W=0$ implies the vanishing of all the other coordinates, leading to
a unique vacuum configuration in which $G$ is broken to $\IZ_2\times\IZ_4$,
and to massless fluctuations described by a superpotential interaction.
This region thus corresponds to the Landau--Ginzburg orbifold.

Finally, in phase IV the excluded regions are
$\{\phi_0=0\}\cup\{\phi_1=\phi_2=0\}$. Here $U_W=0$ implies
$\phi_3=\phi_4=\phi_5=\phi_6=0$, so that $g_1$ is broken to
a discrete subgroup $\IZ_4$. The expectation values of $\phi_1,\phi_2$
parameterize (after setting $D_2=0$ and taking the $G$ quotient) a
moduli space isomorphic to \CP1. The fluctuations of $\phi_3,\phi_4,\phi_5$
are massless; they interact via a superpotential with coefficients
depending upon the point in \CP1 . The model is a so-called hybrid
combining the properties of a sigma model on \CP1\ with those of a
Landau-Ginzburg theory.
\par\nobreak\line{\hrulefill}

The perturbative quantum corrections to this classical picture are
found along the lines of the discussion in section three. The novel
feature is that \esumq\ is satisfied for all $a$. Thus all of the
singularities predicted classically persist in the quantum theory,
and occur (far from the
origin) at $\theta=0$ or $\theta=\pi$. This is consistent with the
absence of an
anomalous $R$-symmetry which could be used to eliminate one of the
$\theta$ angles. The only effect of perturbative corrections is the
finite shift \eref\ of the asymptotes of the singular locus. Away from
the singularities, the low-energy degrees of freedom are the fields
$\Phi_i$; the gauge symmetry is Higgsed at generic points throughout
parameter space.

In subsection {\it 3.5\/} we used the twisted superpotential to compute
the exact singular locus for the model discussed there. The reasoning
presented there can be applied to the model at hand. The results are
simpler to describe because the addition of $\Phi_0$ eliminates the
distinguished direction in $\sigma$-space. Thus, following the
arguments presented there, we find that the singular locus is
given by the consistency conditions of the equations
\eqn\ealborm{
\prod_{i=0}^n \la\delta_i\ra^{Q_i^a} = q_a\ .
}
Notice that these are all homogeneous of degree zero. As in the
previous discussion, this will yield one component
of the singular locus (in the present
context this is called the {\it principal discriminant\/} of the
model; note that this is precisely the component absent from the $V$
model of section three because it necessarily involves what was there
denoted $\sigma_1$). Other components are to be obtained by integrating
out subsets of the set of chiral fields charged under subgroups
$H\subset G$ such that the charges of the complementary set generate
(with positive coefficients) all of $\IR^{n-d-k}$ where $k$ is the
rank of $H$.
Such a component is given by the consistency conditions for the equations
\eqn\ealborbis{
\prod_{i\in I} \la\delta_i|_H\ra^{Q_i^a} = q_a\ , \
a = n{-}d{-}k{+}1,\ldots, n{-}d\ ,
}
where $\{\phi_i\}_{i\in I}$ is the set of charged fields under $H$,
$\{\sigma_a\}_{a = n{-}d{-}k{+}1,\ldots, n{-}d}$ is the set of
massless $\sigma$'s for $H$ (in an appropriate basis), and where
$\delta_i|_H$ is obtained from $\delta_i$ by setting $\sigma_a=0$
for the massive $\sigma_a$'s, i.e., for $a=1,\ldots,n{-}d{-}k$.

We defer a discussion of
the quantum corrections to \eclast\ to subsection {\it 4.3}.

\line{\hrulefill}\nobreak
\noindent{\it Example 1.}\par\nobreak
In this example the singular divisor~--~a point, is correctly
predicted by \eref (modified to include $\Phi_0$), since setting all
$\sigma_a$ but one to zero is the general case. Thus \ealborm\ here
reduces to  $q =
(-5)^{-5}$.

\noindent{\it Example 2.}\par\nobreak
For the principal component,
\ealborm\ reads (here $s_a = \la\sigma_a\ra$ are the expectation
values)
\eqn\ealii{
\eqalign{
q_1 &= (-4s_1)^{-4} s_1^{3}\,(s_1-2s_2)\cr
q_2 &= s_2^2\,(s_1-2s_2)^{-2}\ .\cr
}}
Solutions to these exist when
\eqn\eprindis{
2^{18} q_1^2 q_2 -(1-2^8 q_1)^2 = 0\ .
}
The solutions satisfy $s_2/s_1 = (1-2^8q_1)/2$; this component
interpolates between three of the four asymptotic singular limits
found in the discussion following \eDex .
There are various (rank-1) subgroups of $G$ under which some subset of
the fields is uncharged (these were effectively catalogued in
constructing the phase diagram for the model). Closer inspection shown
that only one of these, however, satisfies the condition that the
charges under it span all of $\IR$ with positive coefficients. This is
the subgroup generated by $g_2$ as found above, and for it
eqn.~\ealborbis\ reduces to
\eqn\emordis{
q_2 = s_2^2\,(-2s_2)^{-2} = 1/4\ .
}
This is the second component of the singular
locus. The corresponding $\sigma$-vacua are at $s_1=0$. They occur for
any value of $q_1$, of course. Figure~3
shows the exact singular locus in the $r$-plane. The curve shows the
singularity for real $q$; singularities at other values of $\theta$ exist for
$r$ in the enclosed region in the center of the figure.\foot{This
figure was first calculated in \rsmall\
for the $B$-model of the mirror partner; here we get it directly
from $A$-model considerations.} Asymptotically at large $r$ the
singular locus approaches
the cones predicted in subsection {\it 3.2\/} above. For finite $r$ there
are instanton corrections as predicted in section two and computed
above; these introduce a
$\theta$ dependence in the singular locus (which is a holomorphic
divisor in $q$ space).

\iffigs
\midinsert
\centerline{\epsfxsize=3.5in\epsfbox{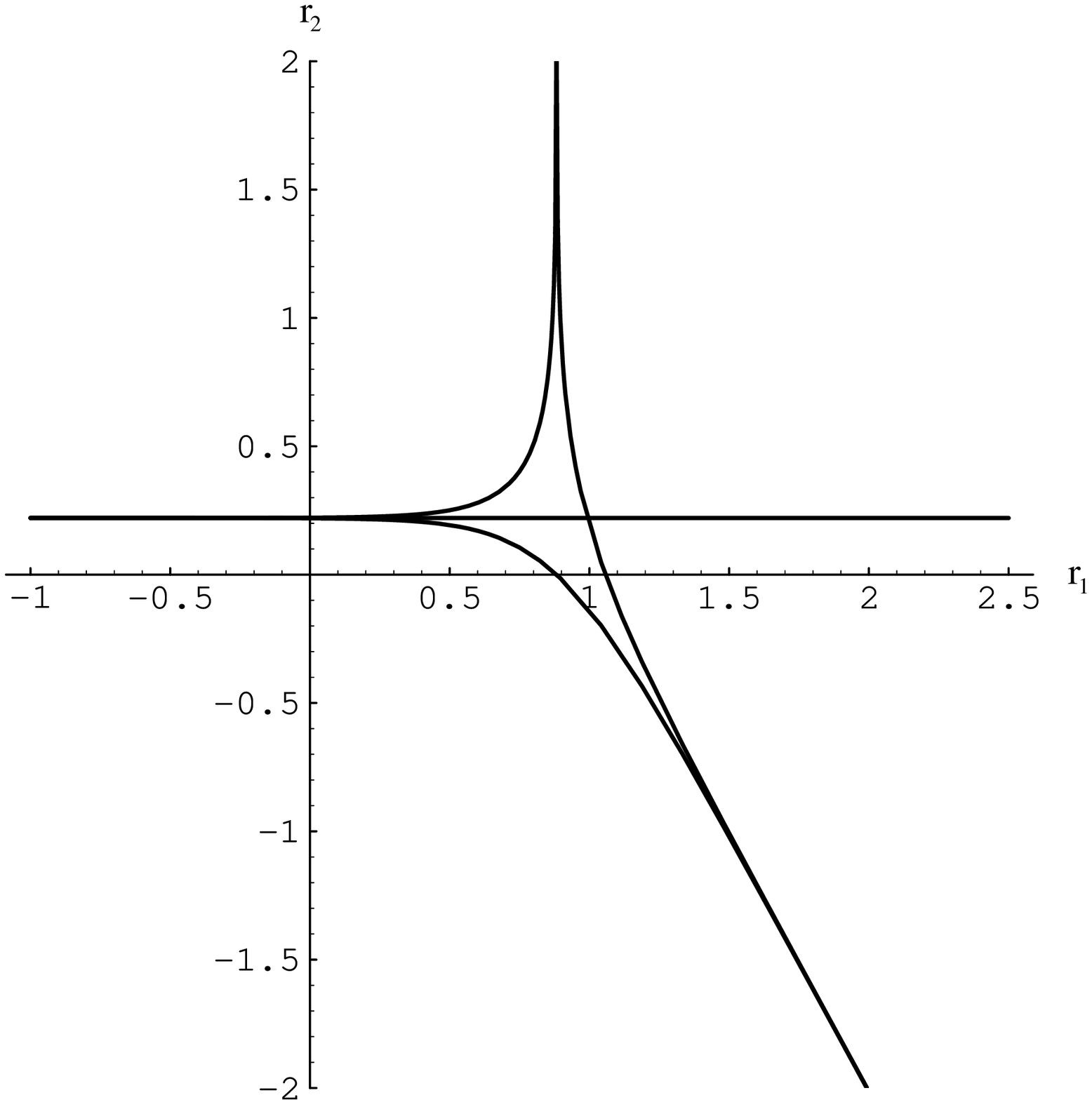}}
\centerline{Figure 3. Fully corrected phase diagram for M.}
\endinsert
\fi
\par\nobreak\line{\hrulefill}

For the remainder of this section we once more restrict
attention to the ``geometric'' phases (values of $r$ in the interior of
$\CK_q$).  (This includes the phases we have labeled I and II in example 2.)

\subsec{The Topological Model}

As in the previous section, our main interest will be in the
topological {\bf A} model. Our earlier conclusions regarding the spectrum of
$Q$-closed local operators are still valid. The quantum cohomology
algebra is generated by $\sigma_a$. This is in line with the
description of the cohomology ring $H^*_V(M)$ in subsection {\it 4.1}.
The correlation functions we compute will (for $r$ in $\CK_V$) be
interpreted as the correlators in the nonlinear sigma model with
target space $M$. We denote the correlation functions in the new model
by
\eqn\ecorM{
X_{a_1\ldots a_s} =
\lal \sigma_{a_1}(z_1)\cdots\sigma_{a_s}(z_s)\rar
\ .}
It is easy to see that the interaction \esup\
is in fact $Q$-exact. Thus our correlation functions will be
independent of the coefficients in the polynomial $P$.

In subsection {\it 3.6\/} we found that the twisted superpotential
$\widetilde{W}(\Sigma)$ computed at one-loop order about the free
gauge theory at large $\sigma$ led to operator relations \ealsol\
which held throughout parameter space by analytic continuation.
In the case at hand, however, this is not {\it a priori\/}
clear, because there is no domain from which to continue analogous to
the interior of the complement of $\CK_c$. When we
compute the correlators, however, we will find that in fact the
relations \ealborm\
hold exactly in all correlators, and in the case
of smooth $V$ suffice to determine them. Of course, away from the
singular locus \ealborm\ are relations between the operators
$\delta_i$ and not their expectation values. This point will be
discussed more fully in section five.

\nref\reva{E. Silverstein and E. Witten, ``Global $U(1)$ $R$-symmetry
and conformal invariance of $(0,2)$ models,'' Phys. Lett. {\bf 328B}
(1994) 307--311, hep-th/9403054.}%

The ghost number conservation law \eanomt\ will be altered by the
superpotential term in the action. The modified $R$-symmetry under
which the superpotential interaction is invariant will assign unit
$R$-charge to $\phi_0$ (of course this assignment is determined up to
the possibility of an accompanying gauge transformation, which will
have no effect on the discussion). Computing the contribution of this
to the gravitational anomaly \eanomt\ we find indeed\foot{A detailed
discussion of the $R$-symmetry in the model and its relation to the
superconformal symmetry of the low-energy limit has been given in
\reva .}
\eqn\eanomtm{
\Delta (Q_R) = -\Delta (Q_L) = {d-1\over 2}\chi(\Sigma)\ ,
}
where the instanton contribution vanishes by \esumq .  This agrees
with our observations in the zero instanton sector where we found a
restriction to $M\subset V$.  The correlation functions neglecting
instanton contributions will give the intersection form on $M$ as
determined from \eclass . To compute instanton corrections to this we
now proceed to study how equations \einst{} are modified.

To begin with, of course, there is an additional field $\phi_0$.
Further, the instanton equations are modified.
The additional condition imposed is
\eqn\einstw{
{\partial W\over\partial\phi_i} = 0\ .
}
This is invariant under local $G_{\IC}$ transformations so the discussion
following \einst{} is still valid. In particular, before imposing
\einstw\ the equations \einst{} (including $\phi_0$) still determine a
toric variety. In fact, this will be precisely $\CM_\vn$, as we now
show. The argument surrounding \enada\ shows that for $r\in\CK_V$, the
moduli space is empty unless $\vn\in \CK_V^\vee$. For these
values of $\vn$ we have $d_0 = \sum_a Q_0^a n_a = -\sum_{i=1}^n d_i\le
0$. Thus $\phi_0$ is identically zero or at most a constant.
But if $\phi_0$ is a nonzero
constant then the image in $V$ of the map (forgetting $\phi_0$) is
constrained by \einstw\ to lie in the critical point set of $P$ (as a
function on $Y$). We will use our freedom to modify
the coefficients of $P$ to choose it such that $P=0$ is a {\it
quasi-smooth}\/ hypersurface in $V$. This condition~--~which holds for
generic values of the coefficients~--~means that the critical
points of $P$ are contained in the set $F$. This however implies
that if $\phi_0$ is nonzero then the image is contained in $F$ and so such
maps are not (complex) gauge transforms of true solutions. On the
other hand, setting $\phi_0=0$ we
recover the original equations \einst{} for $\phi_i$.
In all, the moduli space from which we
obtain the instanton contributions to correlation functions is the
subset satisfying \einstw\ in $\CM_\vn$.

Setting $\phi_0=0$ this becomes the equation
\eqn\einstwii{
P(\phi) = 0
\ .}
The homogeneity properties
of $P$ guarantee that it is a section of a line bundle over $\CP1$ of
degree $-d_0 = \sum_{i=1}^n d_i$, hence described by a homogeneous
polynomial of this degree in $(s,t)$. Solving \einstwii\
identically is the same as requiring that all $1{-}d_0$ coefficients of
this polynomial vanish. These coefficients, in turn, are polynomials
in the homogeneous coordinates $\phi_{ij}$ on $\CM_\vn$. For generic
maps these will be $1{-}d_0$ independent equations in $\phi_{ij}$,
so the formal dimension (an estimate of the
true dimension) of the moduli space of
solutions to the full set of equations (compare \edimm )
is just $d{-}1$, the dimension of $M$,
and independent of $n$ as expected. As before, this
space is a compactification of the space of maps $\Sigma\to M$
obtained by adding maps that intersect $F\cap \widehat M$, where
$\widehat M\subset Y$
is the vanishing locus of $P$ interpreted as a function on $Y=\IC^n$.

The nonvanishing correlation functions are now of the form \ecorM\ with
$s=d-1$. Interpreting the $d{-}1$ inserted operators as restricting
the path integral to the intersection of $d{-}1$ divisors in
$\CM_\vn$, we find a finite number of solutions at each instanton number.
In the model of section three the contribution of the $\vn$th instanton
sector was computed simply by counting these solutions.
The new feature here is that the contribution counts solutions with a
nontrivial sign. In the absence of a superpotential interaction (and
of an expectation value for the $\sigma$ fields) the fermion
determinant is manifestly real and positive because the
$(\lambda^+,\psi^-)$ determinant is the complex conjugate
of the $(\lambda^-,\psi^+)$ determinant. We can choose the
sign of the operator insertions to respect this complex structure. The
superpotential, on the other hand, manifestly reverses the complex
structure, because it leads for example to a $\psi^i_-\psi^j_+$
term. In computing the contribution at instanton number $\vn$ to some
nonzero correlator there are $1{-}d_0$ zero modes of $\psi_0$ which
cannot be soaked up in the operator insertions (which only involve
$\chi$ and $\chibar$, or in other words have positive (negative)
charge under the right-moving (respectively left-moving) $R$-symmetry.
These zero modes are lifted by the superpotential term; the resulting
measure on the remaining zero modes is the product of a positive
quantity (essentially $|\p^2W|^2$ evaluated on the zero modes)
and the sign originating from the reordering of the fermionic integral
required to render the coefficient of this term positive. Thus the
contribution of $\CM_\vn$ is weighted by a sign proportional to
$(-1)^{1-d_0}$. (The two extra $\chi$ zero modes are more subtle to
treat, but will not affect this discussion.) The overall sign of the
series in $q$ is of course a matter of convention, and we fix it to
obtain agreement with the classical result at $q=0$. This leads to the
final result that the contribution of $\CM_\vn$ is weighted by
$(-1)^{d_0}$.

We have mentioned above that the noncompactness of instanton moduli
spaces and the need to carefully account for ``contributions at
infinity'' are a serious obstacle to computing Gromov--Witten
invariants. Our moduli spaces are compact, but these problems reappear
in the current context as a degeneration of the instanton equations at
the compactification subsets of $\CM_\vn$. (This was pointed out in
\rphases .) The na\"{\i}ve dimension counting fails precisely for
``pointlike''
instantons. Heuristically, the common factor in the polynomials which
characterizes such maps means that $P$ is effectively of lower
degree, and the space of maps satisfying $P=0$ is of larger dimension.
This means we cannot think of adding the pointlike instantons to the
space of holomorphic maps $\Sigma\to M$ as a compactification by
adding lower-dimensional spaces; at a given degree the generic map is
pointlike. For this reason we cannot expect the contribution to the
correlation function at any given degree as we compute it to be the
same as the contribution at this degree in the nonlinear sigma
model. However, our arguments show that the low-energy limit of the
theory is a nonlinear sigma model with target space $M$, and that
further the correlation functions computed in the microscopic theory
should be equal to those in the low-energy theory. The resolution to
this apparent contradiction is that including the pointlike instantons
can lead to a renormalization of the couplings in the low-energy
model. In the case at hand these are the $\tau_a$. In other words, we
can expect the correlation functions we compute to reproduce the
correlators in the nonlinear sigma model, but \eclast\ will receive
nontrivial corrections in this case. The finite shift from
perturbative effects allowed for
there is computable in this case. The contributions of smooth
instantons in the nonlinear model are counted with positive signs (for
essentially the reason given above in the absence of superpotential
interactions), whereas we have here found a nontrivial sign. This is
interpreted as a finite shift in $\theta$ correcting \eclast
\eqn\etpert{
t_a = \tau_a + {1\over 2} Q_0^a\ .
}
In addition, there will be corrections due to the contributions of
pointlike instantons. As we have seen, the moduli space of pointlike
instantons of instanton number $\vn$ is isomorphic to the space of
instantons of a lower degree. Adding such lower degree corrections at
degree $\vn$ is the typical effect of corrections to \etpert\
with exponential drop-off at large $r$.
This will be borne out in the examples
in the next subsection; we note that this situation differs from that
encountered in section three. Thus, in terms of the Gromov--Witten
invariants, we will obtain the correct
generating function but not the correct
expansion variable.

In fact, we will not use the instanton moduli space in the form
described above.
We find it more convenient to choose a different way of incorporating
the effects of the pointlike instantons. This will also lead
to nothing more than a renormalization of the couplings in the low-energy
Lagrangian. We simply replace \einstwii\ by the requirement that
$P=0$ hold at some fixed collection of
 points $Q_0,\ldots Q_{-d_0}$ in $\Sigma$ (with $1{-}d_0$ points in all).
This is the same as \einstwii\ for non-pointlike maps.
The advantages of this particular choice are immediately
apparent. We have effectively reduced the computation once more to a
set of intersection calculations in the toric varieties $\CM_\vn$, a
manifestly tractable problem. Writing an instanton expansion for the
correlation functions as before we are claiming that the modified
\eYvnchi\ reads
\eqn\eXvnchi{\eqalign{
X_{a_1\ldots a_s}^\vn &= (-1)^{d_0}
\la (\eta_{a_1})_\vn\cdot(\eta_{a_2})_\vn\cdots\cdot(\eta_{a_s})_\vn
\chi_\vn (-K)_\vn^{1-d_0}\ra_\vn \cr
&= -\la (\eta_{a_1})_\vn\cdot(\eta_{a_2})_\vn\cdots\cdot(\eta_{a_s})_\vn
\chi_\vn (K)_\vn^{1-d_0}\ra_\vn \ , \cr
}}
where $(K)_\vn$ is obtained from \ek\ using the usual lift to
$\CM_\vn$. This equation contains a complete solution to the {\bf A}
model with target space $M$.

The fact that the ghost number anomaly is independent of instanton
number means that \eXvnchi\ gives the Hilbert space the structure of a
graded algebra of length $d{-}1$. Put otherwise, it is clear that \eXvnchi\
vanishes unless $s=d{-}1$; when this holds there will be nonzero
contributions from all instanton sectors in $\CK_V^\vee$.

The correlation functions computed by summing \eXvnchi\ will be
algebraic functions of the $q_a$.  In particular, their only
singularities will be poles. We will explicitly see this in subsection
{\it 4.5}.
This property singles out the particular
set of coordinates and frame in which we are working (as well as the
overall normalization factor~--~the gauge choice~--~arising from the
normalization of the vacuum state). We call these
the {\it algebraic coordinates}\/ and {\it algebraic gauge}.
As anticipated above, these differ from
the canonical coordinates on moduli space. However, the algebraic
coordinates have the
distinguishing property that they are ``good'' coordinates globally.
In terms of these coordinates, one can unambiguously continue the
correlation functions around their poles and into regions of the
moduli space described by ``nongeometric'' phases of the theory. The
analytic properties of correlation functions in topological field
theories away from the singularities mean that these analytic
continuations in fact yield the correct correlation functions
in these regions of parameter space. In section five we
present an approach to the model which allows us to independently compute
the expansions about nongeometric limiting points.

There are also algebraic
coordinates which appear naturally in discussions of {\bf B} models. The
correlation functions of these are naturally expressed in terms of
intrinsically algebraic objects, e.g., local rings of polynomials. In
\rmondiv\ a mapping from these coordinates (in terms of the {\bf B}
model on a manifold $W$) to the canonical coordinates (on the moduli
space of the {\bf A} model on its mirror $M$), valid in the large-$r$
limit, was proposed and named the {\it monomial-divisor mirror map}.
What we have given above is in fact an interpretation of the algebraic
coordinates in the interior of {\bf A} model moduli space. They
represent the set of coordinates and the frame that arise naturally in
the GLSM. To give a more complete understanding of this directly in
terms of the low-energy nonlinear model we should compute the
instanton corrections to \etpert . We believe this should be possible using
the simple recursive structure of $\CM_\vn$, but will not pursue the issue
here.
In what follows we will use the natural GLSM coordinates; to obtain
from these the {\it algebraic coordinates\/} for
the low-energy nonlinear model
(in the terminology of \rmondiv), the shift \etpert\ should be applied.

\subsec{Solving the Examples}

In this subsection we illustrate the construction of the previous
subsection by explicitly solving our two \CY\ examples. Since we have
$d=4$, the nonzero couplings we compute will be the trilinear Yukawa
couplings. In the first example we have precisely one of these; it is
\eqn\eXq{
X_3 = \lal \sigma^3 \rar = \sum_{n\ge 0} X_3^n q^n
\ .}
Here we have $d_0(n) = -5n$ and $K = -5\eta$,
and the coefficients in the expansion are
\eqn\eXnq{
X_3^n = -\la \eta^3 K^{5n+1}\ra_n = -\la \eta^3 (-5\eta)^{5n+1}\ra_n
 = -(-5)^{5n+1}
\ .}
Inserting this we can sum the series to find
\eqn\esolq{
X_3 = {5 \over 1 + 5^5 q}
\ .}
This exhibits the expected singularity at
$\tau = \half + i{5\log 5\over 2\pi}$.
Notice, however, that the singularity is just a simple pole in $q$ and
there is no difficulty in continuing around this pole to define the
correlation function for all $r,\theta$.
The Yukawa coupling in this model has been computed by
Candelas et al.\ \rCdGP\
using mirror symmetry; our result is in complete agreement with theirs
if one sets $q = (-5\psi)^{-5}$ and $X_3 = \kappa_{\tau\tau\tau}$ and
performs the change of variables treating $\kappa$ as a tensor. (Note
that in fact $\kappa$ is a tensor-valued section of a line bundle
$\CL^2$. We have obtained a trivialization of $\CL$ identical to that
which came up naturally in the work of \rCdGP . We will discuss this
issue in section five.)

In the second example there are four different Yukawa couplings, which
we denote by
\eqn\eYj{
X_j = \lal \sigma_1^{3-j}\sigma_2^j\rar
\ .}
Here we have $d_0(n) = -4n_1$ and $K = -4\eta_1$; the expansion
coefficients are thus
\eqn\eYjn{
X_j^\vn = -\la \eta_1^{3-j}\eta_2^j K^{4n_1+1}\chi_\vn\ra_\vn =
2^{8n_1+2} \la \eta_1^{4n_1+4-j}\eta_2^j\ra_\vn\ .
}
Comparing \esoly\ we thus find
\eqn\esolyj{
X_j = X_j^{(0)} +\sum_{n_1,n_2\geq 0; (n_1,n_2)\ne(0,0)}
2^{8n_1+2n_2+3-j}\left( {n_1+1-j\atop 2n_2+1-j}\right)
q_1^{n_1} q_2^{n_2}\ .
}
We have separated the $\vn=0$ term out; this is the classical
intersection calculation on $M$, given by \eclass . Performing the
sums we find
\eqn\eyjs{
\eqalign{
X_0 &= {8\over\Delta}\cr
X_1 &= {4(1-2^8 q_1)\over\Delta}\cr
X_2 &= {8q_2(2^9 q_1-1)\over (1-4 q_2) \Delta}\cr
X_3 &= {4q_2(3072 q_1 q_2 + 2^8q_1 -4 q_2 -1)\over (1-4 q_2)^2 \Delta}\ ,\cr
}}
where $\Delta = (1-2^8 q_1)^2 - 2^{18} q_1^2 q_2$ (compare \eprindis\
and \emordis ).

Comparing with the correlation functions
computed in the mirror model by Candelas et al.\ \rCdFKM, we find that under
the substitution $(-\phi)^{-2} = 4 q_2,\ (-\psi)^{-8} = 2^{24} q_1^2 q_2$
the correlation functions are identical if thought of as the components
of a tensor of rank three on the parameter space.
(This substitution is precisely the one specified by the monomial-divisor
mirror map \rmondiv, as noted in \rCdFKM.)

These results verify our conclusion that we have correctly summed the
instanton series to obtain exact correlation functions. They also show
that these are obtained in the ``algebraic'' coordinates and gauge.

\subsec{Quantum Restriction Formula}

In the previous subsections we have shown how toric geometry allows us
to sum the instanton series and compute exactly the correlation
functions for the topological nonlinear sigma model with target space
$M$. In this subsection we will find a general formula relating the
quantum cohomology of $M$ to that of $V$. In the case of smooth $V$,
where the discussion of subsection {\it 3.9\/} gives an algebraic
computation of the latter, the result below leads to an algebraic
computation for $M$. Essentially we will find the quantum version of
\eclass .

We have almost found this in the formula \eXvnchi . We would like to
interpret this formula as the contribution \eYvnchi\ to some
correlator in the $V$ theory. The difficulty is only that the operator
inserted in \eXvnchi\ seems to depend upon $\vn$ (beyond the simple
lift in \eYvnchi ). The solution to this is quite simple, however.
When \eXvnchi\ is nonzero and $s=d-1$ is the dimension of
$M$~--~in which case the correlator we are
computing receives nonzero contributions from all instanton
numbers~--~we can in fact replace \eXvnchi\ by
\eqn\eXnsum{
X_{a_1\ldots a_s}^\vn =
-\Big\langle (\eta_{a_1})_\vn\cdot(\eta_{a_2})_\vn\cdots\cdot(\eta_{a_s})_\vn
\chi_\vn\sum_{m> 0} K_\vn^m\Big\rangle_\vn
\ .}
Because $d_0\le 0$ one term in the sum over $m$ is \eXvnchi ;
all other terms vanish because the class we evaluate is not of top
degree in $\CM_\vn$. Now we can regroup the sums in $X_{a_1\ldots
a_s}$ to write the correlation function in the $M$ theory as a
particular computation in the quantum cohomology ring of $V$
\eqn\eXqv{
X_{a_1\ldots a_s} = -\sum_{m > 0} \la
\sigma_{a_1}\cdots\sigma_{a_s} K^m\ra
\ .}
Notice that if $s<d-1$ then \eXnsum\ and \eXqv\ still coincide with
\eXvnchi, since all are zero; however, if $s>d-1$ then \eXnsum\ and \eXqv\
can be nonzero even though \eXvnchi\ vanishes.

More mathematically, thinking of the correlation function as defining
an expectation function
on the quantum cohomology algebra, we can write a general result,
the {\it quantum restriction formula}, for the restriction to
an anticanonical hypersurface:
\eqn\eqsthg{
\lal \CO\rar = \la \CO {-K\over 1-K}\ra
}
where $\CO$ is any polynomial in $\eta_a$ of degree at most
$d-1=\dim_{\IC}M$.  This provides a way to compute the quantum
cohomology algebra of $M$ in terms of the quantum cohomology of $V$,
the latter being a
problem which was solved in the previous section.
This quantum restriction formula is to be interpreted as follows: the
sum
\eqn\egeomseries{
\sum_{m=1}^\infty K^m
}
should be expected to
converge in the quantum cohomology algebra of $V$ (with respect to
any Hermitian norm on that algebra, thought of as a complex vector
space).  If it does converge as expected, then $1-K$ must be
invertible in the algebra, and the sum \egeomseries\ must converge
to $-K/(1-K)$.
The correlation functions for $M$ can thus be computed using \eqsthg.
(Although we cannot directly interpret \eqsthg\ as
defining an
expectation function on a quotient algebra, we will see in the next
section that a variant of this calculational procedure does admit
such an interpretation.)

We note that when $-K = p e$ for $e\in \IZ [\eta_1,\ldots ,\eta_{n-d}
]$ then the summand in \eXnsum\ is nonzero (at any $\vn$) only for
$m = 1\, {\rm mod}\, p$,
so  \eqsthg\ can be replaced by
\eqn\eqsthgp{
\lal\CO\rar=\la\CO{-K\over1-K^p}\ra\ ;}
this form is manifestly consistent with the $\IZ_p$ grading and
is useful for explicit computations.  Eqn.~\eqsthg\ (or its variant
eqn.~\eqsthgp) is precisely the desired quantum version of \eclass.

\line{\hrulefill}\nobreak
\noindent{\it Example 1.}\par\nobreak
We work in the quantum cohomology ring of $V=\CP4$, which is generated
by $\sigma$ with relation $\sigma^5=q$; the expectation function is
normalized by
$\la\sigma^4\ra=1$.  Since $-K=5\sigma$, we may use \eqsthgp\ with $p=5$,
and we thus must calculate
\eqn\eqresone{
{-K\over1-K^5}={5\sigma\over1-(-5\sigma)^5}={5\sigma\over1+5^5q}\ .}
{}From this, it is easy to check that the quantum restriction
formula \eqsthgp\ agrees with the
earlier calculation \esolq.

\noindent{\it Example 2.}\par\nobreak
In this example, the quantum cohomology ring of $V$ is generated by
two classes $\sigma_1$ and $\sigma_2$, and the
 ideal of relations is generated by
the two polynomials
\eqn\erelwon{F_1\equiv\sigma_1^3(\sigma_1-2\sigma_2)-q_1\ ,}
\eqn\ereltwo{F_2\equiv\sigma_2^2-q_2(\sigma_1-2\sigma_2)^2\ .}
The normalization of the expectation function can be taken as
$\la\sigma_1^4\ra=2$ or equivalently $\la\sigma_1^3\sigma_2\ra=1$
(as is seen from \erln\ with $\vec{n}=0$), and the anticanonical
class is $-K=4\sigma_1$.  We can set $p=4$ in \eqsthgp; in order to find
\eqn\eqrestwo{
{-K\over1-K^4}={4\sigma_1\over1-4^4\sigma_1^4}}
we must make some calculations with the ideal generated by $F_1$ and $F_2$.
The specific element of that ideal which we need is
\eqn\eidealelt{
\left[(1+4q_2)\sigma_1^4+(2-8q_2)\sigma_1^3\sigma_2
+(4q_1q_2-q_1)\right]F_1+
4\sigma_1^6F_2
=\sigma_1^8-2q_1\sigma_1^4+q_1^2(1-4q_2)
\ .}
{}From the fact that the right side of \eidealelt\ is zero in the quantum
cohomology ring, it is easy to find the inverse of $1-4^4\sigma_1^4$
in that ring and to conclude that
\eqn\eqrestwocon{
{4\sigma_1\over1-4^4\sigma_1^4}=
{4\sigma_1(2^8\sigma_1^4+(1-2^9q_1))\over\Delta}
\ ,}
where $\Delta=(1-2^8q_1)^2-2^{18}q_1^2q_2$ as in eqn.~\eyjs.
A short calculation then verifies that the quantum restriction
formula \eqsthgp\ reproduces eqn.~\eyjs.
\par\nobreak\line{\hrulefill}

\newsec{The $V^+$ Model, Mirror Symmetry and the Monomial-Divisor Mirror Map}

In the previous section we have, essentially, solved the {\bf A}
topological nonlinear sigma model with target space $M$ a hypersurface
in a toric variety $V$. Our computation was based in the K\"ahler cone
of $V$, but we found that the correlation functions
are meromorphic objects, hence naturally determine analytic
continuations to the entire parameter space of the model. This is
certainly natural from the point of view of the mirror theory, since
the space of complex structures on the mirror manifold $W$ is of the
general form $A -  B$ where $A$ and $B$ are subvarieties in
projective space (the r\^ole of $B$ is played by the singular locus, $A$
is essentially complexified
$r$-space). It thus becomes unnatural to treat the
theory in a way that distinguishes the K\"ahler cone. In this section
we reformulate the model so that instanton expansions may be computed
about a limiting point deep in the interior of {\it any}\/ of the cones
in $r$-space, leading to analytic correlation functions throughout the
nonsingular part of the parameter space.

It is intriguing that hypersurfaces in toric varieties are
precisely the class of \CY\ manifolds  for which a detailed conjecture
on the construction
of a mirror manifold $W$, itself realized as a hypersurface in a toric
variety $\varLambda$, was proposed by Batyrev \rbatmir.
Since in some sense at least the
{\bf B} model on $W$ should be easier to solve than the {\bf A} model
on $M$ (the correlation functions are given in terms of classical
geometry) we should be in a position to prove these conjectures, at
the level of topological field theory. Furthermore, since we have
explicit solutions we should be able to shed some light on the mirror
map between the moduli spaces of the two models. In this section we
will partially fulfill these expectations. In particular, in the case
that $V$ is smooth we have an algebraic computation of the
correlators. It is in this case that we will manage to make contact
with {\bf B} model calculations. Our ability to prove the conjecture
in this class of models is limited only by the absence of a general
algebraic solution of the {\bf B} model.

\subsec{Toric Geometry, Once Again}

Given a toric variety $V$ we have seen that there is a canonical way
of producing a Calabi--Yau hypersurface in $V$ (more properly, a
family of hypersurfaces related by deformations). There is another way
of producing a Calabi--Yau space, from any K\"ahler manifold. We
consider not a
hypersurface but the total space of a line bundle over $V$. To have
trivial canonical class this line bundle should be the canonical line
bundle. For a toric variety $V$ this construction was introduced in
section four as $V^+$. This manifold is simple in
that it is a {\it toric}\/ Calabi--Yau manifold. Of course, it is
noncompact. A holomorphic quotient construction of $V^+$ is
given by \evplus .

In terms of the fan $\Delta$ we obtain a toric construction for $V^+$
by considering a fan $\Delta^+$ in $\IR^{d+1}$ given by $(n+1)$
vectors lying in
the affine hyperplane $y_{n+1}=1$. These comprise the vertices of $\Delta$,
promoted to the hyperplane
by the addition of a last component whose value is $1$, as well as the
origin in the construction of $\Delta$, similarly promoted. As is
clear from the affine hyperplane condition, this is not a complete fan, as
expected since $V^+$ is noncompact.

The dimension of $V^+$ is $d+1$.  The problem of computing the
ring structure on the cohomology of $V^+$ is ill-posed
since the manifold is noncompact, but there is certainly one compact
divisor~--~the zero section of the bundle $\{ x_0=0\} = V$. If we only
consider intersection calculations involving this divisor (and other
classes) then the
problem will be well-posed. The class of this divisor is represented
by $\xi_0$, and in practice,
\eqn\etr{
\Tr_{\rm cl}(\CO) = \la \xi_0 \CO \ra_{V^+}
}
defines an expectation function on a quotient algebra
$\IC[\eta_1,\ldots,\eta_{n-d}]/\CJ_{\rm cl}$ (by Nakayama's theorem),
yielding a graded ring of length $d$.
Of course, this will be precisely $H^*(V)$. As in the previous
section, we will look for a useful quantum generalization of this.
What we will find is in fact a modification of the quantum cohomology
of $V$, one which leads very naturally to a solution for the quantum
cohomology of the hypersurface $M$.

The combinatorial condition \rbatmir\ which ensures
that $V$ contains quasi-smooth Calabi--Yau hypersurfaces $M$ is most
easily stated in terms of $V^+$ and $\Delta^+$.  Consider
the cone $\Sigma =|\Delta^+|$
which is the support of $\Delta^+$ in $\bN^+_{\IR}\sim \IR^{d+1}$. This
is a {\it Gorenstein\/} cone (cf.~\rbatbor), which means if
we take its intersection
with the hyperplane $n_{d+1}=1$ we obtain points of the lattice $\bN$ (the
generators of one-dimensional cones in $\Delta$) as vertices of the
resulting polytope ${\cal P}\equiv\Sigma\cap\{n_{d+1}=1\}$.
The condition we need is that $\Sigma$ be a
{\it reflexive\/} Gorenstein cone of index one.\foot{This version of
the definition from \rbatbor\ is equivalent to Batyrev's original
definition in \rbatmir.} This means that if we
construct the dual cone
$\Sigma^\vee = \{m\in \bM^+_{\IR}\ |\ m\cdot n\ge 0\,\forall
n\in\Sigma\}$, it too must be Gorenstein, that is, its intersection
with the hyperplane $m_{d+1}=1$ must be a polytope ${\cal P}^0$ whose
vertices lie in the lattice $\bM$. ${\cal P}^0$ is called the
{\it polar polytope\/}
of $\cal P$.

We now review
the construction, due to Batyrev
\rbatmir , of a dual (to $V$) toric variety $\varLambda$. Batyrev's
conjecture is
that the family of Calabi--Yau hypersurfaces $W$ determined by
$\varLambda$ is related by mirror symmetry to $M$. More accurately, the
facet of the conjecture upon which we focus here states that some subspace
of the space of deformations of $W$ is related to the space of
K\"ahler structures ($r$-space) for $M$ by a mirror map, which will
equate correlation functions in the {\bf A} model for $M$ as computed
above to correlators in the {\bf B} model on $W$.

We can use the
polar
polytope ${\cal P}^0$ to determine $\varLambda$. To do this
we need to find the fan $\nabla$ which will play the r\^ole that
$\Delta$ played for $V$, so we need to see the relation between $\cal
P$ and $\Delta$. The one-dimensional cones in $\Delta$ include the
rays containing the
vertices of $\cal P$, of course. The construction guarantees that both
$\cal P$ and ${\cal P}^0$ have a unique interior lattice point~--~the
origin in $\bN$ and $\bM$ respectively. But there can be points in the
faces of the polytope, and we can choose different fans by including
different subsets of the rays generated by
these in the set of one-dimensional cones. As
described in detail in \rAGM, different choices correspond to
different birational models; in our language different choices for
$\Delta$ correspond to different values for $r$ in the cone $\CK_q$ of
``geometric'' phases. There is thus some ambiguity in determining
$\varLambda$, but for our current purposes this will be irrelevant. The
reason is that we will be interested in the {\bf B} model on a
hypersurface $W\subset\varLambda$. For this model, the parameters $r$ are
irrelevant (they couple to $Q$-exact operators) just as the coefficients of
the polynomial $P$ were irrelevant to the {\bf A} model on $M\subset V$.
We will find it convenient to choose a model for $\varLambda$ in which we
include none of these extra points. This (generically singular) toric
variety has the property that its anticanonical divisor class is
ample. The salient point here is that $\varLambda$ is determined
uniquely by this construction to within irrelevant choices. Then, as
described in section four, $\varLambda$ itself determines a family of
\CY\ hypersurfaces $W\subset\varLambda$ related by deformations. The
deformation space of these will be related by mirror symmetry to the
$r$-space of the {\bf A} model on $M$. The family of all models
corresponding to any point in $r$-space is determined by the
polyhedron $\cal P$ (which is the convex hull of the integral
generators of the one-dimensional cones in
the fan $\Delta$); this is also precisely the information needed to
construct $\Sigma$ and hence $\varLambda$. This shows that the
construction is involutive in the sense that had we started with the
fan $\nabla$ for some model of $\varLambda$ we would have been led to
construct ${\cal P}^0$ and the dual cone would lead to $\cal P$. When
studying the {\bf A} model, we will prefer to find a smooth model for
$V$ (we assume such a model exists) and so will include the rays
generated by {\it all\/} of
the points of $\cal P$ in $\Delta$.\foot{The exception to
this is the case of points in the interior of codimension-one faces of
$\cal P$. These correspond to point singularities in $V$ which
therefore will be disjoint from generic hypersurfaces; we do not need
to include these.}

The rather abstract combinatorics discussed above can be interpreted
in a somewhat more concrete way as follows. As mentioned in
previous sections, points in $\bM$ correspond to monomials in the
homogeneous coordinates which can be interpreted as meromorphic functions
on $V$. Points in $\bM^+$ can (by including the appropriate powers of
$x_0$ with the monomial)
be seen to correspond to sections of powers of the anticanonical
line bundle. The cone $\Sigma^\vee$ contains the {\it holomorphic\/}
sections; thus the set of all points in $\nabla$ is precisely the set
of holomorphic sections of $-K_V$. This was identified in section four as
being precisely the set of possible gauge-invariant monomials linear
in $\Phi_0$ which could be used to construct the superpotential $W$.
The involutive property of the construction discussed above means we
can give the same interpretation to the points in $\cal P$ as
monomials appearing in the superpotential for the GLSM describing
$W\subset\varLambda$. This association of divisor classes ($\xi_i$) to the
corresponding monomials~--~call them $\mu_i$~--~is
the basis for the {\it monomial-divisor mirror
map\/} of \rmondiv , and will appear in our discussion of mirror
symmetry.

\subsec{The Topological Model for $V^+$}

When we studied the gauged linear sigma model leading to the $M$ model we
found that the correlators of this theory are independent of the
details of the superpotential $W$. It is tempting, therefore, to
simplify the model by simply setting $W=0$.\foot{We thank E. Witten
for suggesting that the computations from section four might have an
interpretation in such a theory.}  By the discussion of
section two this is na\"{\i}vely expected (for some range of $r$) to lead to
a topological theory equivalent to the {\bf A} model with target space
$V^+$. Of course, there are difficulties with defining this model,
because of the noncompact space of bosonic zero modes; the point $W=0$
is clearly a singular point in the parameter space. Further, if we
were to obtain the $V^+$ model it would be a topological conformal
field theory (since $V^+$ is \CY ) with central charge (ghost number
anomaly) $d{+}1$~--~the dimension of $V^+$~--~rather than $d{-}1$. We will
see that the two theories, while not identical, are intimately
related.

We thus proceed to apply the formal methods of section three to the
model of the previous section with $W=0$, which we will call the $V^+$
theory. The absence of a superpotential will suggest a more symmetric
treatment of $\phi_0$ and the $\phi_i$. The $Q$-cohomology classes are
still represented by the
$\sigma_a$ which we continue to identify with $\eta_a$.
The path integral will be well-defined at the level of zero
modes for correlation functions containing an insertion of
\eqn\exio{
\delta_0 \sim -\sum_{a=1}^{n-d} Q_0^a \sigma_a = -\sum_{i=1}^n \xi_i
\ .}
The zero section $V\subset V^+$ is a representative of the divisor
class $\xi_0$.  Thus in the
zero mode sector this insertion of $\delta_0$ reduces the integral to $V$.
In the presence of this insertion, the discussion of the model
along the lines of previous sections is straightforward. Once more one
finds singularities along the cones in $r$-space on which a continuous
symmetry is unbroken; perturbative quantum effects are limited to the
finite shift \eref . The computation of the singular locus in section
four made no use of the superpotential so continues to hold here as well.
The new features arise when we consider instanton
effects.

In this model the only modification to \einst{}\ is the addition of
the extra field $\phi_0$. This in effect means that the moduli spaces
are naturally obtained as noncompact toric varieties, at least when
$d_0\geq 0$. If
we restrict attention momentarily to the ``geometric'' phases,
however, the presence of $\delta_0$ in correlation functions will
suffice to reduce the computation to an intersection computation in
the compact moduli spaces $\CM_\vn$ of the previous sections.
The arguments immediately following \einstw\
show that $d_0\le 0$. Thus $\phi_0$ is at most a constant; then the
insertion of $(\delta_0)_\vn$, interpreted as restricting attention to
maps for which $\phi_0$ vanishes at a point, is in effect a
restriction to $\phi_0=0$. Thus in the
``geometric'' phases, the instanton moduli spaces are the $\CM_\vn$
of section three. The correlation functions we compute will be denoted
\eqn\eZs{
Z_{a_1\ldots a_s} = \lall
\sigma_{a_1}(z_1)\cdots\sigma_{a_s}(z_s)\delta_0(z_0)\rarr \ .
}
By the reasoning of previous sections these have an expansion
\eqn\eZex{
Z_{a_1\ldots a_s} =
\sum_{\vn\in \CK^\vee} Z_{a_1\ldots a_s}^\vn
\prod_{a=1}^{n-d} q_a^{n_a}\ .
}
The restriction to
$\CK^\vee$, the dual of the K\"ahler cone of $V$,
follows here using the $\delta_0$
insertion.

The expansion coefficients are once more computed as intersection
calculations in $\CM_\vn$. Since $\phi_0=0$ is here imposed by degree
considerations and
not by \einstw\ we will have ghost zero modes and an associated Euler
class. In addition, due to the absence of a superpotential
the sign that modified the contribution computed
in section four is absent here.
The arguments of section three lead, up to normalization, to
\eqn\echiplus{
\chi^+_\vn = \prod_{d_i<0} (\xi_i)_\vn^{-d_i-1}\ ,
}
where the product includes (where appropriate) $i=0$. As above, we
assume the simplest normalization. In a ``geometric'' phase, this leads to
the following situation. If $d_0=0$, we have $\chi^+ = \chi$. Moreover,
the $\delta_0$ insertion in \eZs\ is absorbed in restricting
to $\CM_\vn$ so that
\eqn\eZvno{
Z_{a_1\ldots a_s}^\vn =
\la (\eta_{a_1})_\vn\cdots(\eta_{a_s})_\vn\chi_\vn\ra_\vn\ .
}
On the other hand, for $d_0<0$, we have $\chi^+ =
\xi_0^{-d_0-1}\chi$ and the insertion is not used to perform the
restriction so that
\eqn\eZvnchi{
Z_{a_1\ldots a_s}^\vn =
\la (\eta_{a_1})_\vn\cdots(\eta_{a_s})_\vn\chi_\vn(\xi_0)_\vn^{-d_0}\ra_\vn\ .
}
Comparing these with \eXvnchi\ we see that for
any monomial $\CO$ in the $\sigma_a$
\eqn\eclear{
\lal \CO \rar = \lall (-\delta_0^2) \CO \rarr\ .
}
Thus the $V^+$ model, modified by the insertion of $(-\delta_0^2)$,
computes the correlators of the $M$ model.

It may thus seem that we have done little more than find a rather
clumsy way to rewrite what we already knew, but we will see that this
expression has some advantages over the previous one. For now, we note
that if we define the quantum generalization of \etr
\eqn\etrt{
\Tr(\CO) = \lall \delta_0 \CO \rarr
\ ,}
then \eclear\ in conjunction with the results of section four shows
that the ring
\eqn\rone{
\CR_0(V)\equiv \IC [ \delta_0,\ldots,\delta_n]/\CJ_0
\ ,}
where
\eqn\eJone{
\CJ_0\equiv \{ {\cal P}\in \IC [ \delta_0,\ldots,\delta_n]\ |\
\Tr (\CO {\cal P}) = 0
\quad \forall\CO \}
\ ,}
is a graded ring of length four. This ring will play a part in the
argument of subsection {\it 5.4}.

The relations in $\CJ_0$ include a set of linear relations, following
from \elinrd\ and \exio , which together can be written as
\eqn\elinjo{
\sum_{i=0}^n \la m^+,v_i^+\ra \delta_i = 0\ .
}
We find additional nonlinear
relations by repeating the argument of subsection {\it 3.9\/}. The
considerations surrounding \etosho\ express properties of the
$\CM_\vn$. In the new model their interpretation \etherel\ is
modified, because instanton contributions are associated to factors of
$\delta_0$. Thus the modified relations read
\eqn\erelone{
\prod_{i=1}^n \delta_i^{d_i^*} = \prod_{a=1}^{n-d} q_a^{n_a^*}
\delta_0^{-d_0^*}\ .
}
The modification renders the relations homogeneous as one
expects for a graded ring. When $V$ is smooth the arguments of Batyrev
as presented in subsection {\it 3.9\/} guarantee that these relations generate
$\CJ_0$. (The key point is the fact~--~noted above~--~that in the
limit as $q_a\to0$ the ring $\CR_0(V)$ approaches the classical
cohomology ring $H^*(V)$.)
As illustrated in appendix B, there can be additional relations
when $V$ fails to be smooth.

In terms of $\CR_0(V)$ we can rewrite
\eclear\ as
\eqn\eclearr{
\lal\CO\rar = \Tr (-\delta_0 \CO )\ .
}
We will think of this as determining the chiral ring of the {\bf A}
model with target space $M$ (i.e., the quantum cohomology ring of $M$)
in terms of $\CR_0(V)$ as being
\eqn\echiral{
\CR_A(M) = \CR_0(V)/\{{\cal P}\ |\ \lal\CO{\cal P}\rar=0\ \forall\CO\}
\ .}
Notice that the form of \eclearr\ is again precisely what one expects
from Nakayama's theorem~--~it defines $\lal\ \rar$ as an expectation
function on a quotient algebra of $\CR_0(V)$.

\subsec{Other Phases}

As discussed in section four, the correlation functions in the $M$
model are rational functions of the parameters in the coordinates and
frame given by our simple choices. This means their only singularities
are poles, and analytic continuation into the complement of the
singular locus in $r$-space is unambiguous. This continuation leads to
predictions for the correlation functions in ``nongeometric'' phases. The
analysis to this point has been restricted to the ``geometric'' phases,
where we could observe that analytically continuing around a singular
locus from one ``geometric'' phase to another leads to correlators which
agree precisely with those computed in the new phase directly (a local
argument showing how this happens is given in \rphases ). It is
therefore very natural to assume that the same procedure will yield
the correlation functions in the ``nongeometric'' phases as well, since
the correlators must be holomorphic objects away from the
singularities. Explicit computations in ``nongeometric'' phases can be
difficult, though, and thus direct verification of this assumption has
been unavailable. (Of course, in addition to the general argument of
holomorphicity one can use mirror symmetry to prove this indirectly.)

One advantage of the symmetric treatment of $\phi_0$ is that
computations in other regions of $r$-space are now
formally as simple as those for the ``geometric'' phases. The
moduli spaces are determined from \einst{} in the same way; the Euler
classes $\chi$ follow from \echiplus .
The correlation functions are now computed as
expansions about various limiting points in $q$ corresponding to the
various semiclassical limits in the model. These expansions will
agree with each other (after analytic continuation).
We will encounter a new subtlety in studying the ``nongeometric''
phases. In some instanton sectors we find an
unbroken discrete subgroups $H\subset G$. As discussed in section four, when
this occurs we should quotient by $H$. In practice
this means that our na\"{\i}ve computation of the contribution to the
path integral from these sectors will overcount by a factor of $|H|$.
We will illustrate this
in our two examples, performing the calculations in all
non-smooth phases. (The orbifold phase of example 2 could have been
discussed in the previous sections, but we treat it with the other
phases now.)

\line{\hrulefill}\nobreak
\noindent{\it Example 1.}\par\nobreak
We now consider the first example, in the Landau--Ginzburg phase $r\ll
0$. In this region we find $\CK^\vee = \{n\le 0\}$ and the expansion is
in ``anti-instantons''. A physical description of these was given in
\rphases . The instantons appear as Nielsen-Olesen vortices. Since
the nonzero expectation value of $\phi_0$ breaks $G=U(1)$ to $\IZ_5$,
we expect the
instanton number to be quantized in ${1\over5}\IZ$. This would lead
to an expansion of correlation functions in powers of $q^{-1/5}$.
Analytically continuing a rational function, however, will never lead
to branch singularities, so we will find, in fact, contributions only
from integer instanton number. First we will need
a description of the moduli spaces. For $n=0$ the $D$ term will
require $\phi_0 \neq 0$ because of the sign of $r$. This however will
never intersect our chosen representative of $\delta_0$, hence we get no
contribution from the zero sector. This is a general feature of
``nongeometric'' phases. For $n<0$ we see that $d_i = n<0$ so $\phi_i=0$,
whereas $d_0 = -5n>0$. Here $Y_n = \IC^{1-5n}$; as before $F_n =
\{0\}$ is the origin. Thus $\CM_\vn = \CP{-5n}$. We also need
$\chi_n = \prod_{i=1}^5 (\xi_i)^{-d_i-1} = (-\xi_0/5)^{-5-5n}$.
The unbroken $\IZ_5$ symmetry in each sector means
we need to divide our na\"{\i}ve result by 5.

We are interested in $X_3$ of \eXq
\eqn\eXplus{
X_3 = \lal \sigma^3 \rar = \lall (-\delta_0^2)\sigma^3\rarr
\ .}
We will of course reexpress $\sigma=-\delta_0/5$
(and $\eta = -\xi_0/5$), and find
\eqn\eXlg{
X_3 = \sum_{n<0} \la (-5)^{5n+1} (\xi_0)^{-5n}\ra_n q^n =
{5\over 1+5^5 q}
\ ,}
in agreement with \esolq\ as expected.

\noindent{\it Example 2.}\par\nobreak
Here there are, in addition to the smooth phase discussed in
section four, three additional phases. We treat these in order.
To compute the contributions of the various instanton numbers we will
use
\eqn\epgen{
X_j^\vn = \la \eta_1^{3-j}\eta_2^j \chi^+_\vn\ra_\vn
\ .}

In phase II we find $\CK^\vee = \{ n_1\geq 0,\ n_1\ge 2n_2 \}$.
As in phase I, there are two regions to study. The region
$n_1,n_2\ge 0,\ n_1\ge 2n_2$ comprises precisely those instanton
configurations studied in the discussion of phase I which still make
sense in the unresolved model. The contributions of these to
correlation functions will be identical to those computed in section
four by the arguments of subsection {\it 5.2}\/ (except for the contribution of
$\vn=0$~--~nontrivial since we are in a ``geometric'' phase~--~which
we include separately by hand). The
new region is $n_1\ge 0,\ n_2< 0$. In this region $d_1=d_2< 0$;
setting $\phi_1=\phi_2=0$
the moduli space is seen to be $\CP{3n_1+2}\times\CP{n_1-2n_2}$ with
the hyperplane section of the first factor corresponding to $\eta_1$
and of the second to $\xi_6$. The unbroken discrete subgroup is
$\IZ_2\subset G_2$. Here
$\chi^+_\vn=\eta_2^{-2-2n_2} (4\eta_1)^{4n_1+1}$ hence
\eqn\epii{
X_j^\vn = 2^{8n_1+2n_2+3-j} \left( {j-2-2n_2\atop n_1-2n_2}\right)
\ .}
This is nonzero for $n_1\le j-2$, and agrees precisely  with \eyjs .

In phase III we have $\CK^\vee = \{  n_1\le 0,\ n_1\ge 2n_2 \}$. This
will lead to the simplest computation of the correlation functions, a
general property of Landau--Ginzburg phases. In this (nongeometric)
phase we see that $\vn=0$ will not contribute~--~once more a constant
$\phi_0$ would need to be nonzero to solve $D=0$. Here $d_i<0$ for
$1\le i\le 5$, and the moduli space is
$\CM_\vn=\CP{-4n_1}\times\CP{n_1-2n_2}$; the hyperplane classes are
$\xi_0,\ \xi_6$ respectively. Also
$\chi^+_\vn=\eta_2^{-2-2n_2}\eta_1^{-3-3n_1}$, and
$H=\IZ_2\times\IZ_4$. We find
\eqn\epiii{
X_j^\vn = - 2^{8n_1+2n_2+3-j} \left( {j-2-2n_2\atop j-2-n_1}\right)
\ .}

In phase IV, $\CK^\vee = \{ n_1\le 0,\ n_2\ge 0\}$; once more $n_1=0$
cannot contribute. Here we have $\CM_\vn =
\CP{-4n_1}\times\CP{2n_2+1}$, with hyperplane classes $\xi_0$ and
$\eta_2$, and $\chi^+_\vn = \eta_1^{-3-3n_1} \xi_6^{2n_2-n_1-1}$,
and $H=\IZ_4\subset G_1$. Hence
\eqn\epiv{
X_j^\vn = 2^{8n_1+2n_2+3-j}\left( {2n_2-n_1-1\atop j-2-n_1}\right)
\ .}

We see that the expansions in all phases agree as expected. The
enclosed region in figure~3 is interesting. It is inherently
strongly-coupled in the sense that none of the instanton expansions
about semiclassical limiting points converges there. Thus we do not
have an understanding of the low-energy degrees of freedom in this
region even in the limited sense in which this exists for, say, phase
IV.
\par\nobreak\line{\hrulefill}

\subsec{Mirror Symmetry}

\nref\rshelperii{P. Berglund and S. Katz,
``Mirror symmetry for hypersurfaces in weighted projective space and
   topological couplings,''
Nucl. Phys. {\bf B420} (1994) 289--314, hep-th/9311014.}%

We now turn to Batyrev's construction of a (conjectured) mirror
manifold to $M$. This is realized in a toric variety $\varLambda$
determined by combinatorial data which are dual in a natural sense to
those of $V$. To prove the conjecture we would need to show that the
superconformal field theories determined by the two manifolds are
isomorphic. In this paper we will do quite a bit less. What we can
attempt with our methods is to prove that the correlation functions
we are able to compute in the {\bf A} model~--~those of the operators
which are related to cohomology classes in $H^*_V(M)$~--~are equal to
suitably defined correlators in the {\bf
B} model on $W$.
The mirror counterpart of $H^*_V(M)$ will turn out to
be the subring of $\oplus H^p(\wedge ^q TW)$ which is generated by
polynomial deformations of the complex structure. We will denote
this subring by $P_\varLambda (W)$. In proving the isomorphism we
will also find a mirror map for the ``toric'' parameter spaces of the two
theories~--~the spaces of deformations generated by the elements in the
ring at degree one. We note that while falling short of a
complete proof of mirror symmetry, this is a much stronger
statement than what has previously been proved, namely the equality of
respective Hodge numbers \rbatmir\ and of the asymptotic ($q=0$)
limits of (some
of) the correlators \refs{\rAGM,\rbator,\rshelperii}. We will manage
to establish the
equality of correlation functions under the following hypotheses: (i)
$M$ is realized as a hypersurface in a {\it smooth\/} toric variety
$V$. This will allow us to use the algebraic solution of
subsection {\it 3.9\/} or, equivalently, \erelone . (ii) The
fan $\nabla$ determining
the model for $\varLambda$ in which the canonical divisor is ample is
{\it simplicial}. This technical requirement means that the cones in
the fan are all cones over simplices. We will need this condition
because various properties of the {\bf B} model on $W$ that we will
use are proved (at present) only when it holds.

\nref\rBatHodge{V. V. Batyrev,
``Variations of the mixed {H}odge structure of affine hypersurfaces
  in algebraic tori,'' Duke Math. J. {\bf 69} (1993) 349--409.  }%

The method of proof will be the following. We will construct from the
data of the {\bf B} model on $W$ a ring $P_0(\varLambda )$ which
we call the {\it extended chiral ring}.  This ring~--~which has been
studied by Batyrev \refs{\rBatHodge,\rbator} in a slightly different
formulation~--~is related to
the mixed Hodge structure on the affine hypersurface $W\cap \CT^\vee\subset
\CT^\vee$, where\foot{The notation $\CT^\vee$ indicates that this
torus is dual to the torus $\CT$ which is contained in the toric
variety $V$.} $\CT^\vee \sim (\IC^*)^d$ is the torus
 contained in the toric variety $\varLambda$.  We will construct a
natural isomorphism~--~the {\it global monomial--divisor mirror
map}\/~--~between $P_0(\varLambda)$ and the ring
 $\CR_0(V)$ of \rone .

\nref\rbatcox{V. V. Batyrev and D. A. Cox, ``On the Hodge structure of
projective hypersurfaces in toric varieties,''
Duke Math. J. {\bf 75} (1994) 293--338, alg-geom/9306011.}
\nref\rcox{D. A. Cox, ``Toric residues,'' Amherst preprint, 1994,
alg-geom/9410017.}

The ring $\CR_0(V)$
comes with an expectation function \etrt . The isomorphism we
establish will allow us to use this as an expectation function on
$P_0(\varLambda )$. Further, under condition (ii) above,
Batyrev and Cox \rbatcox\ and Cox \rcox\ have shown
that the ring $P_0(\varLambda )$ is related to the chiral ring of the
{\bf B} model on $W$ in the following way.  The subring of the chiral
ring which is generated by polynomial deformations of complex structure
will be denoted by $\CR_B(W)$ and referred to as the {\it restricted chiral
ring}. This ring $\CR_B(W)$ is the quotient
of $P_0(\varLambda)$ by an ideal, with the construction corresponding
precisely to \eclearr\ and \echiral, in
terms of a certain expectation function on $P_0(\varLambda )$ defined by
Batyrev and Cox and a certain distinguished element in the ring
(the analogue of $\delta_0$).
We then appeal to the arguments given in section two
regarding expectation functions on graded rings to claim that the two
expectation functions on $P_0(\varLambda )$ are in fact identical
(up to a normalization), hence finally
that the specified subrings of the chiral rings of the two models agree.

The first relationship of this kind was found by Batyrev
\refs{\rBatHodge,\rbator},
who related a part of
the Hodge structure of the affine hypersurface
to the quantum cohomology of the ambient space of the mirror.
Our isomorphism is closely related to his.

Let us begin by noting some facts about the restricted chiral ring
$\CR_B(W)$. We recall
that $W$ is a hypersurface in the dual toric variety $\varLambda$
described in subsection {\it 5.1}. The homogeneous coordinates $y_1,\ldots
,y_u$
of $\varLambda$ correspond to the vertices of the polytope ${\cal P}^0$,
since we will not include in the fan $\nabla$ any other lattice
points. This choice ensures that the canonical class of $\varLambda$
is an ample divisor class. In terms of these, the hypersurface $W$ is
determined by an equation $f(y)=0$, with $f$ a polynomial of
appropriate multidegree.
We write $f$ in the form
\eqn\eQy{
f(y) = \sum_{i=0}^n  c_i \mu_i
\ ,}
for some coefficients $ c_i$, where $\mu_i$ runs over all the monomials
in $y$
of the correct multidegree. The coefficients $ c_i$ parameterize the
``toric part'' of the
deformation space, and can be regarded as homogeneous coordinates
on that space.  The ${\IC^*}^u$ action on the original
homogeneous coordinates $y_j$ induces an ${\IC^*}^d$ action on
$\varLambda$, hence on the coefficients $ c_i$.
The ``toric'' deformation space is a toric variety
of dimension $n{-}d$ which can be described as a quotient of the
space of coefficients by this ${\IC^*}^d$ action
(see \rAGM\ and \rmondiv\ for more details).

Under condition (ii), Batyrev and Cox \rbatcox\ show that
the restricted chiral ring $\CR_B(W)$ can be described as follows.
Start with  the ``affine Jacobian ideal'' of $f$, which is the ideal
$\CJ_0(f)$ generated by the ``affine'' partial derivatives $y_j {\partial
f\over\partial y_j}$, and define $\CJ_1(f)$ to be the ``ideal quotient''
of $\CJ_0(f)$ by the monomial $\mu_0:=\prod y_j$, that is,
\eqn\eidquot{
\CJ_1(f)= \CJ_0(f){:}\mu_0
= \{{\cal P}\ |\ \mu_0 {\cal P} \in \CJ_0(f)\}\ .
}
(In the case of a weighted projective space,
this ideal agrees with the ordinary Jacobian ideal $\CJ(f)$~--~the
one generated by the partial derivatives
$\partial f\over\partial y_j$~--~but in general $\CJ_1(f)$ is larger than
$\CJ(f)$.)
Then $\CR_B(W)$ is the subalgebra of
\eqn\erq{
\CR_1 (f) = \IC[y_1,\ldots ,y_u]/\CJ_1(f)\ ,
}
generated by all elements
whose multidegree is the same as that of $f$.\foot{Batyrev and Cox
actually show that the primitive cohomology of the hypersurface
is isomorphic to the subalgebra of $\CR_1(f)$ which consists of
all elements whose multidegree is a {\it multiple}\/ of that of $f$.
Our ``restricted chiral algebra'' is the subalgebra of this generated
by elements of multidegree precisely $|f|$; when the hypersurface
is a threefold this coincides with the primitive cohomology.}
The algebra $\CR_B(W)$ has natural monomial generators which
are in one-to-one correspondence with the holomorphic
sections of the anticanonical line bundle of $\varLambda$, i.e., with the
points in $\cal P$. Condition (i) then means these are precisely the
generators of the
one-dimensional cones in $\Delta$,\foot{We ignore the subtlety
associated to points in the interior of codimension-one faces. This
can be incorporated at the price of a somewhat more cumbersome
discussion.} as well as the unique interior point of $\cal P$.
To each cone, in turn, we associated one of the homogeneous
coordinates $x_i$ on $V$, and the corresponding divisor $\xi_i$. Let
us make this correspondence more concrete. The generators of
one-dimensional cones
in $\Delta$ were denoted $v_1,\ldots ,v_n$. Let $v_0=0$, and let
$v_i^+$ denote the ``promoted'' $v_i$, a vector in $\bN_{\IR}^+$
whose last component is $1$.
Similarly, let $\lambda_1,\ldots, \lambda_u$ be the generators of
one-dimensional cones in
$\nabla$, and $\lambda_0$, $\lambda_j^+$ as above. Then to $v_i$ we
associate the monomial
\eqn\emd{
\mu_i = \mu_i(y)=\prod_{j=1}^u y_j^{\la\lambda_j^+, v_i^+\ra } =
\prod_{j=1}^u y_j^{1+\la \lambda_j, v_i\ra }\ ,
}
such that $y_0 \mu_i$ is a possible term in the superpotential for the
$W$ model.
It is convenient to include the coefficient $ c_i$ along with
the monomial $\mu_i$, and to present the algebra $\CR_B(W)$ as
\eqn\ePQ{
\CR_B(W) = \IC[ c_0 \mu_0,\ldots, c_n \mu_n]/\CJ_1 '(f)
\ ,}
where $\CJ_1 '(f)$ is the inverse image of $\CJ_1(f)$ under the map
\eqn\emap{
\IC[ c_0 \mu_0,\ldots, c_n \mu_n] \to \IC[y_1,\ldots ,y_u]
}
induced by \emd.

Among the relations in $\CJ_1 '(f)$ are those in the kernel of \emap, which
follow simply from the expression \emd\
for the $\mu_i$. These are generated by relations corresponding to
sets $n^*_1,\ldots ,n^*_{n-d}$ such that $d_i^* = \sum_{a=1}^{n-d} Q_i^a
n^*_a \ge 0\ \forall i>0$, given by
\eqn\enlrelbb{
\prod_{i=1}^n  \mu_i{}^{d_i^*} =
\mu_0{}^{\Sigma d_i^*}\ ,
}
or, written in terms of the
generators of $\IC[ c_0 \mu_0,\ldots, c_n \mu_n]$,
\eqn\enlrelb{
\prod_{i=1}^n ( c_i \mu_i)^{d_i^*} =
\left(  c_0 \mu_0\right)^{\Sigma d_i^*}
\prod_{i=1}^n \left( { c_i\over c_0}\right)^{d_i^*}
\ .}
Further relations in  $\CJ_1 ' (f)$ come from the
``affine'' derivatives
$y_j {\p f\over \p y_j}$. These are of degree $|f|$ and can thus be
themselves written as linear combinations of the $\mu_i$'s,
leading to {\it linear}\/
relations among the latter. There are $u$ such relations, but the
homogeneity of $f$ implies the existence of
Euler equations relating these, so that
precisely $d{+}1$ are independent. It is convenient to write these
relations in
terms of the one-parameter group actions on $\varLambda^+$.
Given $m^+\in \bM^+$, \emd\ and \eQy\ yield as the linear relations in
$\CJ_1 '(f)$ the following:
\eqn\elrelb{
\sum_{i=0}^n \,\la m^+,v_i^+\ra \,  c_i \mu_i = 0,\ \ p=1,\ldots ,d\ .
}
Now we see why the coefficients $ c_i$ were included with the
generators~--~doing so gives the linear relations \elrelb\ a universal form,
independent of the choice of coefficients.

The relations derived in the previous paragraph do not generate the entire
ideal $\CJ_1'(f)$ (because we have used the ``affine'' derivatives
rather than $\p f\over \p y_j$, and because we have not yet included the
``extra'' elements which enlarge $\CJ_1(f)$ beyond the Jacobian ring
$\CJ(f)$).
We let $\CJ_0(f)\subset \IC[y_1,\ldots ,y_u]$ be the ideal
generated by the affine derivatives $y_j {\p f\over \p y_j}$ (note that
$\CJ_0(f)\subset\CJ(f)\subset\CJ_1(f)$), and
let the {\it extended chiral ring}\/ $P_0(\varLambda)$ be the subalgebra of
$\IC[y_1,\ldots ,y_u]/\CJ_0(f)$ generated by all elements whose degree
is divisible by $|f|$.  This extended chiral ring can be presented as
\eqn\ePQextended{
P_0(\varLambda)=\IC[ c_0 \mu_0,\ldots, c_n \mu_n]/\CJ_0'(f)\ ,}
where $\CJ_0'(f)$, the inverse image of $\CJ_0(f)$ under \emap ,  is
generated by \enlrelb\ and \elrelb.
According to \rbatcox, this ring is isomorphic to
the mixed Hodge structure on the $d$th primitive cohomology of
the affine hypersurface
$W\cap T^\vee$
(which had been studied by Batyrev \rBatHodge\ using a different
presentation).

\nref\rGZK{
  I. M. Gel'fand, A. V. Zelevinski\u\i, and M. M. Kapranov,
``Discriminants of
  polynomials in several variables and triangulations of Newton polyhedra,''
  Leningrad Math. J. {\bf 2} (1991) 449--505 (from Russian in
Algebra i analiz {\bf 2}).}%

The isomorphism between $\CR_0(V)$ and $P_0(\varLambda)$ is given by the
simple mapping
\eqn\emondiv{\eqalign{
\delta_i &\leftrightarrow  c_i \mu_i\cr
q_a &\leftrightarrow
\prod_{i=1}^n ( c_i/ c_0)^{Q_i^a}\ ,\cr
}}
under which the relations \erelone\ are mapped to \enlrelb,
and the relations \eok\ and \exio\ are mapped to \elrelb.
This equation is precisely the {\it monomial-divisor mirror map\/}
(up to sign\foot{As we have discussed earlier, although our comparison
between {\bf A} and {\bf B} model correlation functions involves no signs,
some signs are needed
in order to correctly compare linear and nonlinear $\sigma$-models.
We have given a formula for these in \etpert . This differs from a
conjectured form of the signs given in \rsmall,
based upon ideas from \rGZK.}) of \rmondiv ,
and is closely related to maps appearing in \rbator.
In those works
only the asymptotic behavior of the map was
studied in the {\bf A} model; we see here that the $q_a$ give a
precise meaning to this map throughout parameter space. For this reason,
we refer to \emondiv\ as the {\it global}\/ monomial-divisor mirror map.

The fact that the relations for $\CR_0(V)$~--~which were derived from
studying instanton moduli spaces and quantum cohomology~--~are exactly
the same as those for $P_0(\varLambda)$~--~derived from the Jacobian
ring of a polynomial~--~is quite remarkable, in fact,
completely unexpected from a mathematical point of view.
There is an additional remarkable correspondence.  As we have seen in
subsections {\it 3.5}\/ and {\it 4.2}, the set of $q_a$'s for which the
$V^+$ model is singular has a very explicit description, given by
\ealborm\ and \ealborbis . On the {\bf B} model side
there is an equally explicit description of the singularities \rGZK.  The
hypersurface with equation $\sum c_i\mu_i=0$ fails to be quasi-smooth
precisely when $E_A(c_i)=0$, where $E_A$ is a polynomial called
the {\it principal $A$-determinant}\/ of the
set $A:=\{\lambda_0, \lambda_j^+\}$.  Each irreducible component of
the set $E_A=0$ is associated to a face $\Gamma$ of $\cal P$ (including
the ``face'' $\Gamma={\cal P}$), and coincides with the zero-locus of
another polynomial called the {\it $A$-discriminant}\/
$\varDelta_{A\cap\Gamma}$.  We saw a similar structure on the
${\bf A}$ model side in subsection {\it 3.5}, in which faces
of the polytope ${\cal P}$ were associated to components of the singular locus.

\nref\rhorn{J. Horn, ``Ueber die Convergenz der hypergeometrischen Reihen
zweier und dreier Ver\"anderlichen,'' Math. Ann. {\bf 34} (1889) 544--600.}%
\nref\rKap{
M. M. Kapranov, ``A characterization of $A$-discriminantal hypersurfaces
in terms of the logarithmic Gauss map,'' Math. Ann. {\bf 290} (1991)
277--285.}%

In addition, the structure of the
components $\varDelta_{A\cap\Gamma}=0$ is known in detail.
The main idea goes back to a 19th century paper of Horn \rhorn,
but it has been put into modern form by Kapranov \rKap: taking
as coordinates on the parameter space the expressions
$p_a:=\prod_{i=1}^n(c_i/c_0)^{Q_i^a}$ (which correspond to $q_a$
under the monomial-divisor mirror map), the zeros of the principal
discriminant
$\varDelta_{A}(p_a)$ are precisely the $p_a$'s for
which
\eqn\epa{
p_a = \prod_{i
} \left(
\sum_{b=1}^{n-d} Q_i^b\ell_b\right)^{Q_i^a}
}
for some values of $\ell_b$.
The other components (with defining polynomials $\varDelta_{A\cap\Gamma}(p_a)$)
admit a similar description: in appropriate coordinates,
they are the $p_a$'s for which
\eqn\epabis{
p_a = \prod_{i
\in I
} \left(
\sum_{b=n{-}d{-}k{+}1}^{n-d} Q_i^b\ell_b\right)^{Q_i^a} \ ,\
a = n{-}d{-}k{+}1,\ldots, n{-}d\ ,
}
for some values of $\ell_b$,
where $\{\mu_i\}_{i\in I}$ is the set of monomials in the face $\Gamma$
of ${\cal P}$, where $Q_i^a=0$ for all $i\not\in I$ and $a > n{-}d{-}k$, and
where $\vec{Q}^{ n{-}d{-}k{+}1}$, \dots, $\vec{Q}^{ n{-}d}$ span the
set of all $\vec{Q}$'s for which $\vec{Q}_i=0$ $\forall$ $i\not\in I$.
But this exactly corresponds to the description of the components
of the singular locus we found on the ${\bf A}$ model side!

We turn now to the expectation functions on the algebra
$\CR_0(V) \cong P_0(\varLambda)$.  This is a graded Frobenius algebra,
and as discussed in section two, has a graded expectation function which
is unique up to scalar multiple.  Each of our descriptions of this
algebra comes equipped with an expectation function: the ``trace''
map $\Tr$ given by \etrt\ defines a graded expectation function on $\CR_0(V)$,
and the ``toric residue'' $\Res$ of \rcox\ defines a graded expectation
function on $P_0(\varLambda)$.  The scalar multiple which relates these
may in principle depend on $q$, so using the
isomorphism \emondiv\ we have
\eqn\etraces{
\Tr(\CO) = s(q)\, \Res(\CO)\ ,
}
for some scaling function $s(q)$.

\nref\rHY{T. H\"ubsch and S.-T. Yau,
``An $SL(2,\IC)$ action on certain Jacobian rings and the mirror map,''
in {\sl Essays on Mirror
Manifolds\/}, S.-T. Yau, ed. International Press, Hong Kong, 1992,
pp.~372--387.}%

Under the global monomial-divisor mirror map \emondiv, the special
class $\delta_0$ is mapped to the special monomial $c_0\mu_0=c_0\prod y_j$.
The quantum cohomology ring of $M$ is determined by insertion of
$(-\delta_0)$ as described in \eclearr; remarkably, when $\varLambda$ is
simplicial it follows from results
in \refs{\rbatcox,\rcox} that the expectation
function $\lal\,\rar_B$ on the restricted chiral ring of $W$
is given by\foot{The r\^ole of the ``fundamental monomial'' $c_0\mu_0$
in understanding the chiral ring of a hypersurface was first
considered by H\"ubsch and Yau \rHY.}
\eqn\ecbc{
\lal \CO \rar_B = \Res\left( (-c_0\mu_0)\CO \right)\ .
}
In fact, as mentioned earlier the restricted chiral ring of $W$ is described in
\refs{\rbatcox,\rcox}
as the subalgebra of
$\IC[y_1,\ldots ,y_u]/\CJ_1(f)$ generated by all elements whose degree
is divisible by
$|f|$, where
$\CJ_1(f)$ is the ideal quotient $\CJ_0(f){:}\mu_0$ defined in \eidquot.
When $c_0\ne0$
this is clearly the same as modding out by the annihilator of $c_0\mu_0$,
as we expect from Nakayama's theorem.

Comparing \ecbc\ to \eclearr\ we conclude that under the
monomial-divisor mirror map we have
\eqn\mirror{
\lal \CO \rar_A = s(q) \lal \CO\rar_B\ ,
}
where the subscript on the left-hand side is added for clarity.

Mirror
symmetry na\"{\i}vely predicts $s(q)=1$. But of course this ignores the
fact that correlation functions are sections of a line bundle over
parameter space and thus defined as functions only after a choice of
trivialization. In this sense \mirror\ is a proof of mirror symmetry
and the mirror map used (even if $s(q)$ is nontrivial).
In fact, we claim that the mirror map can be
extended to mirror choices of trivializations as well; in other words,
we can give an interpretation in the context of the {\bf B} model of
the gauge choice we made implicitly in the {\bf A} model. The gauge in
question is what we will call {\it algebraic gauge\/} and is defined
(in part) by the property
that correlation functions be meromorphic functions.
(This is not possible in arbitrary coordinate systems but is manifestly
possible in the $q$ coordinates.) This property restricts the unknown
function $s(q)$
to being a meromorphic function. We can use the compactification of
parameter space as a toric variety \rmondiv\ to claim that such a
function should be determined by its singularities. Moreover, we have found
the exact location for the singularities on both the ${\bf A}$ and
${\bf B}$ model sides, and they correspond under our map.
This means that $s(q)$ is a meromorphic function whose zeros and poles
are all located along the components of the singular locus (since a zero
or pole elsewhere would cause a mismatch between singularity sets of
the two theories).  It
seems likely that $s(q)$ must in fact be a constant, but we have
no general proof of this.  In any case, though, the monomial-divisor
mirror map determines a natural gauge on the ${\bf B}$ model side,
which corresponds to the ${\bf A}$ model's algebraic gauge under
the monomial-divisor mirror map.


Note that our hypotheses (i) and (ii) have been used in the following
way:  (i) allowed us to give a relatively simple
 algebraic description of the ring $\CR_0(V)$ and the correlation
function $\lal\ \rar_A$, and (ii) allowed
us to give a similar a description of the extended chiral ring
$P_0(\varLambda)$ and the correlation function $\lal\ \rar_B$.
We would need extensions of these descriptions on both sides
in order to verify the isomorphism in a wider context.

There is a class of examples in which we can make the preceding
discussion much more concrete. This is the case in which the polynomial
$f$, considered as a function $\IC^u\to\IC$, can be chosen to be
{\it nonsingular}\/ away from the origin in $\IC^u$. This condition
must hold, for example, if the $W$ model has a Landau--Ginzburg phase.
In this case the toric residue $\Res$ coincides with
the Grothendieck residue formula, and provides a natural
expectation function on $\CR_B(W)$ of the form
\eqn\egrot{
\lal\CO\rar_B = \oint \prod_j dy_j\, {\CO(y)\over \prod_j {\p f\over\p
y_j}}
\ .}
This is manifestly in algebraic gauge.
Given the explicit formula \egrot\ we see that this is obtained by an
application of \etraces\ (with $s(q)$ constant)
from an expectation function  on
$P_0(\varLambda)$ given by
\eqn\egrotone{
\Tr (\CO) = -{1\over c_0}
 \oint \prod_j dy_j\, {\CO(y)\over \prod_j y_j {\p f\over\p
y_j}}\ .
}

\newsec{Open Problems}

The results in this paper are an application of the proposal of
\rphases\ that the nonlinear sigma model for target spaces related to
toric varieties can be studied with the aid of the massive Abelian
gauge theory which reduces to it at extremely low energies. We have
seen that for the twisted {\bf A} model with target space $V$ a toric
variety or $M\subset V$ a \CY\ hypersurface these ideas suffice to
lead to an essentially complete solution. The solution is
unsatisfactory only in being naturally obtained in a set of
coordinates which, in the hypersurface case, differ from the canonical
set. To complete the picture one should compute the required change of
coordinates using the methods described. Essentially, the coordinate
change is encoded in the combinatorics of the relation between the
compactified moduli spaces of instantons used here and the actual
mapping spaces obtained in the nonlinear model. Since this relation is
explicitly known it should be possible to extract the coordinate
change.

At a technical level this work raises quite a few questions. The two
formulations of sections four and five yield two expressions for the
quantum cohomology of $M$ in terms of the quantum cohomology of $V$.
Consistency of the construction requires agreement of these two
formulas, but we have not explicitly showed this. The
generator-relation presentation of the quantum cohomology algebra for
a smooth $V$ should have a generalization to a general toric variety.
It appears the relations are not integrable into a twisted
superpotential as in the smooth case. Integrating them would thus
require the introduction of auxiliary fields (in addition to
$\sigma$). It would be of interest to characterize these and perhaps
their physical interpretation.

We have obtained useful results for the case of \CY\ hypersurfaces in
toric varieties. The GLSM can describe in the deep infrared nonlinear
models on arbitrary complete intersections of such hypersurfaces, and
it should be a simple matter to extend our results to this case. In
particular, the description of the instanton moduli spaces should be
no harder to obtain. We note, however, that the arguments leading to
the quantum restriction formula would not hold in this case;
presumably the validity of this is restricted to hypersurfaces. We see
no such obstruction, however, to applying the methods of section five.
Similarly, we have restricted our attention to \CY\ subvarieties
because these are the ones relevant for string theory. The methods of
section four however should lead to an explicit computation of the
quantum cohomology of {\it any\/} hypersurface in a toric variety.

The massive GLSM was crucial for obtaining all of the explicit results
on which this work is based. It is however not clear how many of the
qualitative conclusions depend upon the relation to toric varieties.
For example, the arguments leading to \eqsthg\ could probably be
reproduced for a \CY\ hypersurface $M$ in an arbitrary Fano variety
$X$, leading to a computation of the quantum cohomology of $M$ in
terms of the quantum cohomology of $X$. It would be interesting to
make this construction explicitly, as well as to verify to what extent
the same holds for the methods of section five. One can imagine adding
a superpotential interaction to the action for the {\bf A} twisted
model with target space $X^+$ the total space of the canonical line
bundle over $X$. This could be chosen so that
the space of classical vacua reduces to $M\subset X$.

Beyond the ability to solve particular examples the interest in mirror
symmetry arises from the hope that it reflects a general property of
string theory. Our current state of understanding of the
phenomenon~--~we know a few examples of mirror pairs~--~does not allow
us to address this question at all. To this end the results of the
present work are at best a small step. A more meaningful one would be
a manifestly mirror-symmetric construction of the models. The
description of the dynamics of $\Sigma_a$ directly in terms of the
twisted superpotential is a promising hint in this direction, which we
will pursue further in \rdoit . Of course, this is at present limited
to models related to toric varieties.

\bigbreak\bigskip\bigskip\centerline{{\bf Acknowledgements}}\nobreak
We thank Edward Witten for many discussions concerning both \rphases\ and
this paper, for providing us with his computations for example 1, and
for suggesting that we try another example.
We also thank P. Aspinwall, P. Berglund, D. Cox, J. Distler, B. Greene,
K. Intriligator, S. Katz,
G. Moore, E. Rabinovici, and N. Seiberg for useful discussions.
D.R.M. thanks the
School of Natural Sciences, Institute for Advanced Study for hospitality
during the preparation of this paper, and M.R.P. thanks the Rutgers
theory group for hospitality and useful discussions.  The work of D.R.M.
was supported by National Science Foundation grants DMS-9304580 and
DMS-9401447, and by an American Mathematical Society Centennial Fellowship.
The work of M.R.P. was supported by  National Science Foundation grant
PHY-9245317 and by the W.M. Keck Foundation.

\appendix{A}{Some Details Concerning Example 2}

We want to compute
\eqn\eyjna{
Y_j^{(n)} = 4^{4n_1+1} \la y^j z^{4n_1+4-j}\ra _{\CM_n}\
}
using the recursion relations \erecf\ that follow from
\eqn\enlrelma{
\eqalign{
y_1^{n_2+1} y_2^{n_2+1} &= y^{2n_2+2} = 0\cr
y_6^{n_1-2n_2+1}y_3^{n_1+1}y_4^{n_1+1}y_5^{n_1+1} &=
z^{3(n_1+1)}(z-2y)^{n_1-2n_2+1} =0\ .\cr
}}
So denote as always
\eqn\efaa{
\varphi_a = 2^{-a} \la y^{2n_2+1-a} z^{4n_1-2n_2+3-a}\ra _{\CM_n}
}
with $\varphi_a=0$ for $a<0$ and $\varphi_0=1$. Then \enlrelma\ implies
\eqn\erecfa{
2^{-a}\varphi_a + \sum_{j=1}^{n_1-2n_2+1} (-2)^j \left( {n_1-2n_2+1\atop
j}\right)
2^{a-j} \varphi_{a-j} = 0}
for $1\leq a\leq 2n_2+1$.
These relations determine the $\varphi_a$ completely. We claim this is solved
by
\eqn\esolfa{
\varphi_a = \left( {n_1-2n_2+a\atop a}\right)\ .
}
The proof is trivial. We need to show
\eqn\eprfa{
\left( {m+a\atop a}\right) + \sum_{j=1}^{m+1} (-1)^j\left( {m+1\atop j}\right)
\left( {m+a-j\atop a-j}\right) = 0.
}
Note that (a) we can incorporate the first term into the sum as $j=0$;
(b) this is the same as what we need as indeed $\left( {m\atop n}\right)$
vanishes for $m>0,\ n<0$. With the (true) observation that $n!$
diverges for $n<0$ we rewrite this as
\eqn\eprfb{
{1\over m!} \sum_{j=0}^{m+1} \left( {m+1\atop j}\right)
{(m+a-j)!\over(a-j)!}
}
where all negative factorials are to be understood as Gamma functions.
However, if we let
\eqn\eprfc{
F(x) = x^{m+a}(1-{1\over x})^{m+1} = x^{a-1}(x-1)^{m+1}
}
then expanding yields
\eqn\eprfd{
F(x) = \sum_{j=0}^{m+1} (-1)^j \left( {m+1\atop j}\right) x^{m+a-j}
}
hence
\eqn\eprfe{
{d^m F\over dx^m} = \sum_{j=0}^{m+1} \left( {m+1\atop j}\right)
{(m+a-j)!\over(a-j)!} x^{a-j}
}
(with same understanding as above). Setting $x=1$ we have a proof of
\eprfa.

\appendix{B}{A Singular Example}

In this appendix we perform the instanton sum calculation in an
example for which $V$ is not smooth. Thus Batyrev's algebraic solution to
the quantum cohomology problem is not valid. We show that the
instanton computations are well-defined, and sum to the result
predicted by mirror symmetry, in this case as well. The mirror
manifold was studied in \rHKTY .

We start with \WCP{4}{1,2,2,3,4} (in which a degree 12 hypersurface
is \CY). This has a codimension two set of $\IZ_2$ singularities at
$x_1=x_3=0$. It also has codimension four singularities at
(0,0,0,1,0) ($\IZ_3$) and (0,0,0,0,1) ($\IZ_4$). The first we resolve
easily; the resulting fan is spanned by
\eqn\efan{\eqalign{
v_1 &= (-2,-2,-3,-4)\cr
v_2 &= (1,0,0,0)\cr
v_3 &= (0,1,0,0)\cr
v_4 &= (0,0,1,0)\cr
v_5 &= (0,0,0,1)\cr
v_6 &= (-1,-1,-1,-2) = 1/2(v_1+v_4)\ .\cr
}}
Find the kernel to compute the charges
\eqn\eQQ{
Q = \left(\matrix{-1&1&1&0&2& 3&-6\cr
                   1&0&0&1&0&-2& 0\cr}\right)\
}
(adding the seventh $x_0$ column).

Using this find the phase structure straightforwardly.
There are five phases, as follows (we give cone in $r$ space as well as
the indices of the primitive collections of coordinate hyperplanes)

\item{I} CY, $r_1>0,\ r_2>0$; (14),(2356).

\item{II} \CP4, $r_2<0,\ 3r_2+2r_1>0$; (6),(12345).

\item{III} \CP4 with exoflop, $r_1<0,\ r_2+r_1>0$; (10),(23456).

\item{IV} Hybrid, $r_2+r_1<0,\ r_2>0$; (0),(14).

\item{V} LG, $r_2<0,\ 3r_2+2r_1<0$; (0),(6).

To see the failure of the algebraic solution we concentrate on the
smooth \CY\ phase; the classical relations are (from $F$)
\eqn\enonll{\eqalign{
\xi_1\xi_4 &=0\cr
\xi_2\xi_3\xi_5\xi_6 &=0\ .\cr
}}

We have
$d=(n_2-n_1,n_1,n_1,n_2,2n_1,3n_1-2n_2,-6n_1)$ (again include
$d_0$ last). The cone ${\cal K}^+$ of
instantons such that all $d_i\geq 0$ is thus spanned by
(1,1) and (2,3) which lead to the relations (eqn.~\erelone )
\eqn\eqnonl{\eqalign{
\delta_2\delta_3\delta_4\delta_5^2\delta_6 &= q_1q_2\delta_0^6\cr
\delta_1\delta_2^2\delta_3^2\delta_4^3\delta_5^4 &= q_1^2 q_2^3 \delta_0^{12}\
.\cr
}}
Dividing second by square of first indeed get $\delta_1\delta_4 =
q_2\delta_6^2$
as expected. However, these do not suffice to obtain the quantum
cohomology of $M$.

To compute the correlators we will work in the Landau--Ginzburg phase.
This leads to the simplest computation and by analytic continuation
determines the correlators in all phases.
We have ${\cal K}_{\rm LG}^\vee = \{n_1<0,\ 3n_1-2n_2>0\}$ (the form
of $F$ here
shows that we need the strong inequalities). From $d$ we see that the
moduli space is $\CM_\vn = \CP{3n_1-2n_2}\times\CP{-6n_1}$ with the
hyperplane sections of the two factors given by $\xi_6,\xi_0$
respectively. The general correlation function is given by
\eqn\one{
\lal\CO\rar = \sum_{\vn\in\CK^\vee} (-q_1)^{n_1}q_2^{n_2}
\la\CO\xi_0^2\xi_1^{n_1-n_2-1}
\xi_2^{-n_1-1}\xi_3^{-n_1-1}\xi_4^{-n_2-1}\xi_5^{-2n_1-1}\ra_\vn
\ .}
Now rewrite everything in terms of $\xi_0,\xi_6$ using linear
relations, and use $\CO_j = \xi_0^{3-j}\xi_6^j$ as a basis, find
\eqn\two{
\lal\CO_j\rar = 2^5 3^4\sum_{\scriptstyle{m_1>0}\atop\scriptstyle
2m_2>3m_1} r_1^{m_1} r_2^{m_2}
\la\xi_0^{4m_1-j+2}\xi_6^j
(\xi_0+2\xi_6)^{m_2-1}(\xi_0+6\xi_6)^{m_2-m_1-1}\ra_\vn
\ }
where $r_1 = {1\over 27 q_1},\ r_2 = {1\over 48  q_2}$.
This now needs evaluation. From the expression for $\CM_\vn$ we can
rewrite it as
\eqn\three{
\lal\CO_j\rar = 2^5 3^4 \sum_{\scriptstyle{m_1>0}\atop\scriptstyle
2m_2>3m_1} r_1^{m_1} r_2^{m_2}
\oint dx x^{3m_1-2m_2+j-1} (1+2x)^{m_2-1}(1+6x)^{m_2-m_1-1}
\ .}
Now exchange the sum with integral and sum the geometric series that
appear to get
\eqn\four{
\lal\CO_j\rar = 2^5 3^4\oint dx \, x^{j-1}
{ x^{-2}r_1^2r_2^4(1+2x)^3(1+6x)+x^{-1}r_2^2(1+2x)\over
\left[ 1-r_2x^{-2}(1+2x)(1+6x)\right]
\left[ 1-r_1^2r_2^3(1+2x)^3(1+6x)\right] }
}
where the two terms in numerator arise from odd and even $m_1$
respectively. For $j=0$ the integrand has a pole at $x=0$. For any
$j$, however, there are poles at the roots of $x^2=r_2(1+2x)(1+6x)$.
These are of order $r_2$ and hence should be considered encircled by
our contour. The correct Yukawas (which match the {\bf B} model
ones calculated in \rHKTY ) follow from simply
adding up the residues about these poles.

\listrefs

\bye